\documentclass[a4paper,12pt]{article}
\usepackage{jheppub}
\usepackage[english]{babel}
\usepackage{amsmath}
\usepackage{amssymb}
\usepackage{caption}
\usepackage{subcaption}
\usepackage{braket}
\usepackage{hyperref}
\usepackage{multirow}
\usepackage[toc,page]{appendix}
\usepackage{float}
\restylefloat{table}
\usepackage{lipsum}
\usepackage{graphicx}
\usepackage{xcolor}
\usepackage{float}
\usepackage{hyperref}
\usepackage{slashed}
\usepackage{cleveref}
\usepackage{dsfont}
\usepackage{tikz-feynman}
\usepackage[labelfont=bf]{caption}
\usepackage{soul}
\usepackage{float}
\usepackage{pdflscape}

\def\be{\begin{equation}}
\def\ee{\end{equation}}
\def\tr{\operatorname{tr}}

\newcommand{\ra}[1]{{\textcolor{red}{ a_#1}}}

\newcommand{\bb}[1]{{\textcolor{blue}{ b_#1}}}

\newcommand{\gc}[1]{{\textcolor{green}{ c_#1}}}

\newcommand{\nn}{\nonumber}
\newcommand{\hlambda}{{\hat \lambda}}
\newcommand{\bPhi}{\bar \Phi}
\newcommand \tkappa{\tilde \kappa}

\newcommand{\vphi}{\varphi}
\renewcommand{\tr}{{\rm tr}}

\title{Extremal fixed points and Diophantine equations}

\date{}
\author{Christopher P.\ Herzog$^a$, Christian B. Jepsen$^b$, Hugh Osborn$^c$ and Yaron Oz$^d$}
\affiliation{${}^a$ Department of Mathematics, King's College London, Strand, London, WC2R 2LS, UK
\\
${}^b$ School of Physics, Korea Institute for Advanced Study, Seoul 02455, Korea
\\
${}^c$ Department of Applied Mathematics and Theoretical Physics, Wilberforce Road, Cambridge CB3
0WA
\\
${}^d$ School of Physics and Astronomy, Tel Aviv University, Ramat Aviv 69978, Israel}

\abstract{ 
\noindent The coupling constants of fixed points in the $\epsilon$ expansion at one loop are known to satisfy a quadratic bound due to Rychkov and Stergiou. We refer to fixed points that saturate this bound as \emph{extremal fixed points}. The theories which contain such fixed points are those which undergo a saddle-node bifurcation, entailing the presence of a marginal operator. Among bifundamental theories, a few examples of infinite families of such theories are known. A necessary condition for extremality is that the sizes of the factors of the symmetry group of a given theory satisfy a specific Diophantine equation, given in terms of what
we call the extremality polynomial. 
In this work we study such Diophantine equations and employ a combination of rigorous and probabilistic estimates
to argue that these infinite families constitute rare exceptions. The Pell equation, Falting's theorem, Siegel's theorem, and elliptic curves figure prominently in our analysis.
In the cases we study here,  more generic classes of multi-fundamental theories saturate the Rychkov-Stergiou bound only in sporadic cases or in limits where they degenerate into simpler known examples.
}

\begin{document}

\maketitle

\section{Introduction}

A fundamental dilemma in theoretical physics is that computable models of the real world usually have a high degree of symmetry that the real world lacks.  The concept of universality provides a resolution to this dilemma for those fortunate properties of physical systems that are insensitive to the microscopic, symmetry breaking details.  Conformal field theories provide an excellent example of universality.  The dilemma is resolved because conformal symmetry emerges in the real world near second order phase transitions and gives rise to universal behavior that can be modeled by conformal field theories.  In this work, exploring the landscape of conformal field theories, we consider an epicycle of this dilemma that occurs when the conformal field theory has additional global symmetries.  We ask, which properties of the conformal field theory are artefacts of looking at simple systems with large amounts of global symmetry and which are generic?

The task of charting out the full landscape of possible renormalisation group (RG) fixed points, or conformal field theories,  is an ongoing field of research, and finding all fixed points in general theories is not easily achieved. For further exploration of the fixed point {\it terra incognita} we consider here scalar theories in the $\epsilon$ expansion. Many examples of fixed points for scalar theories in $4-\epsilon$ spatial dimensions are known at one loop, but even though determining the existence of fixed points in this context reduces to solving algebraic equations, a complete analysis of all such fixed points is far from realised. The known examples either have low values of the number of scalar fields $N$, or they come in infinite families but enjoy a high degree of global symmetry, possessing typically but one or two quartic operators. As these latter cases furnish a special computationally tractable subset of the more generic fixed points,  it is in this context that we ask the question that ended the previous paragraph.  Within this subset, we are able to decrease the amount of global symmetry and consider the effects.

More specifically, we direct this question towards the specific property of a fixed point to be \emph{extremal}. Our aim then is to identify fixed points where the couplings $\lambda_{ijkl}$  to lowest nontrivial order in the $\epsilon$-expansion saturate the Rychkov-Stergiou bound 
\cite{Rychkov:2018vya}, which (after an appropriate rescaling) reads
\begin{align}
\label{RychkovStergiou}
|| \lambda ||^2 \equiv \lambda_{ijkl}\,\lambda_{ijkl} \leq \frac{1}{8}N\,.
\end{align}
Such families with extremal fixed points are physically significant because they undergo a fold-bifurcation (see e.g. \cite{Gukov:2016tnp}), and possess an associated marginal operator: extremality implies marginality.\footnote{%
The converse is not true. Although an operator becomes marginal, so that the scaling dimension is exactly the spatial dimension $d$, at a bifurcation where fixed points are created or annihilate, marginal operators can occur more generally to leading order in the $\epsilon$-expansion. If $O(N)$ is reduced to a subgroup $G$, there are $\dim O(N) - \dim G$ zero modes for the full anomalous dimension matrix $\frac{\partial}{\partial \lambda} \beta$ defined over the space of all couplings and evaluated at the fixed point. Although these operators are marginal at leading order in the epsilon expansion, they generally  gain an anomalous dimension at higher orders, since they become descendants of the broken symmetry currents defined by Lie algebra generators in $\mathfrak{o}(N)$ not in $\mathfrak {g}$ \cite{Rychkov:2018vya}, and  whose scaling dimension is no longer constrained by conformal symmetry to be $d-1$.
 }
 
Within a family of theories defined via its type of symmetry group, finding the subset of fixed points that realise the Rychkov-Stergiou bound is more tractable mathematically than identifying the full set and may potentially deepen our understanding of the possibilities of fixed points more generally. The known examples all lead to fixed point couplings which are rational, unlike the numerical searches of RG fixed points for low values of $N$, where in most cases the results for the couplings are irrational. The arguments in this paper rely on an interplay between polynomial conditions in a theory and symmetries present in the underlying model and are of a general nature that can hopefully be applied to other perturbative expansions. Beyond lowest order in the $\epsilon$-expansion, extremal fixed points do gain higher order contributions, which shift the bifurcation point, but we have little to say about them here.

For scalar theories with $N$ real scalar fields $\phi^i$ in $4-\epsilon$ dimensions the maximal symmetry is  $O(N)$ but this may be reduced to a subgroup $G \subset O(N)$ by adding classically marginal quartic monomials  to the Lagrangian, where each term corresponds to an additional coupling. In this work, to keep the analysis controlled, we assume $G$ is a product  of $O(N_i)$, $U(N_j)$,  and $Sp(N_k)$ groups or a product of one of these with the permutation group $\mathcal{S}_n$. Before describing our findings, let us review the known examples of extremal fixed points. For the $O(N)$ model itself, the perturbative fixed point has a marginal operator in the case when $N=4$ \cite{Brezin:1973jt}. There is also a solution with $N=5$ which has a discrete hypertetrahedral symmetry. For the bifundamental theories, where the scalars belong to the product of the fundamental representations of $G_{N_1},G_{N_2}$ and the symmetry group is $G_{N_1}\times G_{N_2} \subset O(N)$,  there exist three infinite families of values $(N_1,N_2)$ for which the bound is saturated, and there is a corresponding marginal operator \cite{Osborn:2017ucf,Rychkov:2018vya,Osborn:2020cnf,Kousvos:2022ewl,Osborntalk}. Such theories are realised with just two couplings.

We search for new extremal fixed points by investigating theories with more than two couplings and with more factors $G_{N_i}$ in the symmetry group. It might be hoped these theories would lead to further solutions of the extremality condition so that they would proliferate with increasing numbers of the integer parameters $N_i$. In fact the opposite appears to be true, as we will demonstrate in the examples we explore here, and argue for more generally. Extremal fixed points are apparently a peculiarity of the theories with one, two, and occasionally three quartic couplings. For more generic families with less symmetry, the number of extremal fixed points is with overwhelming probability equal to zero, except in limits where these more complicated theories degenerate to simpler ones.

 In addition to identifying extremal fixed points as a kind of lamppost effect, our study of complicated, more generic theories also sheds light on properties of the stability matrix $\frac{\partial}{\partial \lambda} \beta$. The eigenvalues $\{\kappa \}$ of this matrix at any fixed point are characteristic of the fixed point and independent of how the fixed point is realised as an end point of RG flow. Marginal operators of course have $\kappa=0$.  Previously \cite{Osborn:2020cnf}
  it appeared that all $\kappa$ at lowest order were in the range $[-1,1]$ but here we illustrate examples with $\kappa>1$ and also $\kappa<-1$.  The values $\kappa=\pm 1$ have broader significance.
 Our fixed points always have at least one eigenvalue $\kappa=1$ corresponding to perturbing the fixed point by an eigenvector parallel to the fixed point couplings.  (Two eigenvectors with $\kappa=1$ signal the presence of decoupled theories at the fixed point.)  On the other hand $\kappa=-1$ occurs when a decoupled free sector is present at the fixed point, see e.g.\ \cite{Osborn:2020cnf}.

Analysing theories with less symmetry is more challenging mathematically but also provides us with the opportunity to introduce to the physics literature new results from algebraic geometry. The condition for extremality that (\ref{RychkovStergiou}) is saturated, along with the conditions that the lowest-order beta function for each coupling constant vanishes, furnish a set of algebraic equations in the coupling constants and group sizes $N_i$. These equations can be manipulated via a Gr\"obner basis by eliminating all the couplings leaving a single condition on the group sizes $N_i$: $P(N_1,N_2,...)=0$ where $P(N_1,N_2,...)$ is a polynomial, which we refer to as the \emph{extremality polynomial}, that can be chosen to have integer coefficients. In other words, the condition on the group sizes becomes a Diophantine problem. Some deep results in number theory, in particular  Faltings' and Siegel's theorems, can in some cases we study tell us if the number of solutions is finite. 

Determining the actual solutions for a given family of theories is a more elusive endeavour.  For plane curves, i.e.\ $P(N_1, N_2)=0$, the genus, which is determined by the degree and the singularities, plays a  crucial role \cite{genus}.\footnote{We discuss the curve genus in more detail in section \ref{sec:Genus}, where we provide an explicit formula.}
For genus zero, we may use the Pell equation to characterise all solutions. 
For plane curves of genus one, we use instead standard technology from elliptic curves. For higher genus curves and polynomials of more than two variables, we fall back on estimates based on the degree of the polynomial, the sizes of the coefficients, and the number of variables, using a type of scaling arguments also invoked by mathematicians \cite{Poonentalk}.  These estimate nonetheless  allow us to conclude with a high degree of confidence that we have found all the integer solutions. The upshot is that for theories with three quartic couplings there are some new sporadic extremal fixed points which are described in this paper, but for more general examples  with reduced symmetry, the number of instances of extremality is, with overwhelming probability, equal to zero, save in limits where these more complicated theories degenerate to the simpler ones previously found.  In such examples then by choosing appropriate linear combinations of the couplings the RG flow is constrained to a lower dimensional subspace,  where fixed points can be found more simply. Such reductions appear generically in multi-coupling theories. They can arise when the fixed point has a higher symmetry group but this need not always be the case. 

For theories with more than three couplings the extremality polynomials are very large, but for the theories we study with three or fewer couplings we list the polynomials in Table~\ref{tab:polynomials}. The tabulated polynomials are referred to as the largest irreducible factors because the table does not include, for each theory, additional polynomial factors associated to sectors of the theory with enhanced symmetry.

Diophantine equations have showed up before in the study of renormalisation group flow.  For example,
solutions of the Markov equation 
 \[
 \frac{a^2}{\alpha} + \frac{b^2}{\beta} + \frac{c^2}{\gamma} = a b c
 \]
 are known to describe fixed points of a class of triangular quiver theories with ${\mathcal N}=1$ supersymmetry in 4 dimensions \cite{Cachazo:2001sg,Wijnholt:2002qz,Herzog:2003dj}.
 Here $a$, $b$ and $c$ count the number of bifundamental chiral fields while $\alpha$, $\beta$, and $\gamma$ govern the
 number of gauge groups of the same rank.  The actual ranks of the gauge groups can then be related in a simple way to $a$, $b$, $c$, $\alpha$, $\beta$, and $\gamma$.  

 From the perspective of a physicist, an interesting fact about Diophantine equations is, that they provide an explicit setting to observe violations of naturalness. In the absence of any type of scale or ratio of scales, posing questions involving only $\mathcal{O}(1)$ numbers can oftentimes lead to answers involving unfathomably large numbers.\footnote{On a historical note, it may be remarked that the first person to present a clearcut instance of unnaturalness in numbers was Archimedes, who likely authored the anonymous Greek poem, rediscovered in 1773 by Gotthold Ephraim Lessing in a library manuscript, which formulates the puzzle of the oxen of the Sun god. The puzzle asks the reader to determine the numbers of cattle of different colours and genders, subject to constraints that can be reformulated as a Diophantine equation, in fact a Pell equation. Although the constraints are listed in the poem through the mention of only the numbers 1, $\tfrac{1}{2}$, $\tfrac{1}{3}$, $\tfrac{1}{4}$, $\tfrac{1}{5}$, $\tfrac{1}{6}$, and $\tfrac{1}{7}$, the smallest solution to the problem has the herd of Helios numbering about $8\cdot 10^{206\,544}$ oxen \cite{archimedes1897works}.} In the present context, unnaturalness manifests itself already at an earlier stage. Even in writing down the polynomial that determines extremality, a consummate explosion in complexity ensues on advancing from bifundamental to trifundamental theories. For bifundamental theories the polynomials are quite tractable and there
are  infinitely  many  integer roots.
But the polynomial $P_{OOO}(N_1,N_2,N_3)$ that provides the condition for extremality in the 
$O(N_1)\times O(N_2)\times O(N_3)$ model contains $85\,807$ terms, whose coefficients have a median absolute value of about $9.3\cdot 10^{24}$.

\begin{table}
\centering
\scalebox{0.8}{
\renewcommand{\arraystretch}{1.5}
\begin{tabular}{|c|c|c|} 
\hline
theory & 
largest irreducible factor in marginality polynomial & $\#$ of integer roots
\\
\hline
$O(N)$ & $N-4$ & 1
\\
\hline
$Sp(N_1)\times Sp(N_2)$ &
$13+2N_1+N_1^2+2N_2-10N_1N_2+N_2^2$
& $\infty$
\\
\hline
$U(N_1)\times U(N_2)$ &
$24+N_1^2-10N_1N_2+N_2^2$
& $\infty$
\\
\hline
$O(N_1)\times O(N_2)$ & 
$52 - 4 N_1 + N_1^2 - 4 N_2 - 10 N_1 N_2 + N_2^2$
& $\infty$
\\
\hline
$Sp(m)\times {\mathcal S}_n$ &
$11 + 27 m + 9 m^2 + m^3 - 10 n - 28 mn - 10 m^2 n - n^2 + m n^2$
& 11
\\
\hline
$U(m)\times {\mathcal S}_n$ & 
$\begin{matrix}{}-32 + 44 m^2 + 12 m^3 + m^4 + 32 n + 8 mn 
\\[-4pt]
{} - 40 m^2 n - 10 m^3 n + 4 n^2 - 4 m n^2 + m^2 n^2\end{matrix}$
& 3
\\
\hline
$O(m)\times {\mathcal S}_n$ &
$
\begin{matrix}
64 + 224 m + 132 m^2 + 20 m^3 + m^4 + 64 n - 96 mn 
\\[-4pt]
{} - 84 m^2 n - 10 m^3 n + 16 n^2 - 8 m n^2 + m^2 n^2
 \end{matrix}
 $
& 6
\\
\hline
$Sp(N_1)\times O(N_2)$ & 
$
\begin{matrix}
16 N_1^4 + N_2^4- 136 N_1^3 N_2 + 81 N_1^2 N_2^2 - 34 N_1 N_2^3 
\\[-4pt]
{}+ 144 N_1^3 -  468 N_1^2 N_2 + 234 N_1 N_2^2  - 18 N_2^3 + 420 N_1^2 
\\[-4pt]
{}- 408 N_1 N_2   + 105 N_2^2+ 216 N_1   - 108 N_2 + 36 
\end{matrix}
 $
& 3
\\
\hline
$U(N_1)\times O(N_2)$ &
$
\begin{matrix}
352 + 512 N_1 + 276 N_1^2 + 56 N_1^3 + 4 N_1^4 - 352 N_2 
\\[-4pt]
{}- 488 N_1 N_2 - 212 N_1^2 N_2 - 28 N_1^3 N_2 + 140 N_2^2 
\\[-4pt]
{} + 144 N_1 N_2^2 + 29 N_1^2 N_2^2 - 20 N_2^3 - 14 N_1 N_2^3 + N_2^4
\end{matrix}
$
& likely 6
\\
\hline
\end{tabular}
}
\caption{Largest irreducible factor for the extremality polynomials of the one- and two-parameter theories studied in this paper. The listed numbers of integer roots include positive and negative roots. Only some of these correspond to extremal fixed points.
 } 
    \label{tab:polynomials}
\end{table}

\subsection{Overview of paper}

The remainder of the paper is organised as follows:
Section~\ref{sec:OneLoop} reviews general facts about one-loop beta functions in the $\epsilon$ expansion, their fixed points and anomalous dimension, and the Rychkov-Stergiou bound.
Section~\ref{Groebner} presents a gentle introduction to the method of Gr\"obner bases and Siegel's theorem for physicists not familiar with these tools and also tabulates facts about the extremality polynomials studied in this paper.
Section~\ref{sec:simpbif} reviews the two-coupling bifundamental $O(N_1)\times O(N_2)$, $U(N_1)\times U(N_2)$, and $Sp(N_1)\times Sp(N_2)$ theories and relates their extremal fixed points to the Pell Diophantine equation.
Section~\ref{sec:UO} studies the mixed bifundamental three-coupling theories with $U(N_1)\times O(N_2)$ and $Sp(N_1)\times O(N_2)$ symmetry and identifies the new extremal fixed point contained in the former theory.
Section~\ref{sec:MultiConical} studies the three-coupling theories with $Sp(m)\times \mathcal{S}_n$, $U(m)\times \mathcal{S}_n$, and $O(m)\times \mathcal{S}_n$ symmetry and presents the four new sporadic extremal fixed points contained in the $Sp(m)\times \mathcal{S}_n$ theory.
Section~\ref{sec:Trifund} turns to tensorial theories and presents statistical arguments for why the trifundamental four-coupling $U(N_1)\times U(N_2)\times U(N_3)$ and five-coupling $O(N_1)\times O(N_2)\times O(N_3)$ theories are very unlikely to contain new extremal fixed points. 
Section~\ref{sec:Discussion} closes the main part of the paper by discussing our results and possible follow-up work.
Appendix \ref{App:BoundBreaking} discusses examples where the eigenvalues of the stability matrix exceed $| \kappa | > 1$. 
Appendix \ref{sec:closerlook} presents some facts about the seven-coupling theory of two quaternionic vectors and examples of seemingly distinct fixed points related by a field redefinition.
The lengthy extremality polynomial for the trifundamental $U(N)$ model is presented in Appendix \ref{appendix}.
Finally, Appendix \ref{sec:rationalpoints} gives the first few rational solutions for the rank one elliptic curves that appear
for the $Sp(N_1) \times O(N_2)$ and $U(m) \times {\mathcal S}_n$ models.

\section{One Loop RG Equations and Fixed Points}
\label{sec:OneLoop}
The action of a general multiscalar quartic theory with real scalars $\phi^i$, $i = 1, \ldots, N$, with a spatial dimension $d=4-\epsilon$,  has the form
\begin{align}
S = \int {\rm d}^{d }x\,\bigg(\frac{1}{2}(\partial_\mu\phi^i)^2  +\mu^\epsilon V (\mu^{-\frac12\epsilon}\phi)  \bigg)\, , \end{align}
and after a rescaling  $V(\phi)\to 16 \pi^2 \epsilon V(\phi)$,  the one-loop beta function is given by \cite{analogs}
\be
\beta_V(\phi) = V(\phi)  - \tfrac{1}{2}\, \phi^i V_i(\phi) + \tfrac{1}{2} \, V_{ij} (\phi) V_{ij}(\phi) \, ,
\label{Vone}
\ee
where $V_i(\phi)=\frac{\partial}{\partial \phi^i}V(\phi)$ and $V_{ij}(\phi)=\frac{\partial^2}{\partial \phi^i\partial \phi^j}V(\phi)$.
For quartic potentials
\be
V(\phi)= \frac{1}{24}\lambda_{ijkl}\,\phi^i\phi^j\phi^k\phi^l \, ,
\ee
where  $\lambda_{ijkl} $ is a symmetric real tensor with
\begin{align}
P_N=\tfrac{1}{24} N(N+1)(N+2)(N+3) \,,
\end{align}
independent components, then
\be
\beta_V (\phi) = \frac{1}{24}\beta_{{ijkl}} \,\phi^i\phi^j\phi^k\phi^l \, .
\ee
The one-loop beta function \eqref{Vone}  gives \cite{Wallace:1974dy}
\begin{align}
\label{renormbeta}
\beta_{{ijkl}}=-\lambda_{ijkl}
+\lambda_{ijmn}\,\lambda_{mnkl}
+\lambda_{ikmn}\,\lambda_{mnjl}
+\lambda_{ilmn}\,\lambda_{mnjk}\, .
\end{align}
The corresponding RG fixed  point given by
\begin{align}
 \beta_{{ijkl}}=0 \,  \ , 
 \label{fp1}
\end{align}
is covariant under the action of $O(N)$, so that there are $P_N - \frac12 N(N-1)$ independent equations. For any theory restricted by a symmetry group, it is necessary to solve \eqref{fp1} before setting up an $\epsilon$-expansion with higher loop contributions. The number of equations increases rapidly with $N$, and becomes impossible to analyse in general for $N\ge 4$.\footnote{A fully general discussion for $N=3$ where there are 12 essential equations, was described just recently in \cite{Rongtalk}.} Recent discussions of these equations are \cite{Hogervorst,Osborn:2020cnf} and in a mathematical context \cite{overflow}. At a perturbative fixed point, the equation $\beta_{{ijkl}}=0$ relates a linear term to a quadratic term and a simple scaling shows that there are bounds on the magnitude  of the components  $\lambda_{ijkl}$; the Rychkov-Stergiou bound \eqref{RychkovStergiou} constrains the quadratic norm
\begin{align}
\label{norm}
||\lambda ||^2 = \lambda_{ijkl}\,\lambda_{ijkl} \, .
\end{align}

Following \cite{Hogervorst}, it is useful to analyse the fixed point equation \eqref{fp1}
by decomposing the coupling $\lambda_{ijkl}$ into irreducible components under the action $O(N)$
so that
\begin{align}
V(\phi)= \tfrac18\, \hlambda \, (\phi^2)^2 +  \tfrac14 \, \phi^2 d_{2,ij}\phi^i \phi^j 
+ \tfrac{1}{24}\, d_{4,ijkl}\,\phi^i \phi^j\phi^k \phi^l  \, , 
\end{align}
with $d_{2,ij}, \, d_{4,ijkl}$ symmetric traceless tensors. With this basis
\begin{align}
||\lambda  ||^2 = 3N(N+2) \hlambda^2 + 6(N+4) \, ||d_2||^2 &+ || d_4||^2 \, , \nn \\
\quad\quad\quad \mbox{where} \quad
&|\lambda | \equiv \lambda_{iijj} = N(N+2) \hlambda\, .
\label{modl}
 \end{align}
 Projecting $\beta_{ijkl}$ on to its spin zero component, 
 $|\beta| = N(N+2) \beta\raisebox{-1.5 pt}{$\scriptstyle {\hat \lambda}$} $, gives
 \begin{align}
\beta\raisebox{-1.5 pt}{$\scriptstyle {\hat \lambda}$}  = - \hlambda + (N+8) \hlambda^2 + \tfrac{(N+4)(N+16)}{N(N+2)}\,  \, ||d_2||^2  +  \tfrac{2}{N(N+2)} \,  || d_4||^2 \, .
\label{betalh}
\end{align}
Imposing $\beta\raisebox{-1.5 pt}{$\scriptstyle {\hat \lambda}$}  =0$, and eliminating $ ||d_4||^2 $ gives directly
 \begin{align}
||\lambda||^2 = \tfrac{1}{8}N  - \tfrac{1}{2N} \big ( |\lambda | - \tfrac12 N \big )^2 - \tfrac{(N+4)^2}{2} \, ||d_2||^2  \, .
\end{align} 
Hence $||\lambda||^2 $ lies below a parabolic curve in $|\lambda |$ and at the maximum, where
the bound is attained, we must have
\begin{align}
\label{eq:absLambda}
|\lambda| = \tfrac{1}{2}N \, , \qquad  ||d_2||^2 =0  \, , \qquad  ||d_4 ||^2 = \frac{N(N-4)}{8(N+2)} \, .
\end{align} 
Fixed points realising these conditions are extremal. Numerical searches up to $N=7$ \cite{Osborn:2020cnf} show that there are very many fixed points lying below the parabolic curve but relatively few on the curve where $ ||d_2||^2 =0 $, and these were all examples from scalar theories previously considered with particular symmetry groups. 

We now explain the connection between saturating the Rychkov-Stergiou
bound and the existence of a marginal operator. The problem of computing anomalous dimensions or equivalently finding eigenvalues of the stability matrix at lowest order amounts to solving
\begin{align}
6\, \lambda \vee v = (\kappa + 1) \, v \, .
\label{keig}
\end{align} 
with the nonassociative product on four-index symmetric tensors \cite{Michel}: 
\begin{align}
6 ( \lambda \vee v)_{ijkl} = 6 (v \vee  \lambda)_{ijkl} =&  \lambda_{ijmn}\, v_{mnkl}
+\lambda_{ikmn}\, v_{mnlj} +\lambda_{i lmn}\, v_{mnjk  } + \lambda_{k j mn}\, v_{mn l i } \nn \\
&{} +\lambda_{ljmn}\, v_{mnik } +\lambda_{klmn}\, v_{mnij } \, .
\end{align} 
(With a different notation this product was discussed more recently in \cite{Sundborg}.) In this notation $\beta_{ijkl} = 0$ is just $3\, \lambda \vee \lambda = \lambda$ and clearly for $v_{ijkl} \to \lambda_{ijkl} $, $\kappa=1$.

Following  \cite{Rychkov:2018vya},  we may verify the presence of a marginal operator at an extremal fixed point. Defining\footnote{%
Note ${\mathds 1}$ is not the same as the identity operator used in (\ref{keig}).
}
\begin{align}
{\mathds 1}_{ijkl} = \delta_{ij} \delta_{kl} +  \delta_{ik} \delta_{lj} + \delta_{il} \delta_{jk} \, ,  \qquad
{\mathds 1}\vee {\mathds 1} = \tfrac13(N+8)\, {\mathds 1}\, , \quad {\mathds 1}\vee d_4 = 2\, d_4 \, ,
\end{align} 
and at an extremal fixed point, where $\lambda= \frac{1}{2(N+2)} \, {\mathds 1} + d_4$, then
\begin{align}
d_4 \vee d_4 = \tfrac{N-4}{12(N+2)^2} \, {\mathds 1} + \tfrac{N-4}{3(N+2)} \, d_4\, ,
\end{align} 
which follows from $3\, \lambda \vee \lambda = \lambda$. For a stability matrix eigenvector $ v=  v_0 \, {\mathds 1}  + v_4\, d_4 $, then \eqref{keig} reduces to a $2\times 2$ matrix eigenvalue equation
\begin{align}
  \begin{pmatrix} \frac{6}{N+2} &\  \frac{N-4}{2(N+2)^2} \\ 12 & \ \frac{N-4}{N+2}  \end{pmatrix} 
 \begin{pmatrix} v_0 \\ v_4   \end{pmatrix} = \kappa \,  \begin{pmatrix} v_0 \\ v_4   \end{pmatrix} \, .
\end{align} 
The eigenvalues are one and zero, one corresponding to the fixed point value $v=\lambda$ and zero demonstrating marginality. Of course the full stability matrix has many more eigenvalues and requires looking at additional directions in coupling space.

\subsection{The \texorpdfstring{$O(N)$}{O(N)} theory}

The theory with maximal $O(N)$ symmetry has $ ||d_2||^2 = ||d_4 ||^2 = 0$, and just a single coupling $\hlambda$.
Solving $\beta\raisebox{-1.5 pt}{$\scriptstyle {\hat \lambda}$}  = 0$ from \eqref{betalh} is trivial.
From \eqref{modl} at the $O(N)$ fixed point $|| \lambda ||^2$ becomes
\begin{align}
\label{OOO}
|| \lambda ||^2 = \frac{3N(N+2)}{(N+8) ^2 } = \frac{N}{8} \biggl ( 1 - \Big ( \frac{N-4}{N+8} \Big)^2 \bigg )  \, .
\end{align}
Clearly the bound is saturated when $N=4$, and the extremality polynomial in this case is  just
\begin{align}
\label{PO}
P_O(N) \equiv N - 4\,.
\end{align}
The condition $P_O(N)=0$ is necessary and sufficient for the $O(N)$ symmetry to reach the bound. For $M$ complex fields there is a  natural maximal  $U(M)$ symmetry, but by decomposing into real and imaginary parts  in the single-coupling theory, the $U(M)$ symmetry is  promoted to  $O(2M)$. 
The extremality polynomial becomes
\begin{align}
\label{PU}
P_U(N) = M-2\,.
\end{align}
Similarly, in the single-coupling theory $Sp(M)$ is promoted to $O(4M)$ so that the extremality polynomial 
is
\begin{align}
P_{Sp}(M) = M-1\,.
\end{align}

In the space of all couplings the stability matrix eigenvalues
at one loop
and their associated degeneracies are
\be
\label{ONeigs}
1 \ (1) \, , \quad \tfrac{8}{N+8} \ 
\big ( \tfrac12 (N-1)(N+2)\big  ) \, , 
\quad - \tfrac{N-4}{N+8} \ 
\big (\tfrac{1}{24}N(N-1)(N+1)(N+6) \big) \, .
\ee

\subsection{Theories with two couplings}

A large number of theories with different symmetry groups and realising a  large variety of
fixed points can be described by scalar theories with two couplings.  Besides $\hlambda$ the additional 
coupling $g$ couples to a $d_4$ tensor which is assumed unique under the symmetric group $G$.
The $g$ coupling becomes relevant for $N>4$. The lowest order $\beta$-functions can then be 
expressed in a canonical form
\begin{align}
\beta\raisebox{-1.5 pt}{$\scriptstyle {\hat \lambda}$} = - \hlambda + ( N+8) \hlambda^2 + a\, g^2 \, , \qquad
\beta_g = - g + 12 \, \hlambda \, g + b \, g^2 \, ,
\label{betatwo}
\end{align}
with $\frac12 N(N+2) \, a\,  g^2 = || d_4||^2$. In this case
\begin{align}
|| \lambda ||^2 = 3 N(N+2) \big ( \hlambda^2 + \tfrac16 a \, g^2\big ) \, , \qquad | \lambda | = N(N+2) \hlambda \, .
\label{modl2}
\end{align}
Necessarily $a>0$ and by a choice of the sign of $g$ we can assume $b>0$.

Finding nontrivial fixed points reduces to solving a simple
quadratic equation. The nontrivial solutions are
\begin{align}
\hlambda_\pm=  \frac{b \pm \sqrt{R}} { 2(N+2)b \pm 12 \sqrt{R}} \, , \quad
g_\pm=  \frac{N-4} { (N+2)b \pm 6 \sqrt{R}} \, ,
\label{sol2}
\end{align}
for
\be
 R = 4(4-N) a + b^2 \, .
 \label{Rdef}
 \ee
 For real fixed points it is necessary that $R>0$ and $R=0$ is a square root bifurcation. 
 Up  to an overall factor, $R$ is then the extremality polynomial in this case.
 
For the $2\times 2$ stability matrix for this theory, one eigenvalue is 1 and otherwise
\begin{align}
\kappa_\pm =\pm  \frac{(N-4)\sqrt{R}} { (N+2)b \pm 6 \sqrt{R}}  \, , \qquad || \lambda_\pm ||^2
= \tfrac18 N ( 1- \kappa_\pm{\!}^2 ) \, .
\end{align}
For $N>4$ and $R>0$, $\sqrt{R}< b$ so that
$\kappa_+ < \frac{N-4}{N+8}$ and also $\kappa_->-1$.

\section{Gr\"obner Basis and the Extremality Polynomial}
\label{Groebner}

For theories with several couplings then in the $\epsilon$-expansion it is first necessary to require
vanishing of  the one loop  $\beta$-functions. For  $S$ quartic coupling constants $g_1$ to $g_S$, with associated beta functions $\beta_1$ to $\beta_S$, then  there are $S$ equations determining fixed points with, at one
loop, potentially $2^S$ solutions. Of course 
general theories always contain the trivial Gaussian fixed
point as well as the $O(N)$ fixed point with maximal symmetry but many other nontrivial  fixed points with reduced symmetry are possible. 
We investigate in this paper extremal solutions where the bound \eqref{RychkovStergiou} is saturated and there is therefore an additional constraint. Such cases are associated with bifurcations and the existence of a marginal operator.
 In this more generic setting, the theories in the given family may be labelled by multiple integer parameters $N_1, \dots,N_r$, indicating the sizes of the subgroups of the full $O(N)$  symmetry group.   For extremal solutions there are $S+1$ equations with $S+r$ variables.
However restricting all  $N_i$ to be positive integers provides a very severe additional constraint and there
need be no possible solutions in any particular case.

Barring limits of patience or computational power, it is always possible to derive a necessary condition for extremality in the form of a Diophantine equation in the parameters $N_1$ to $N_r$,
\begin{align}
\label{polynomialCondition}
0=G(N_1,...,N_r)\,,
\end{align}
where $G(N_1,...,N_r)$ is a multivariable polynomial with integer coefficients, which we will refer to as the \emph{extremality polynomial}. In applications considered $G$ may have very large degree and involve
enormous coefficients.

The general way to derive the condition \eqref{polynomialCondition} is via the method of Gr\"obner bases. This type of basis is a useful practical tool in solving general systems of polynomial equations in multiple unknowns
\begin{align}
\nonumber
0&= p_1(x_1,x_2,...)\,,
\\
0&= p_2(x_1,x_2,...)\,,
\\ \nonumber
&\hspace{2mm}\vdots
\end{align}
The basic idea is to choose an ordering of the variables, say $x_1$, $x_2$, ..., and then re-express the equations into an equivalent set of polynomial equations, but where the first equation depends only on $x_1$, the second only on $x_1$ and $x_2$, and so on,
\begin{align}
\nonumber
0&= g_1(x_1)\,,
\\
0&= g_2(x_1,x_2)\,,
\\ \nonumber
&\hspace{2mm}\vdots
\end{align}
One can then proceed to solve the system iteratively, first solving $0=g_1(x_1)$ for $x_1$, and then plugging the value into the second equation $0=g_2(x_1,x_2)$ and solving for $x_2$, and so on. It can happen that $g_{i+1}(x_1,...)$ introduces more than one new variable compared to $g_{i}(x_1,...)$, or that no new variable is introduced, but the hierarchy always persists where the variables of $g_{i}(x_1,...)$ are a subset of the variables entering into $g_{i+1}(x_1,...)$. Several algorithms exist for computing the Gr\"obner basis for any set of polynomial equations, and have been incorporated into standard mathematical software like Mathematica 
\cite{Mathematica}. 

For the purpose of our present application of investigating extremality, the way to proceed is by computing the Gr\"obner basis of the polynomial ring
\begin{align}
\Big\{\beta_1,\,\beta_2,\,...,\,\beta_s,\,8\,||\lambda ||^2-N\Big\}
\end{align}
in the variables $\{N_1,g_1,g_2,...,g_s\}$, while treating $N_2$ to $N_r$ as fixed parameters. Labelling the Gr\"obner basis as
\begin{align}
\Big\{G_1,\,G_2,\,...,\,G_k\Big\}\,,
\end{align}
the choice of lexicographic ordering where $N_1 \prec g_i$ implies that $G_1$ does not depend on the coupling constants, but only on the variable $N_1$ and the remaining parameters $N_2$ to $N_r$. We therefore define the extremality polynomial $G$ in \eqref{polynomialCondition} to be the first element in the Gr\"obner basis: $G=G_1$. The overall normalisation of $G$ is of no significance and can be chosen freely.

Let us illustrate the above discussion with the simplest possible example: the $O(N)$ model, whose beta function is given by
\begin{align}
\label{betaON}
\beta\raisebox{-1.5 pt}{$\scriptstyle {\hat \lambda}$}
=-\hat{\lambda}+(8+N)\hat{\lambda}^2\,.
\end{align}
To search for extremal fixed points we need to set to zero $\beta\raisebox{-1.5 pt}{$\scriptstyle {\hat \lambda}$}$, 
and also either impose saturation of \eqref{RychkovStergiou} or, equivalently by \eqref{eq:absLambda}, impose $|\lambda|=\frac{N}{2}$. We will proceed with the latter option. Defining 
\begin{align}
f\equiv |\lambda|-\frac{N}{2} = N(N+2)\hat{\lambda}-\frac{N}{2}\,,
\end{align}
the Gr\"obner approach is to eliminate $\hat{\lambda}$, so as to obtain a condition on $N$ for extremality. To this end, we subtract off from $\beta_{\hat{\lambda}}$ a suitable polynomial in $f$ to obtain a function of $N$ alone that must vanish for extremality to hold,
\begin{align}
\label{nolambda}
0=\beta\raisebox{-1.5 pt }{$\scriptstyle {\hat \lambda}$}-\frac{N+8}{N^2(N+2)^2}f^2-\frac{6}{N(N+2)^2}f
=\frac{4-N}{4(N+2)^2}\,.
\end{align}
Disregarding the denominator, we have re-derived the extremality condition $0=P_O(N)$, with the linear extremality polynomial given in equation \eqref{PO}.

\subsection{Diophantine equations, genera, and Siegel's theorem}
\label{sec:Genus}
Identifying instances of extremality within the $\epsilon$ expansion amounts to solving the Diophantine equation $G(N_1, N_2, \ldots, N_r) = 0$.
Outside of special cases,
determining the set of solutions for a Diophantine equation is a notoriously difficult problem. Even innocuous examples can prove challenging and defy attempts at solution.\footnote{For example, the problem of determining whether 42 can be written as a sum of three cubes,
\begin{align*}
42 = X^3+Y^3+Z^3\,,
\end{align*}
with $X,Y,Z\in\mathbb{Z}$, was only solved (in the affirmative) in 2019 \cite{booker2021question}.} Hilbert's tenth problem consists in devising a general algorithm to determine whether any given Diophantine equation has solutions in the rationals, and this problem was proven to be unsolvable by Matiyasevich \cite{matiyasevich1970diophantineness}, building on earlier work by Davis, Putnam, and Robinson \cite{davis1961decision}. 

In this paper we limit our study to the cases $r=2$, where $G(N_1, N_2)=0$ cuts out an algebraic curve, and $r=3$ where the locus of points is a surface.  
In the case of an algebraic curve, the concept of genus is central in deciding existence of integer solutions.  Siegel proved that the number of integer points is finite (possibly zero) if the genus $g>0$  \cite{Siegel}.\footnote{%
 Faltings' theorem \cite{faltings1983endlichkeitssatze,faltings1984endlichkeitssatze} is often cited as superseding Siegel's Theorem for $g>1$.  Indeed, Faltings proved that curves with $g>1$ have at most a finite set of rational solutions.  However, here we are interested in integer solutions, for which Siegel's result is sufficient. An example of a Diophantine equation with no integer solutions is $2x^4 + 3 y^4 = 1$.} Given a curve for which all the singular points are double-points, the genus of the curve can be computed as 
\be
\label{genusformula}
g = \tfrac{1}{2}(n-1)(n-2) -\mu \ , 
\ee
where $n$ is the degree of the polynomial defining the curve 
and $\mu$ is the number of nodes, i.e.\ the number of points where $\{G, \partial_{N_1} G, \partial_{N_2} G\}$
vanish simultaneously. A double point occurs when a curve intersects itself such that the two branches of the curve have distinct tangent lines.  More general types of singularities can occur, in which case the genus formula above must be modified.

Except for the $O(N)$ model, in all the theories we study the extremality polynomial factorises as $G(N_1,N_2)=\prod_i P_i(N_1,N_2)$, with most of the irreducible polynomials $P_i(N_1,N_2)$ being associated to a sector of the theory with enhanced symmetry. For the purpose of determining the roots of $G$, we can consider each factor $P_i$ separately and compute its genus.

Generalising Siegel's result to higher dimensional surfaces, the Bombieri-Lang conjecture posits that for an algebraic variety of general type, the set of rational points does not form a dense set in the Zariski topology.   
In the special case of a curve, the Bombieri-Lang  conjecture reduces to Faltings' theorem, i.e.\ if a curve has Zariski dense rational points, then
its genus is zero or one.\footnote{%
 A stronger form of the conjecture is F.5.2.3 of \cite{hindry2013diophantine}
 where
if from the variety $X$ the special closed subset $Sp_X$ (defined above F.5.2.3 as the Zariski closure of the union of all the images of nontrivial rational maps from abelian varieties to $X$) is removed then 
 for $U = X - Sp_X$, the set of rational points is finite.
 }
Loosely, the conjecture can be thought of as stipulating that polynomials of higher degrees have fewer solutions. 
An intuitive way to conceptualise this statement, which extends to more general equations in $d$ unknowns, is to think of a large $d$-dimensional box of side length $L$. Now, consider the question of how many integer points within the box satisfy the equation
\begin{align}
\label{Pzero}
0 = P(X_1,...,X_d)\,, 
\end{align}
where $P(X_1,...,X_d)$ is a polynomial of degree $n$ with integer coefficients. The set of values that $P$ ranges over within the box scales as $L^n$, and if the coefficients of $P$ do not force $P$ to be positive, and if modular arithmetic does not force $P$ to be non-zero, we expect that a fraction $L^{-n}$ of the points near the surface of the box will satisfy \eqref{Pzero}. Since the total number of points near the boundary scales a $L^{d-1}$, the expected number of solutions to \eqref{Pzero} within some fixed narrow width of the boundary of the box is generically expected to scale as $L^{d-n-1}$. If we integrate over solutions near the surface of the box as we take the box size $L$ to infinity, we get a convergent result for $n>d$, so that we expect a finite number of solutions. And the larger the value of $n-d$ is, the smaller we expect the number of solutions to be.

The theories we study in this paper along with the degrees of their extremality polynomials are listed in Table \ref{tab:counting}.  Naively, none of our examples has 
a polynomial $G(N_1, \ldots N_r)$ of an appropriate degree to host more than a finite number of integer solutions.  
For $O(N)$, the polynomial is linear in a single variable, and there can be but one solution.  
For our $r=2$ examples, genus zero in smooth cases typically requires degree one or two, while for all the cases we consider
the degree is at least four.  For $r=3$, the degree of the polynomial is larger than fifty.

The number of integer solutions, however, is sensitive also to whether the polynomial factors and/or has singular points.
An interesting aspect of the extremality polynomial is that, as mentioned above, with the exception of the $O(N)$ case,
 it is never of general type in the examples we consider; $G(N_1, \ldots, N_r)$ always factors into polynomials of lower degree.  
The $U(N_1) \times O(N_2)$ theories for example have fixed points where the symmetry is enhanced to $U(N_1) \times U(N_2)$
and to $O(2 N_1 N_2)$, reducing the degree of the maximal factor to four.
Furthermore, the maximal factor typically has
nodal points.  The $Sp(N_1) \times O(N_2)$ case for example has
two nodal points while $U(N_1) \times O(N_2)$ has one, reducing the genus to one and two respectively.  

In short, we will see below that the global symmetry groups 
$O(N_1) \times O(N_2)$, $U(N_1) \times U(N_2)$, and $Sp(N_1) \times Sp(N_2)$
all give rise to extremality polynomials which have a genus zero factor, and all allow for an infinite sequence of integer solutions.
The more complicated examples further down  Table \ref{tab:counting}
all have degenerate limits in which they regain $O \times O$, $U \times U$, or $Sp \times Sp$ symmetry,
and support these same infinite sequences of solutions. These limits correspond to lower degree factors of the extremality polynomial.
The maximal factor of the extremality polynomial in these more complicated 
cases however supports only a finite number of integer solutions.
 Indeed, even the case $Sp(m) \times {\mathcal S}_n$, which has a genus zero maximal factor, supports only a finite number of integer solutions.
 Note that Siegel's result is not effective: it provides sufficient criteria to guarantee a finite number of solutions but offers no means of ascertaining at what point all integer solutions have been found.
The genus one examples are particularly interesting as they are elliptic curves, and we can use some recently developed techniques \cite{stroeker1994solving,stroeker2003computing}
to show that we have in fact determined all the integer solutions.  For the genus two curve and the surface cases, we resort to probabilistic arguments.

The probabilistic arguments we use rely on the precise forms of our specific polynomials.
Among the type of data that characterises a polynomial,
 two pieces of information are the height and the length of the polynomial. For a polynomial given by
\begin{align}
P(X_1,...,X_d)=\sum_{m_1,...,m_d=0}^n c_{m_1,...,m_d}\,X_1^{m_1}\,...\,X_d^{m_d}\,, 
\end{align}
the height and the length are defined as
\begin{align}
\text{height }&=\hspace{1mm}\underset{m_1,...,m_d}{\text{max}} \hspace{3mm}|c_{m_1,...,m_d}|\,,
\\[3pt]
\text{length }&=\sum_{m_1,...,m_d=0}^n |c_{m_1,...,m_d}|\,.
\end{align}
In Table~\ref{tab:counting} we list this data along with the degrees for the extremality polynomials of the theories we study in this paper. As is manifest from the table, a veritable explosion in complexity sets in for the tri-fundamental theories. In consequence, the probability estimates we perform will indicate that these theories almost certainly contain no new instance of extremality.

\begin{table}
\centering
\scalebox{0.74}{
\renewcommand{\arraystretch}{1.5}
\begin{tabular}{|c|c|c|c|c|c|c|} 
\hline
theory
&
$\begin{matrix}
\text{\# of}
\\[-6pt]
\text{operators}
\end{matrix}$
&
$\begin{matrix}
\text{degree of full}
\\[-6pt]
\text{polynomial}
\end{matrix}$
&
$\begin{matrix}
\text{degree of}
\\[-6pt]
\text{maximal}
\\[-6pt]
\text{factor}
\end{matrix}$
&
$\begin{matrix}
\text{genus}
\\[-6pt]
\text{where}
\\[-6pt]
\text{applicable}
\end{matrix}$
&
$\begin{matrix}
\text{height of}
\\[-6pt]
\text{maximal}
\\[-6pt]
\text{factor}
\end{matrix}$
&
$\begin{matrix}
\text{length of}
\\[-6pt]
\text{maximal}
\\[-6pt]
\text{factor}
\end{matrix}$
\\
\hline
$O(N)$
&
1
&
1
&
1
&
-
&
4
&
5
\\
\hline
$Sp(N_1)\times Sp(N_2)$
&
2
&
4
&
2
&
0
&
13
&
29
\\
\hline
$U(N_1)\times U(N_2)$
&
2
&
4
&
2
&
0
&
24
&
36
\\
\hline
$O(N_1)\times O(N_2)$
&
2
&
4
&
2
&
0
&
52
&
72
\\
\hline
$Sp(m) \times {\mathcal S}_n$
&
3
&
8
&
3
&
0
&
28
&
98
\\
\hline
$U(m) \times {\mathcal S}_n$
&
3
&
9
&
4
&
1
&
44
&
188
\\
\hline
$O(m) \times {\mathcal S}_n$
&
3
&
9
&
4
&
1
&
224
&
720
\\
\hline
$Sp(N_1)\times O(N_2)$
&
3
&
14
&
4
&
1
&
468
&
2425
\\
\hline
$U(N_1)\times O(N_2)$
&
3
&
14
&
4
&
2
&
512
&
2628
\\
\hline
$U(N_1)\times U(N_2)\times U(N_3)$
&
4
&
54
&
36
&
-
&
$\approx 6.29 \cdot 10^{13}$
%$62\,882\,616\,180\,736$
&
$\approx 1.91 \cdot 10^{14}$
%$190\,906\,067\,910\,330$
\\
\hline
$O(N_1)\times O(N_2)\times O(N_3)$
&
5
&
129
&
108
&
-
&
$\approx 2.27\cdot 10^{48}$
&
$\approx 4.01\cdot 10^{50}$
\\
\hline
\end{tabular}}
\caption{Numbers of operators and degrees of extremality polynomials for examples we study. Here 
${\mathcal S}_n$ is the permutation group  on $n$ letters.  
The degree of the full polynomial suffers from a mild ambiguity depending on the ordering chosen in computing the Gr\"{o}bner basis.
}
    \label{tab:counting}
\end{table}

\section{Bifundamental Theories with Two Couplings}
\label{sec:simpbif}

We consider a class of examples where the global symmetry groups, up to discrete quotients,\footnote{In all three cases, there
 is a diagonal ${\mathbb Z}_2$ which acts trivially on the field content. In the bi-unitary theory, the ${\mathbb Z}_2$ is part of a trivially acting $U(1)$.} are $O(N_1) \times O(N_2)$, $U(N_1) \times U(N_2)$ or $Sp(N_1) \times Sp(N_2)$.\footnote{%
  By $Sp(N)$ we mean the unitary group on quaternion numbers,  $U(N,\mathbb{H})$, which is sometimes also called $USp(2N)$ -- the compact group which is the intersection of the noncompact $Sp(2N, {\mathbb C})$ with the unitary group $U(2N)$.}
 That we get an infinite sequence of fixed points which saturate the Rychkov-Stergiou bound in these cases is known \cite{Osborn:2017ucf,Rychkov:2018vya,Osborn:2020cnf,Kousvos:2022ewl,Osborntalk}.  
 It seems conceivable that in fact these three examples may provide the only infinite such sequences \cite{Osborntalk}. 
 Here we give a unified treatment of all three theories;
 the analysis of the bifurcation-node using the Pell equation is to our knowledge new.

We assume the
scalars are $N_1\times N_2$ matrix valued real, complex or quaternionic fields:
\be
\Phi_{ra} = \vphi^i{\!}_{ra}e_i \, , \qquad r=1, \dots , N_1 \, ,  \ a=1, \dots , N_2 \, , 
\ee
where the three possibilities for $e_i$ are bases for real, complex and quaternionic numbers
\be
\{e_i\} = \{1\} \, , \qquad \{e_i\} = \{1, i \} \, , \qquad \{e_i\} = \{1,i,j,k \} \, ,
\ee
which satisfy
\be
e_i {\bar e}_i = \nu \, , \quad e_i  {\bar e}_j e_i = (2-\nu) e_j \, , \quad  \nu = 1,2,4 \, ,
\ee
where there is implicit summation over repeated indices.
There are a total of 
\be
N = \nu \, N_1  N_2 \, 
\label{resN}
\ee
 $\vphi^i{\!}_{ra}$ fields.
 
The essential scalar interaction is then\footnote{%
The products $\Phi \bPhi$ and $\Phi \bPhi \Phi \bPhi$ are hermitian,
and for hermitian quaternionic matrices the trace is the 
usual sum over real diagonal elements.
}
\be
8\, V(\Phi) = \lambda \, \ \big ( \tr( \Phi \bPhi ) \big )^2 + g \,   \tr( \Phi \bPhi \Phi \bPhi )  \, ,
\label{Vphi}
\ee
with 
\be
\bPhi_{ar} = {\tilde \vphi}^i{\!}_{ar}{\bar e}_i \, , \qquad  {\tilde \vphi}^i{\!}_{ar} =  {\vphi}^i{\!}_{ra} \, , 
\ee
and  where ${\bar e}_i$ is obtained  by the usual complex or quaternionic 
conjugation so that $\bPhi$ is the hermitian conjugate of the real, complex, or quaternionic $\Phi$. 
Manifestly $\tr( \Phi \bPhi )= \tr( \vphi^i {\tilde \vphi}^i)= \sum_{a,r,i}  (\vphi^i{\!}_{ra})^2$ is $O(N)$ invariant. 

 From the general formula  \eqref{Vone} and making the shift
\be
{\hat \lambda} = \lambda + \tfrac{1}{N+2}\,  \big  (\nu ( N_1+N_2-1)  + 2 \big )  \, g 
\ee
to eliminate a $\lambda g$ cross term in $\beta_\lambda$ the beta functions reduce to  the universal two-coupling form \eqref{betatwo} with
\begin{align}
& a =  \tfrac{3}{(N+2)^2} \, \nu (N_1-1)(N_2-1) (\nu\, N_1+2)(\nu\, N_2+ 2) \, , \qquad b =  
\tfrac{1}{N+2}\,D_{\nu}(N_1,N_2) \,, 
\nn\\
& D_{\nu}(N_1, N_2) = \nu^2  N_1 N_2 (N_1 + N_2-4) +2 \,\nu ( 4 N_1 N_2 -5N_1-5N_2 +2 ) -8  \, .
\label{Dmn}
\end{align}
These can be directly applied in \eqref{modl2} and from \eqref{Rdef}
\be
R_{\nu}(N_1 ,N_2) =  \nu^2 \big ( N_1{\!}^2 + N_2{\!}^2 - 10\, N_1N_2+ 4(N_1+N_2+1) \big) - 
8\, \nu (N_1+N_2-4) + 16   \, .
\label{Rmn}
\ee
In Table~\ref{tab:polynomials}  the extremality  polynomials for the bifundamental theories are
then 
\begin{align}
P_{\nu}(N_1 ,N_2) = \frac{1}{\nu^2}R_{\nu}(N_1 ,N_2) \,,
\label{bifundPnu}
\end{align}
and we may also define
\begin{align}
P_{OO}(N_1,N_2) = {}& P_1(N_1,N_2) \,, \qquad
P_{UU}(N_1,N_2)\hspace{2mm} = P_2(N_1,N_2) \,, \nn \\
P_{SpSp}(N_1,N_2) ={}& P_4(N_1,N_2) \,.
\label{Pbifund}
\end{align}

As a check on the results so far, recall that formally at least, the transformation $N \to -N$ is expected to leave the $SU(N)$ results invariant while exchanging the $O(N)$ and $Sp(N)$ groups.  Indeed, we find that in the $SU(N_1) \times SU(N_2)$ case, the beta functions are invariant under the transformation $N_i \to -N_i$ and $g \to -g$.  The beta function in the $Sp(N_1) \times Sp(N_2)$ case on the other hand is transformed into the $O(N_1) \times O(N_2)$ case under
$N_1 \to -N_1/2$, $N_2 \to -N_2/2$, $g \to - g/2$.

Setting to zero the two beta functions and imposing saturation of \eqref{RychkovStergiou}  produces a system of three equations in four unknowns: $\lambda_1$, $\lambda_2$, $N_1$, and $N_2$. Manipulations of this system of equations according to the general procedure  discussed in Section~\ref{Groebner}, leads to a necessary condition for extremality that is independent of the coupling constants: $G(N_1,N_2)=0$, with $G$ being a polynomial with integer coefficients that factorises as 
\begin{align}
\label{Gnu}
G(N_1, N_2) = P_O(\nu N_1 N_2) P_{\nu}(N_1, N_2) \ ,
\end{align}
where the first factor is given by the linear polynomial defined in \eqref{PO} and represents the fact that the 
$O(\nu N)$ model is contained inside the bifundamental models, while the second factor is given by \eqref{bifundPnu}.
The factorisation of \eqref{Gnu} illustrates a general feature of extremality polynomials. 
If a theory $T$ contains
a theory $T^\prime$  with a reduced number of couplings which is closed under RG flow, then the zeroes of the extremality
polynomial $G_{T^\prime}$ must also be contained in $G_T$. This is often realised simply by $G_{T^\prime}$ being a factor in $G_{T}$ -- but not always, see equation \eqref{POOO2} below.

\subsection{The Pell equation}

Here we discuss the solutions of $R_\nu(N_1, N_2)= 0$  (\ref{Rmn}) over the integers
where $\nu = 1$, 2 or 4.  

As a first step we try to solve the quadratic in $N_1$ over the reals and find
\be
\label{quadeq}
N_1 = 5N_2 - 2+ \frac{4}{\nu} \pm 2 \sqrt{6} \sqrt{N_2^2 + N_2 \left(\frac{2}{\nu}-1 \right) - \frac{2}{\nu}} \ .
\ee
Thus a necessary and sufficient condition for a solution over the integers is that 
%\CH{adjusted}
\be
y \equiv 2 \sqrt{6} \sqrt{ N_2^2 + N_2 \left(\frac{2}{\nu}-1 \right) - \frac{2}{\nu}}
\ee
and $N_2$ itself be integers. 
The condition on the square root can be cast in a variant form of the Pell equation:\footnote{ We refer the reader to \cite{jacobson2009solving} for a comprehensive review of the Pell equation and the methods of solving it.}
\be
\label{myPell}
x^2 - 6\, y^2 = c^2 \, ,
\ee
where
\begin{align}
x =  3  \left(  4N_2 - 2 \left(1 - \frac{2}{\nu}\right) \right)\ , \quad
c = \frac{6(2+\nu)}{\nu}\ .
\end{align}
For the values $\nu$ that we are interested in, $c$ is integer, equal to $18$, $12$, and $9$ for $\nu = 1$, 2 and 4 respectively.  Further, the shift $2 \left(1 - \frac{2}{\nu}\right)$ is equal to $-2$, 0 and 1.  If we find an $x$ that is an integer multiple of 3, we may be in good shape to find an integer solution for $N_2$.  In fact, it's clear that since $c$ is a multiple of three and $6 y^2$ is a multiple of three, any integer solution for $x$ must also contain a factor of three.

Since\footnote{%
 In certain cases, there may be a different way of obtaining a factorisation of $c^2$ in equation \eqref{myPell}.  For the $Sp(N) \times {\mathcal S}_n$ theory considered later, the fact
 that $(7 - 2 \sqrt{6})(7 + 2 \sqrt{6}) = 5^2$ yields a second set of possible solutions (see section \ref{sec:SpNxSpv2}).
}
\be
\big  ( 5+2\sqrt{6} \big ) \big   (5-2\sqrt{6} \big ) = 1 \ ,
\ee
then defining $x_p,y_p$ by
\begin{align}
x_p  \pm \sqrt{6}\, y_p =  c \big   (5 \pm  2 \sqrt{6} \big )^p   \ ,  \quad p=0,1,2 ,\dots
\end{align}
ensures $x_p,y_p$ satisfy \eqref{myPell} for any $p$. It is easy to see that $x_{p+1}= 5 x_p + 12y_p, \
y_{p+1} = 2 x_p + 5 y_p$ with $x_0=c, \ y_0=0$.
Equivalently $x_{p+1} = 10 x_p - x_{p-1}$. In general $x_p$ is an odd integer multiple of $c$, $y_p$ 
is an even multiple.

Solving the  Pell equation then gives solutions for $N_2$, 
\be
N_{2}^{(p)} =  \frac{1}{12} \, x_p + \frac12 \left  (1 - \frac{2}{\nu} \right) \, , \quad p=0,1,2,\dots  \ .
\label{npgeneral}
\ee
From \eqref{quadeq}, choosing the $+$ sign,
\be
N_{1}^{(p)} =  \frac{5}{12} \, x_p +y_p+  \frac12 \left  (1 - \frac{2}{\nu} \right)  = N_{2}^{(p+1)} \, .
\ee 
These solutions satisfy the recursion relation\footnote{This relation can also be derived directly from (\ref{quadeq}). Taking the difference between the two roots eliminates the square root and provides a relation between distinct integer roots of $R_\nu(N_1,N_2)=0$.}
\be
N_2^{(p+1)} + N_2^{(p-1)} = 10 N_2^{(p)} -4 + \frac{8}{\nu} \ .
\ee
Thus neighbouring values $\{N_1, N_2 \} = \{ N_2^{(p+1)} , N_2^{(p)} \}$ give the desired solutions of the original  equation $R_\nu(N_1, N_2)= 0$.
By this recursion relation, it is easy to see that if $N_2^{(p)}$ and $N_2^{(p-1)}$ are integer, then so is $N_2^{(p+1)}$ when
$\nu = 1$, 2, 4, or indeed 8.  
It turns out these solutions are integer for $\nu = 1$, 2 and 4.
In closed form, the solutions are given by
\begin{align}
N_2^{(p)}=\frac{\nu-2}{2\nu}
+\frac{\nu+2}{4\nu}\Big((5-2\sqrt{6})^p+(5+2\sqrt{6})^p\Big)\,.
\end{align}
  
For the first few $p = 0, 1, 2, \ldots$, we find\footnote{%
While both $\nu = 3$ and $\nu=8$  generate some integer solutions, neighbouring pairs are not both integer and so the original equation $R_\nu(N_1,N_2)=0$  will fail to have integer solutions.  
 $N_1 = \frac{7}{2}$ and $N_2 = 31$ for $\nu = 8$ form a solution pair, and the octonionic case will in general thus involve half integers.
The case $\nu=3$ similarly involves integers divided by three.
}
\begin{align}
\nu=1 : \; \; \; N_2 = {}& 1 , 7 , 73, 727, 7201, 71287 , \ldots \\
\nu=2 : \; \; \; N_2 = {}& 1 , 5 , 49 , 485, 4801, 47525  , \ldots \\
\nu=4 : \; \; \; N_2 = {}& 1 , 4, 37 , 364, 3601, 35644 , \ldots 
\end{align}
which matches the known results \cite{Osborn:2017ucf,Rychkov:2018vya,Osborn:2020cnf,Kousvos:2022ewl,Osborntalk}.

Choosing the  opposite sign for $c$ leads to negative rank gauge groups, which naively we should discard.
However, recalling the symmetry of the beta functions under $N \to -N$ discussed above, we 
recognise some expected patterns.
Only the choices $\nu=1$ and 2 lead to integer solutions:\footnote{%
For $\nu=2$ the integers are odd, while for 
$\nu=1$, they alternate between being divisible by two and four.
These parity and divisibility properties follow from the recursion relation, given the first two entries.
}
\begin{align}
\nu=1 : \; \; \; N_2 = {}& -2, -8, -74, -728, -7202 , -71288 
\ldots \\
\nu=2 : \; \; \; N_2 = {}& -1, -5, -49, -485, -4801 , -47525 
\ldots 
\end{align}
Note that the $\nu=1$ solutions are twice the positive $\nu=4$ solutions, in line with the usual identification that
$O(-2N)$ is $Sp(N)$.  Under $N \to -N$ for the $\nu=2$ solutions, we do not get anything new, reflecting that 
unitary groups are mapped to themselves under this transformation.

%%%%%%%%%%%%%%%%%%%%%%%%%%%%
%%  
%%%%%%%%%%%%%%%%%%%%%%%%%%%%

\input{bigtable.tex}

%%%%%%%%%%%%%%%%%%%%%%%%%%%%
%%
%%%%%%%%%%%%%%%%%%%%%%%%%%%%

\section{Mixed Bifundamental Theories}
\label{sec:UO}

We now consider two main examples where the bifundamental theories are 
extended to  $Sp(N_1) \times O(N_2)$ and $U(N_1) \times O(N_2)$  global symmetry.

To obtain such theories an additional term is added  to \eqref{Vphi} so that there are now three couplings. 
The total number of scalars remains as in \eqref{resN} but the potential becomes
\begin{align}
8\, V(\Phi) ={}&   \lambda \, \big ( \tr( \Phi \bPhi ) \big )^2 + g \,  \tr( \Phi \bPhi \Phi \bPhi )  
+  h \, \Phi_{ra} (\bPhi \Phi)_{ba} \bPhi_{br} \nn \\
= {}&  \lambda \,  \big ( \tr( \vphi^i {\tilde \vphi}^i) \big )^2 + g \, 
\tr \big ( \vphi^i {\tilde \vphi}^j  \vphi^k{\tilde  \vphi}^l  \big )f_{ijkl}
 + h \,  \tr \big ( \vphi^i {\tilde \vphi}^j  \vphi^k{\tilde  \vphi}^l  \big )f_{ikjl} \, , \nn \\
&  f_{ijkl} = \tfrac14 \big ( e_i {\bar e}_j e_k{ \bar e}_l + e_j {\bar e}_i e_l{ \bar e}_k+ e_k {\bar e}_l e_i{\bar e}_j
 + e_l {\bar e}_k e_j{\bar e}_i \big )  \, .
\label{Vphi2}
\end{align} 
There remains a single quadratic invariant $ \tr( \vphi^i {\tilde \vphi}^i )$. 
For $\nu=1$,  $V$
reduces to the $O(N_1)\times O(N_2)/{\mathbb Z}_2$ potential with, besides $\lambda$,  a single coupling $g+h$ 
so that  $h$  is redundant.
For $\nu = 2$, the symmetry is  $U(N_1)\times O(N_2)/{\mathbb Z}_2$. 
For $\nu=4$, the symmetry is  reduced to $Sp(N_1)\times O(N_2) /{\mathbb Z}_2$.   We henceforth focus on the $\nu=2$ and $\nu=4$ cases although we can write expressions for general $\nu$.  
 
 In terms of the couplings $\lambda$, $g$, $h$,
 the beta functions are
 \begin{align}
\beta_\lambda = {}& - \lambda + (N+8) \lambda^2 + 2\big ( (\nu \, N_1  + 2 )(g+h) 
+ (N_2-1) (\nu\,  g + (2-\nu )h)\big ) \lambda \, , \nn  \\
\noalign{\vskip - 2pt}
&{} +  3\,  \nu\,   g^2  +  (4-\nu )( 2\, g  +  h)h \, , \nn \\
\beta_g = {}& -g + 12 \, \lambda\, g + \big ( \nu( N_1+N_2- 4 ) +8 \big )g^2 \nn \\
\noalign{\vskip - 2pt}
&{} + 2 \big ((2-\nu) (N_2 -3 ) + 2 \big ) g\,  h +  \big ( \nu ( N_1  + N_2 - 2) - 2\, N_2 + 8  \big )  h^2   \, , \nn \\
\beta_h = {}& -h + 12 \, \lambda \, h + 2\big ( \nu (N_1-1) +6 \big )  g\,  h + 2 \big (  N_2  + \nu - 2 \big )  h^2  \, .
\label{mixedBetas}
\end{align}
 After shifting $\lambda$ to remove the $\lambda\, g$ and $\lambda \, h$ terms
 \be
{\hat \lambda} = \lambda + \tfrac{1}{N+2}\,  \big  ((\nu ( N_1+ N_2-1)  + 2) g 
+ (\nu ( N_1- N_2 +1)  + 2\, N_2 ) h  \big )  \, ,
\ee
 the $\beta$-functions can be expressed similarly to \eqref{betatwo} where $a\to a_{gg}, \, a_{gh}, \, a_{hh}$
 and $b \to b^g{\!}_{gg}, \,  b^g{\!}_{gh},\,  b^g{\!}_{hh}, \,  b^h{\!}_{gh}, \,  b^h{\!}_{hh}$. The necessary
 $a$ coefficients are
\begin{align}
a_{gg}  ={}&  \tfrac{3}{(N+2)^2} \, \nu (N_1-1)(N_2-1) (\nu\, N_1+2)(\nu\, N_2 + 2) \, , \nn \\
a_{gh}  ={}&  -\tfrac{6}{(N+2)^2} \, \nu (N_1-1)(N_2-1)(\nu\, N_1 + 2)  ((\nu-2) N_2 - 2)\, ,  \\
a_{hh}  ={}&  \tfrac{3}{(N+2)^2} \, (N_2-1)(\nu\, N_1 +2) (\nu^2 N_2(N_1-1) + 2\, \nu \nn
(N_1+ 2\, N_2  +3)-4\,N_2 -8) \, .
\end{align}
and the extremality condition can be written in terms of
 \begin{align}
 || \lambda ||^2 = S_{\nu}(N_1,N_2,\lambda,g,h) =3 N (N+2) \big ( {\hat \lambda}^2 
+ \tfrac16 ( a_{gg} \, g^2+ a_{gh}\, g\, h  + a_{hh} \, h^2 ) \big )\, .
\label{Sgh}
  \end{align}

By computing the Gr\"obner basis for $\{ \beta_{\hat \lambda}, \beta_g, \beta_h, 8|| \lambda ||^2 - N \}$ for general values of $\nu$, 
we find the extremality polynomial given by
\begin{align}
\label{Gnu2}
G_\nu = (N-4)^3\,  P_\nu(N_1, N_2)\, P_{OO}(\nu N_1, N_2) F_\nu(N_1, N_2) \ .
\end{align}
The first three factors in \eqref{Gnu2} signify that the three coupling bifundamental theory 
has fixed points of enhanced symmetry which are equivalent to the fixed points of the $O(N)$ model;
the $U(N_1) \times U(N_2)$ or $Sp(N_1) \times Sp(N_2)$ model depending on whether $\nu=2$ or 4; and 
the $O(\nu N_1) \times O(N_2)$ model respectively.
Novel instances of extremality correspond to roots of the last factor $F_\nu(N_1, N_2)$, which is given by
\begin{align}
\nu^2& F_\nu(N_1, N_2) \nn \\
= {}&
  \nu ^4 N_1{\!}^4 +\nu ^2 N_2{\!}^4+\nu ^2 (\nu ^2+20 \nu -15) N_1{\!}^2 N_2{\!}^2  \nn \\    \label{Fnu} 
& {}+ 
2 (3-5 \nu ) \nu^2 ( \nu N_1{\!}^3 N_2   + N_1 N_2{\!}^3   )  
+4 \nu ^3 (\nu +5) N_1{\!}^3  -8 \nu  (2   \nu +1) N_2{\!}^3  \nn  \\ \nonumber
& {}+
 4 \nu ^2 (\nu ^2-38 \nu +19) N_1{\!}^2 N_2 +36 \nu (\nu ^2+3 \nu -2) N_1 N_2{\!}^2  \\
& {}+
4 \nu ^2 (\nu ^2+12 \nu+41) N_1{\!}^2+8(9 \nu ^2+16 \nu +2) N_2{\!}^2
+8 \nu (9 \nu ^2-95 \nu +32) N_1 N_2   \nonumber  \\ 
&{}-
16 \nu  (5 \nu ^2 -25
   \nu -34) N_1 +32 (4 \nu ^2-29 \nu -2) N_2  -16 (11\nu ^2-40 \nu -52) \ .
\end{align}
The specific $\nu=2$ and $\nu=4$ cases that we are interested in are listed in Table \ref{tab:polynomials} where
\be
F_2(N_1,N_2) =  P_{UO}(N_1, N_2)\, , \qquad F_4(N_1,N_2) = P_{SpO}(N_1, N_2)\, .
\ee

Before studying the possible solutions of $P_{UO}(N_1, N_2)$ and $P_{SpO}(N_1, N_2)$ in detail,
let us first understand precisely how the enhanced symmetry fixed points can come about.
Clearly if $g=h=0$, the theory has $O(N)$ symmetry.  Also, if $h=0$ we return to the two coupling
bifundamental theories considered previously.  Less clear is that when $g=h$, there is an enhanced
$O(\nu N_1) \times O(N_2)$ symmetry.  This enhancement occurs because of the identity 
\be
\label{gequalsh}
\tr( \Phi \bPhi \Phi \bPhi )+  \Phi_{ra} (\bPhi \Phi)_{ba} \bPhi_{br} = 
2\,{\textstyle \sum_{i,j=1}^\nu} \, \tr(  {\tilde \vphi}^i \vphi^i\,  {\tilde \vphi}^j \vphi^j) \, ,
\ee
which follows from the fact that $\bar e_i e_j + \bar e_j e_i = 2 \delta_{ij}$. Consistent with this enhancement is the fact that 
\be
 S_{\nu}(N_1,N_2,\lambda,g,g) = S_{1}(\nu N_1,N_2,\lambda,g,g)=  S_{1}(\nu N_1,N_2,\lambda,2g,0)\, .
 \ee

The new fixed points associated with the zeroes of $F_\nu(N_1, N_2)$  come from an effective
two coupling system.  By restricting the couplings to the submanifold
\be
 h = - g + \frac{2(\nu-1)(N_2-2)}{\nu\,  (N_1+ N_2  -2) -2\, N_2 + 8} \, g\, ,
 \label{hg}
 \ee
 the $\beta$-functions can be brought into the two-coupling form  \eqref{betatwo} where
 \begin{align}
a ={}&  \frac{12  (\nu-1)(N_2-1)( \nu \, N_1+2)}{(N+2)^2( \nu (N_1+N_2 - 2)  - 2\, N_2 +8 )^2}\nn \\
&{}\times \Big (
{\genfrac{}{}{0pt}{3}{\scriptstyle{\nu^3  N_1{\!}^2(N_1-1)N_2 - \nu^2 N_1 (2\, N_1 N_2{\!}^2  
- N_2{\!}^3 - 13 \,N_1\, N_2  + 2 \,N_2{\!}^2  -2\, N_1 + 16\, N_2 -16)}}  
{\scriptstyle{-  2\, \nu (5 N_1 N_2{\!}^2 -24\,N_1\, N_2 - N_2{\!}^2 -4\,N_1 + 20\, N_2 -20) + 16 (N_2+2)} }}\Big ) 
\, , \nn \\
\noalign{\vskip 4pt}
 b ={}&  \frac{2 }{(N+2)(\nu (N_1+N_2 - 2)  - 2\, N_2 +8) }\nn \\
&{}\times \Big (
{\genfrac{}{}{0pt}{3}{\scriptstyle{\nu^3 N_1{\!}^2 N_2 (N_1+N_2-4) - 
\nu^2  N_1 (3\, N_1 N_2{\!}^2  - N_2{\!}^3 - 16 \,N_1\,N_2 -4 \,N_2{\!}^2  -
2\,N_1 + 42\, N_2 -28)}}
{\scriptstyle{-  2\, \nu (8\, N_1 N_2{\!}^2 -37\,N_1\, N_2 - N_2{\!}^2 +2\,N_1 + 32 \, N_2-28) + 8 (5\, N_2+2)} }} \Big ) \, .
\end{align} 
and the polynomial \eqref{Fnu} is given by
\be
\nu^2 F_{\nu}(N_1, N_2) = 
\tfrac14\big ( \nu (N_1+N_2 - 2)  - 2\, N_2 +8 )\big )^2\big ( 4 (4-N ) {a} + {b}^2 \big ) \, .
\ee

The precise form of this shift (\ref{hg}) suggests theories with $N_2=2$ are special in some way.
Indeed, in the $N_2=2$ case, the identity $\epsilon_{ab} \epsilon^{cd} = \delta^c_a \delta^d_b - \delta^c_b \delta^d_a$
can be used to rewrite the potential when $h=-g$.  For example in the $\nu=2$ case we have
\be
\tr \big ( \Phi \bPhi \Phi \bPhi \big )  - \tr \big ( \Phi \Phi^T \bPhi^T \bPhi \big  )  = 4 \,
\big ( \tr ( \Phi_1 \epsilon \Phi_2{\!}^T )\big )^2 \, ,
\qquad \Phi= \Phi_1 +  \Phi_2 \,i \, , \quad \epsilon = 
\left ( \begin{smallmatrix} 0 & 1\\ -1 & 0 \end{smallmatrix} \right ) \, .
\label{Udec}
\ee
Defining
\be
\Phi_\pm = \tfrac{1}{\sqrt{2}}\big ( \pm \Phi_1 + \Phi_2 \epsilon \big ) \, ,
\ee
so that $\Phi_{\pm,ra}$ are two $2N_1$-component real fields, then from \eqref{Vphi2} with
$\Phi_\pm{\!}^2 = \tr( \Phi_\pm \Phi_\pm{\!\!}^T  )$,
\be
V(\Phi)\big |_{h=-g} = \tfrac18(\lambda+ g) 
\big( (\Phi_+{\!}^2)^2+  (\Phi_-{\!}^2)^2 \big )  
+  \tfrac14(\lambda- g) \,  \Phi_+{\!}^2\, \Phi_-{\!}^2 \, , 
\label{Vbi}
\ee
 which in this form has an evident $O(2N_1)\times O(2N_1) \rtimes{\mathbb Z}_2$ symmetry.
 The potential in \eqref{Vbi} is identical with that for a biconical theory.\footnote{%
 A recent discussion
 is given in appendix B of \cite{Osborn:2017ucf}.}
 This defines a decoupled theory when $g= \lambda$
 and saturates the bound for $N_1=2$ corresponding to two decoupled $O(4)$ theories. Setting $\nu=2$, and $N_1$ or $N_2$ equal to 2  
 \begin{align}
 F_2(N_1, 2) ={}&  4 (N_1-2)^2 (N_1+2)^2\,, \nn \\
 F_2(2, N_2) ={}&  (N_2-2) (N_2-34) (N_2{\!}^2 -12 N_3 +44)\,.
 \end{align}

For $\nu=4$ and $h=-g$ there is a similar decomposition in terms of  complex fields $\Psi_1, \, \Psi_2$
by writing
\be
\Phi = \Psi_1 + \Psi_2\, j \, , \qquad {\bar \Phi} = {\bar \Psi}_1- j\, {\bar \Psi}_2 \, ,
\ee
and then \eqref{Udec} can be extended to
\be
 \tr( \Phi \bPhi \Phi \bPhi )  - \Phi_{ra} (\bPhi \Phi)_{ba} \bPhi_{br} = - 
 \big ( \tr(  \Psi_1 \epsilon {\bar \Psi}_1) -  \tr( \Psi_2 \epsilon {\bar \Psi}_2) \big )^2
 -4\, \tr(  \Psi_1 \epsilon {\bar \Psi}_2)  \,  \tr( \Psi_2 \epsilon  {\bar \Psi}_1) \, .
 \ee
For $(N_1,N_2)= (m,2)$ the symmetry group is then $U(2m)\times U(m)/U(1)$ \ and
 \be
 S_4( M,2,\lambda,g,-g)  = S_2(2M,2,\lambda-g,2g,0) \, .
 \ee

Through an exhaustive computer search for all values of $N_1$ and $N_2$ with $2 \leq N_1,N_2 \leq 1\,000\,000$, we have explicitly checked that $P_{UO}(N_1,N_2)$ only has two roots with max$(N_1,N_2)\leq 10^6$, namely the following:\footnote{If we allow for values of $N_1$ and $N_2$ less than two, we find another four integer roots: $(-5,-1)$, $(-4,-2)$, $(-2,2)$, and $(1,5)$.}
\begin{align}
\label{PUOroots}
&  P_{UO}(2,2)  = P_{UO}(2,34)=0 \,.
\end{align}
While the above two equations show that the pairs $(2,2)$ and $(2,34)$ satisfy a necessary condition for extremality, it can be straightforwardly verified that extremal fixed points indeed are present at these values.
The $N_1 = N_2 = 2$ case is the pair of decoupled $O(4)$ models just discussed, while the
$N_1 = 2$ and $N_2 = 34$ is our first genuinely new extremal fixed point.
 A third root is situated at $(N_1,N_2)=(1,5)$.  Note however that when $N_1=1$, the potential terms all become equivalent. The nontrivial fixed point is that of the $O(10)$ theory, which does not satisfy the Rychkov-Stergiou bound, illustrating an example where vanishing of the extremality polynomial is not sufficient to guarantee an extremal fixed point.\footnote{In fact, if one were to compute the Gr\"obner basis at fixed $N_1=1$ or fixed $N_2=5$, then $(N_1,N_2)=(1,5)$ would not be a root of the first element. This is something that can occur more generally for Gr\"obner bases and is unrelated to physics: a Gr\"obner basis computed at specific parameter values sometimes deviates from the generic answer in such a way that roots to the first element of the generic basis, which do not uplift to roots of the full basis, disappear.}
The $N_1 = 2$ and $N_2 = 34$ theory can be realised in a variety of other ways.  We will realise it as a particular case of a trifundamental theory later.  It can also be realised in a theory with a pair of identically transforming bifundamental fields -- either two $O(2) \times O(34)$ fields or two $U(2) \times U(17)$ fields.

Beyond the existence of the above roots, the only rigorous statement we are able to make is that $P_{UO}$ has at most finitely many integer roots. Let us relabel the arguments of $P_{UO}$ as $X$ and $Y$ instead of $N_1$ and $N_2$ to encourage a more abstract geometric view of the condition $0=P_{UO}$. The fact that the integer roots are finite follows from Siegel's theorem because the curve $P_{UO}(X,Y)$ has genus two as follows from (\ref{genusformula}):  it is a degree  four polynomial with one nodal point $(-2,2)$.  We will pursue a probabilistic analysis later on, but first we return to the $Sp(N_1) \times O(N_2)$ case.

Searching through the first few thousand values of $N_1$ and $N_2$ for $P_{SpO}$ 
yields only the integer solutions $(-2, -2)$, $(-1, 2)$ and $(1,4)$.  As in the $U(N_1) \times O(N_2)$ case, the solution with $N_1=1$ is degenerate; the potential terms are degenerate 
and the fixed point corresponds to the $O(16)$ model, which does not saturate the bound.  
The cases with negative values of $N_i$ do not saturate the bound either, and are difficult to make sense of physically.  
That said, the polynomial is invariant under $N_1 \to - N_2/2$ and $N_2 \to -2 N_1$,
reinterpreting $Sp(-N_2/2) \times O(-2N_1)$ as $O(N_2) \times Sp(N_1)$.  

Remarkably, $P_{SpO}(N_1, N_2)$ has two nodes, one at the integer solution $(-1,2)$ and the other at the rational solution 
$\left( -\frac{1}{5}, \frac{2}{5}\right)$.  The genus is thus one, and $P_{SpO}$ is secretly an elliptic curve.  
We will use this fact in the following to prove that the three solutions $(-2, -2)$, $(-1, 2)$ and $(1,4)$ are the only integer
solutions to $P_{SpO}(N_1, N_2) = 0$.

\subsection{Absence of other extremal fixed points for genus one}
\label{sec:StrTz}

It is a straightforward procedure to generate a large number of rational solutions to the equation $P_{SpO}(N_1, N_2)=0$ using the 
rational solutions found above.  For example, drawing a line through a regular rational point and a nodal rational point the line will intersect the quartic at one further point, which must also be rational.  Alternately, it is possible to  construct a conic 
$a_1 N_1^2 + a_2 N_1 N_2 + a_3 N_2^2 + a_4 N_1 + a_5 N_2 + a_6 =0$ that passes through the two nodal points and three regular (rational) points. Such a conic will intersect the quartic at one additional rational point.  In fact an infinite number of rational solutions can be constructed in this way.  As $P_{SpO}(N_1, N_2) = 0$ is an elliptic curve, by Mordell's Theorem, the rational points of
$P_{SpO}(N_1, N_2) = 0$ are a finitely generated Abelian group.  We will see below they correspond to the points on the lattice
$P = n P_1 + s T$ where $n \in {\mathbb Z}$ and $s \in \{0, 1 \}$.  (Below, 
we take the point $P_1 = (\frac{7}{10},\frac{8}{5})$ and $T = (1,4)$.)  
 What is less clear is that the only integer solutions are those we have found already: $(-2, -2)$, $(-1, 2)$ and $(1,4)$.

To demonstrate that we have in fact found all the integer solutions, we use the algorithm 
proposed in refs.\ \cite{stroeker1994solving,stroeker2003computing}.  Observationally (see Appendix \ref{sec:rationalpoints}), as $n$ gets large,
$P$ has rational coordinates with larger and larger numerators and denominators.  Number theorists define the naive height of a rational number to be $h(p/q) \equiv \operatorname{max}(\log|p|, \log|q|)$.   As $n$ gets large, the height of $P$ becomes so large that in fact it becomes hopeless to ever find an integer value again.  Following the algorithm of refs.\ \cite{stroeker1994solving,stroeker2003computing},  an integer solution can only be found for points $P$ with $|n| \leq 6$.  

Adapting the notation  to \cite{stroeker2003computing}, we change variables, $N_1=v$ and $N_2=u$,
and rename $P_{SpO}$:
\begin{align}
P_{SpO} \rightarrow f(u,v) = {}& 36 + 216 v + 420 v^2 + 144 v^3 + 16 v^4 - 108 u - 408 uv - 
 468 v^2 u \nonumber \\
 & {} - 136 v^3 u + 105 u^2 + 234 v u^2 + 81 v^2 u^2 - 18 u^3 - 
 34 v u^3 + u^4 \ .
\end{align}

The next step is to find a birational transformation that puts this quartic polynomial in Weierstrass form, 
$y^2 = x^3 + A x + B$ where $A$ and $B$ are integers.
The algcurves package in Maple \cite{maple} immediately gives
\be
\label{Weierstrassform}
y^2 = x^3 +680244480 x + 66648721563648
\ee
where
\begin{align}
x =& -\frac{1296}{(5u-2)(u+2)(u-2)}\Big(792-1500 u + 954 u^2-89 u^3 + 3264 v
- 4032 u v
\nonumber \\ &\hspace{7mm}
+1104 u^2 v+1152 v^2 - 1152 u v^2+128v^3\Big)
 \ , \nonumber \\[-12pt]
 \\ \nonumber
y =& - \frac{2239488}{(5u-2)(u+2)^2(u-2)}  \Big(-216+528 u-306 u^2+38 u^3 + 19u^4-672 v  \\
& \hspace{7mm} 
+1200 uv-528 u^2 v+36 u^3 v-264 v^2 +504 u v^2-258 u^2 v^2 - 32 v^3 + 32 u v^3 \Big) \ , \nonumber
\end{align}
using the option $[-2,-2,1]$ that pushes the integer point $(-2,-2)$ off to infinity. 
%We will refer to these relations as $x = X(u,v)$ and $y = Y(u,v)$.  
More complicated but similar looking relations express $u = U(x,y)$ and $v=V(x,y)$.  
Note, after the birational transformation, the solutions $(u,v) = (-2,-2)$ and $(2,-1)$ have to be handled separately.
The point $(u,v) = (-2,-2)$ is mapped to infinity in the $(x,y)$ plane while the nodal points $(2,-1)$ and $(2/5, -1/5)$ have been
blown up and resolved.

For an elliptic curve in Weierstrass form
$
y^2 = x^3 + A x + B
$,
the determinant and $j$-invariant are defined to be
\[
\Delta = -16(4A^3 + 27 B^2) = - 2^{38} 3^{36} 47 \ , \; \; \;
j_E = -\frac{2^{12} 3^3 A^3}{\Delta} = \frac{3375}{188} \ .
\]
Using SageMath \cite{sagemath}, we can find out straightforwardly some further facts about this elliptic curve.  It has rank one.\footnote{The rational solutions of an elliptic curve are given by a discrete Abelian group with two kinds of generators. There are torsion generators which, raised to a suitably high power, give  the identity and  there are also free generators with no such condition. The rank of an elliptic curve is the number of free generators \cite{curveRank}.}  
Its torsion generator is $T=(-34992,0)$ which is also the sole root of the curve on the real axis.  
The generator of the free part of the group can be taken to be $P_1 = (-11664, 7558272)$.  
Sage also produces a list of the integer solutions which include the two points above along with 
\[
(-11664, -7558272), (28512, \pm 10450944), (151632 , \pm 60466176) \ .
\]
Of these, only $T$ corresponds to an integer solution $(4,1)$ in the original $(u,v)$ coordinate system.  

$T$ and $P_1$ generate the Abelian group ${\mathbb Z} \times {\mathbb Z}_2$, a general element of which
looks like $P = n P_1 + s T$ where $n \in {\mathbb Z}$ and $s \in \{ 0, 1\}$.   The elements of this
group constitute all rational solutions of the elliptic curve and, with the exceptions
of the nodal points and the point at infinity, also of our original polynomial $f(u,v)$.  
The challenge now is to try to establish that none of these rational solutions are also integer solutions,
aside from the ones we have already found.

The reader is referred to \cite{stroeker1994solving,stroeker2003computing} for the technical details.  In broad outlines,
the idea is to map the abelian group of rational solutions to certain integrals over the elliptic curve, sometimes called elliptic logarithms:
\be
{\mathcal L}(P) = \int_{x(P)}^\infty \frac{dx}{\sqrt{x^3 + A x + B}} \ .
\ee
On the one hand, ${\mathcal L}(P)$ will be bounded above by $c |x(P)|^{-1/2}$ for $c$ an 
appropriately chosen constant.  If $x(P)$ is an integer, its naive height is simply $\log |x(P)|$, and so the integral
gives an upper bound on the height of integer points.  
On the other hand, there exists also a lower bound on ${\mathcal L}(P)$ due to a theorem of David \cite{david1995minorations} 
for elliptic logarithms
which generalises earlier work of Baker for ordinary logarithms \cite{baker1968linear}.
Putting the lower bound and upper bound together yields an upper bound on the values of $n$ which could conceivably 
give integer solutions.  This upper bound starts out extremely large, $|n| < 10^{26}$.  There is then a refinement procedure
which involves redefining the lattice that generates the rational solutions using extremely large lattice generators, of order
$(10^{26})^2$, that brings the bound down to a much more manageable $|n| < 15$.   
Running the refinement procedure a second time reduces the bound to $|n|<8$.  Using SageMath, we tabulated all of these
rational solutions, many of which are listed in  Appendix \ref{sec:rationalpoints}, establishing that we have indeed found all of the integer solutions.

\begin{figure}
    \centering
\begin{align*}
\hspace{-6mm}
\begin{matrix}\text{
\includegraphics[scale=0.65]{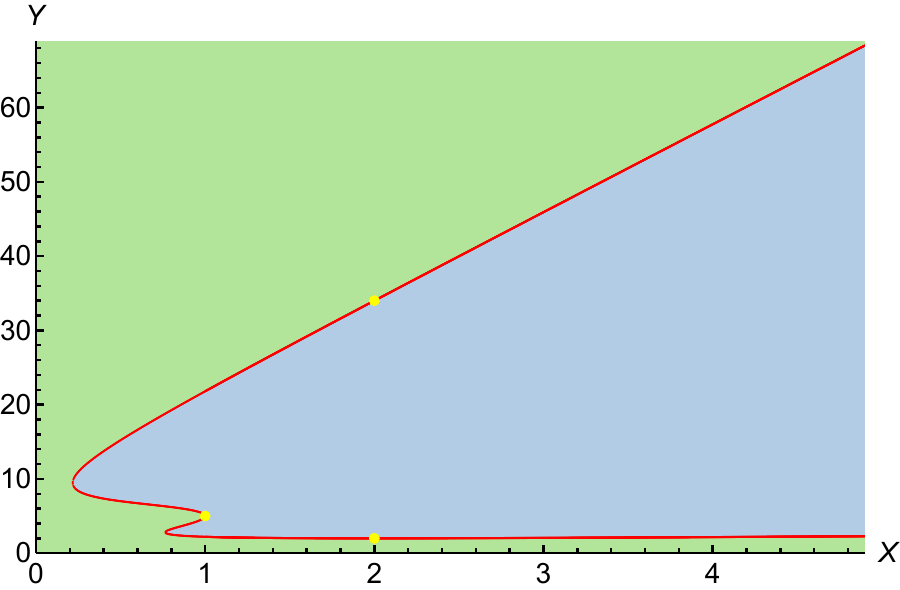}
}
\end{matrix}
\end{align*}
    \caption{Solution curve in red to the equation $P_{UO}(X,Y)=0$, separating the green region where $P_{UO}(X,Y)$ is positive from the blue region where $P_{UO}(X,Y)$ is negative. The three yellow points are the integer roots, situated respectively at $(X,Y)$ values given by $(1,5)$, $(2,2)$, and $(2,34)$. At large $X$, the curve converges on two straight lines whose precise forms are given in \cref{footnote} on page \pageref{footnote}.} 
    \label{fig:UOcurve}
\end{figure}

\subsection{Likely absence of other extremal fixed points for genus two}
\label{UOabsence}

We cannot say with certitude whether there are additional extremal fixed points in the $U(N_1) \times O(N_2)$ theory beyond those at \eqref{PUOroots} for $N_1$ or $N_2$ greater than $10^6$. However, adopting an approximate probabilistic reasoning, we will now argue that it is highly unlikely that $P_{UO}(X,Y)$ has any further integer roots. 

Figure~\ref{fig:UOcurve} shows a plot of the solution curve to $0=P_{UO}(X,Y)$. At large values of $X$, the curve converges to two straight lines\footnote{% 
The precise form of these lines is \label{footnote}
\begin{align}
Y = {}& \frac{1}{2}\left(7 + 2 \sqrt{6} \pm \sqrt{65+28 \sqrt{6}}\right)X + 5 + \sqrt{\frac{3}{2}} \pm \sqrt{\frac{3}{958}(3385 + 1858 \sqrt{6})} \ ,
\end{align}
where $r_l = 1.70525$, $r_u = 11.7285$, $b_l = 1.23954$ and $b_u = 11.2099$.
}
\be
Y_l = b_l + r_l X \ , \; \; \; Y_u = b_u + r_u X \ ,
\ee
with slopes $r_{l/u}$ and intercepts $b_{l/u}$. 
The subscripts $l$ and $u$ stand for ``lower" and ``upper".
For the specific irrational values of the slopes $r_{l/u}$ and intercepts $b_{l/u}$, it can be seen that the asymptotic straight lines intersect no integer points, but for any fixed distance, however small, there will be integer points within that distance from the asymptotic lines. The question then is whether the corrections to the solution curve compared to the asymptotic lines, arising from lower-order terms in the polynomial, ever bring the actual solution curve to pass through integer points.

Our probabilistic approach consists in looking at the integer points that lie close to the solution curve and thinking of $P_{UO}(X,Y)$ as an integer-valued random variable on these points. The solution curve passes through an integer point whenever this random variable assumes the value zero. We will argue that that the probability for this event to occur at a given value of $X$ scales as $X^{-3}$, and summing over all values of $X$, we find the expected number of integer roots. This kind of reasoning can be applied to both the upper and the lower branch of the solution curve. For explicitness, let us first look at the integer points close to the lower branch.

\begin{figure}
    \centering
\begin{align*}
\hspace{-6mm}
\begin{matrix}\text{
\includegraphics[scale=0.65]{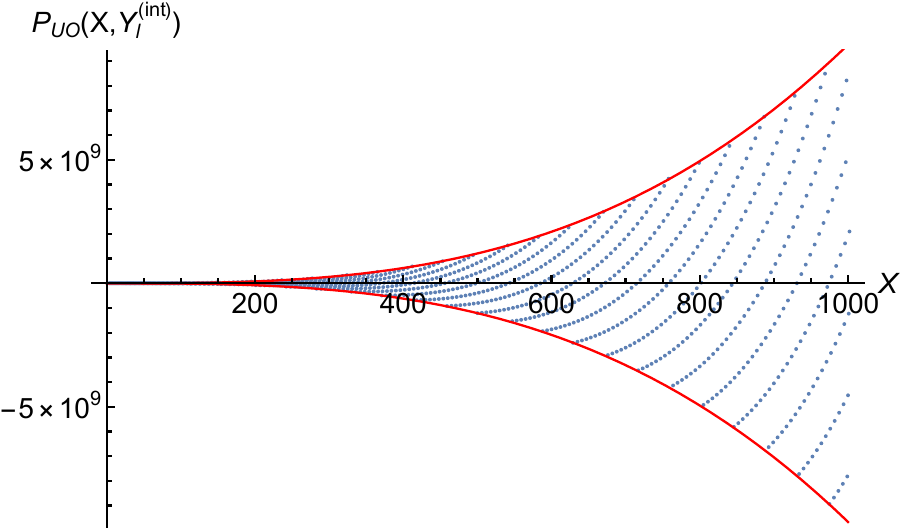}
}
\end{matrix}
\end{align*}
    \caption{Plot of the polynomial $P_{UO}(X,Y)$ when evaluated as a function of $X$ at $Y$ equal to the integer $Y_l^{\text{(int)}}$ that is the closest to the lower part of the solution curve for $P_{UO}(X,Y)=0$. A similar plot can be drawn for the upper part of the solution curve. The two red curves that form an envelope to the plot points are given by plus/minus one half the range given in equation \eqref{eq:range}, which grows as $X^3$.}
    \label{fig:intVals}
\end{figure}

For any large integer value of $X$, let $Y_l(X)$ be the smallest of the two real-valued solutions to the equation $P_{UO}(X,Y)=0$, and let $Y_l^{(\text{int})}$ be an integer nearest to $Y_l(X)$. Furthermore, let us define
\begin{align}
\delta Y(X) \equiv   Y_l^{(\text{int})}(X) - Y_l(X)\,.
\end{align}
From the definition of $Y_l^{(\text{int})}$ it follows that $\delta Y\in (-1/2,1/2)$. Consider now the polynomial $P_{UO}(X,Y)$ when evaluated at the integer point $(X,Y_l^{(\text{int})})$. Depending on the value of $X$, the evaluation can assume a range of values, which we can estimate as follows:
\begin{align}
\label{range}
\text{Range}\Big[P_{UO}(X,Y_l^{(\text{int})})\Big] \approx \underset{\delta Y\in(-1/2,1/2)}{\text{max}}P_{UO}(X,Y_l+\delta Y)
\,\,\,-\underset{\delta Y\in(-1/2,1/2)}{\text{min}}P_{UO}(X,Y_l+\delta Y)\,.
\end{align}
At large values of $X$, $Y_l$ scales as $X$. Since $P_{UO}(X,Y)$ is a fourth-order polynomial, an expansion in $Y$ around a root gives a range \eqref{range} that scales as $X^3$. A more precise estimate, obtained using the leading terms of $P_{UO}(X,Y)$, gives
\begin{align}
\label{eq:range}
\text{Range}\Big[P_{UO}(X,Y_l^{(\text{int})})\Big] \approx 2\,|2r_l^3-21r_l^2+29r_l-14|\,X^3\,.
\end{align}
Figure~\ref{fig:intVals} displays a plot of $P_{UO}(X,Y_l^{(\text{int})})$ as a function of $X$, where the $X^3$ growth of the range of values manifests itself in the growth of the envelope inside of which the plot points are contained.

As a crude way of finding out how many integer roots to expect, we make the assumption that as a heuristic, average viewpoint, we can treat all integer values of $P_{UO}(X,Y_l^{(\text{int})})$ within the allowed range as equally likely. This means that the probability $\text{Pr}_l(X)$ to get a zero is given roughly by\footnote{Even a quick glance at Figure~\ref{fig:intVals}, makes it is clear that it is not a reasonable approximation to assume that the values of $P_{UO}(X,Y_l^{(\text{int})})$ are independently distributed: the values of neighbouring points are strongly correlated and differ by only a small fraction of the allowed range. But, by the linearity of expectations, this fact makes no difference for the purpose of counting expected number of integer roots.}
\begin{align}
\text{Pr}_l(X) \approx \frac{1}{\text{Range}\Big[P_{UO}(X,Y_l^{(\text{int})})\Big]}\,.
\end{align}
We can go through the same steps to estimate the probability $\text{Pr}_u(X)$ that the upper branch of the solution curve passes through an integer point at a given value of $X$. To estimate the expected number $n$ of integer roots for $X$ greater than some value $X_0$, we sum over the probabilities for both parts of the curve with $X>X_0$: 
\begin{align}
\left<n \right>_{X> X_0}  = \sum_{X=X_0+1}^\infty \text{Pr}_l(X)
+\sum_{X=X_0+1}^\infty \text{Pr}_u(X)\,.
\end{align}
The fact that the above sums converge accords with the stipulation of Siegel's theorem that this polynomial has a finite number of integer roots. In our case, for the value $X_0=10^6$, we arrive at the following estimate for the number of integer roots
\begin{align}
\left<n \right>_{X> X_0}  &\approx  
\frac{1}{2}\bigg(
\frac{1}{|2r_l^3-21r_l^2+29r_l-14|}
+
\frac{1}{|2r_u^3-21r_u^2+29r_u-14|}
\bigg)
\bigg(\zeta(3)-\sum_{X=1}^{X_0}\frac{1}{X^3}\bigg)
\nonumber \\
&\approx 3\cdot 10^{-14}\,.
\end{align}
As this value is much less than one, we wager that $P_{UO}(X,Y)$ has no integer roots in addition to those given in equation \eqref{PUOroots}.

\section{Theories with Three Couplings and Permutation Symmetry}
\label{sec:MultiConical}

Interesting examples of theory with potentially relevant fixed points can be obtained by extending
the bifundamental examples by introducing an additional interation which reduces the overall symmetry to
$G\times {\cal S}_n$ with $G=O(m), \, U(m), \, Sp(m)$. 
With fields as before $\Phi_{ra}$, $a=1,\dots, m, \  r=1,\dots , n$, where $\Phi_{ar}$ takes real, complex o
quaternionic  values so that  $N=\nu \, m n$, the potential involves three couplings
\be
 8\,   V(\Phi) =  \lambda \, \big ( {\textstyle \sum_r} \Phi_r {\bar \Phi}_r \big )^2
     +   g \, {\textstyle \sum_{r,s}} \Phi_r {\bar \Phi}_s \, \Phi_s {\bar \Phi}_r 
 +  h \, {\textstyle \sum_{r}}( \Phi_r {\bar \Phi}_r)^2 \,. 
 \label{Vthree}
\ee
For $h=0$ this corresponds to the bifundamental theories already considered. For $g=0$ 
this reduces the so called $M\! N$ theory which has a symmetry group with symmetry $O(\nu m)^n \rtimes {\cal S}_n$ \cite{Pelissetto}.

The essential lowest order $\beta$-functions, after our conventional rescaling of the couplings, 
for the various cases are just
\begin{align}
\beta_\lambda = {}& - \lambda + (N +8) \lambda^2 
+ 2\big (  \nu (m+n-1) + 2 \big )g\, \lambda  + 2 (  \nu \, m+  2 ) h \, \lambda + 2\, \nu\,g\, h 
+ 3\, \nu \,g^2   \, , \nn \\  
\beta_g = {}& - g + 12\, g\, \lambda + \big  ( \nu(m+n-4)+8 \big ) g^2  + 4 \,g \,h\, , \nn \\
\beta_h = {}& - h + 12\, h\, \lambda + 2\big ( \nu(m-1)+6 \big )  \, g\, h +  (  \nu \,  m +8 ) h^2 \, .
\label{Bsym}
\end{align}
With the shift
\be
 {\hat \lambda}  = \lambda +\tfrac{1}{N+2} \Big (   \big(\nu (m+n-1) + 2 \big )g + (\nu \, m +2) h \Big )  \, , 
 \label{hlam}
 \ee
 the canonical form \eqref{betatwo} is achieved and
 \begin{align}
 a_{gg} ={}& \tfrac{3}{(N+2)^2}\, \nu \, (m-1) (n-1)(\nu\, m+2)(\nu\, n +2)\, , \nn  \\
 a_{gh} = {}& \tfrac{12}{(N+2)^2}\, \nu \, (m-1) (n-1)(\nu\, m+2) \, ,  \nn \\
 a_{hh} = {}&  \tfrac{6}{(N+2)^2}\, \nu \, m (n-1)(\nu\, m+ 2) \, ,
 \end{align}
 and
 \be
  || \lambda ||^2 = 3 N (N+2) \big ( {\hat \lambda}^2 
+ \tfrac16 ( a_{gg} \, g^2+ a_{gh}\, g\, h  + a_{hh} \, h^2 ) \big ) \, .
\ee
 
 Fixed points can be easily found by reducing the RG flow to two couplings. Besides
 the trivial Gaussian fixed point and the one with maximal $O(N)$ symmetry when $g=h=0$, there
 are the $M\!N$ fixed points obtained for $g =0$,
 \begin{align} 
 \lambda={} & 0  \, , \quad h = \frac{1}{\nu \, m+8}\, , \qquad \kappa = \frac{\nu\, m -4 }{\nu\, m+8} \, , \nn \\
 \lambda= {}& \frac{4-\nu \, m}{N (\nu \, m+8)- 16(\nu\, m-1)}\, , \ \
  h = \frac{N-4}{N (\nu \, m+8)- 16(\nu\, m-1)}\, , \\
  & \kappa =- \frac{(4- \nu \, m)(4-N)}{N (\nu \, m+8)- 16(\nu\, m-1)} \, \nn .
 \end{align}
 The first case corresponds to $n$ decoupled $O(\nu\,m)$ theories. The extremality bound
 is saturated when the stability matrix eigenvalue vanishes but when this is achieved for
 $\nu\, m=4$, the two fixed points coincide and this is then $n$ decoupled $O(4)$ theories.
 Fixed points with $h=0$ are just those of the bifundamental theories.
  
 The lowest order RG flow can also be restricted to two dimensions by imposing
 \be
 h= - \frac{\nu (m-n+2)+4 }{\nu \, m+4}\, g \, .
 \ee
 With this assumption and from \eqref{hlam}
 \be
 {\hat \lambda}  = \lambda +\frac{\nu \, m (2\, n -3 )+ 6 n -8 }{(\nu\, m+4)(N+2)} \,\nu\,  g \, , 
 \ee
 $\beta\raisebox{-1.5 pt}{$\scriptstyle {\hat \lambda}$},\ \beta_g $ take the form given in \eqref{betatwo} with
 \begin{align}
 a= {}& \frac{3\, \nu (n-1)(\nu\, m+2)}{(\nu\, m+4)^2(N+2)^2} 
 \Big ({\genfrac{}{}{0pt}{3}{\scriptstyle{\nu^3 m^2 n (m-1)+ 2\, \nu^2 m (n^2+4\, m \,n -10\,n  + m + 8)}}  
{\scriptstyle{ + 16\, \nu (m\, n +m -2\, n + 2) +32 } }}\Big ) \, , \nn \\
b= {}& \frac{1}{(\nu\, m+4)(N+2)} 
 \Big ({\genfrac{}{}{0pt}{3}{\scriptstyle{\nu^3 m^2 n (m+n-4)+ 2\, \nu^2 m (4\, n^2+4\, m \,n+ m  -23\,n + 14)}}  
{\scriptstyle{ + 8\, \nu (2\, m\, n + 2\, m -7\, n  + 6) +32 } }}\Big ) \, .
 \end{align}
 The extremality polynomial is then given by
 \be
 \nu^4 F_\nu(m,n) = (\nu\, m+4)^2 \big ( 4 ( 4-N) a + b^2 \big ) \, .
 \ee
 For the allowed values of $\nu$,
 \begin{align}
 F_1(m,n)={}& 64 + 224\, m + 64\, n + 132\, m^2 -96 \, m\, n  + 16\, n^2 + 20\, m^3-  84 \, m^2 n - 8\, m \, n^2\nn \\
 \noalign{\vskip -5pt}
 &{}   + m^4 - 10\, m^3 n + m^2 n^2 \, ,  \nn \\
  F_2(m,n)={}& -32 + 32\, n + 44\, m^2  + 8 \, m\, n  + 4\, n^2 + 12\, m^3-  40 \, m^2 n - 4\, m \, n^2\nn \\
 \noalign{\vskip -5pt}
 &{}   + m^4 - 10\, m^3 n + m^2 n^2 \, ,   \nn \\
 F_4(m,n)={}& (m-1){\hat F}_4(m,n) \, , \nn \\
 \noalign{\vskip -2pt}
  {\hat F}_4(m,n) = {}& 11 +27 \, m  -10 \, n + 9\, m^2  - 28 \, m\, n  - n^2 + m^3 -  10 \, m^2 n + m \, n^2 \, .
 \end{align}
 
\begin{figure}
    \centering
\begin{align*}
\hspace{-6mm}
\begin{matrix}\text{
\includegraphics[scale=0.65]{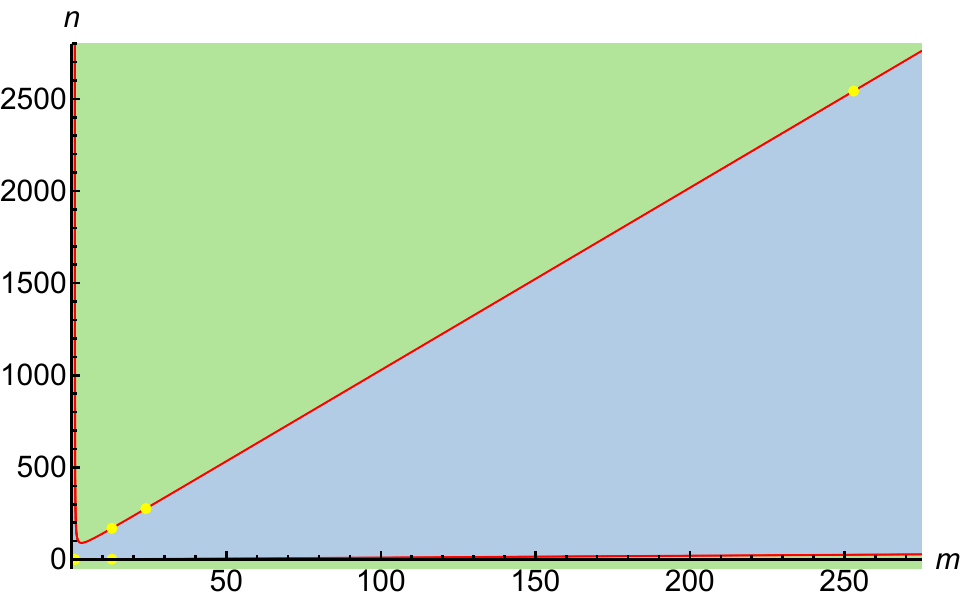}
}
\end{matrix}
\end{align*}
    \caption{Solution curve in red to the equation ${\hat F}_4(m,n)=0$, separating the green region where 
    ${\hat F}_4(m,n)$ is positive from the blue region where ${\hat F}_4(m,n)$ is negative. The five yellow points are the integer roots as given in \eqref{fourExtremes} along with a physically uninteresting root at $(1,1)$.} 
    \label{fig:SpmSnCurve}
\end{figure}

 ${\hat F}_4$ has genus zero and has integer solutions with $m,n>1$ given by
 \be
\label{fourExtremes}
{\hat F}_4(m,n) = 0 \ \ \mbox{for} \ \  (m,n)= (13, 2), (13 ,170), (24,277), (253,2542) \ .
\ee
The solution curve for ${\hat F}_4(m,n)=0$ is plotted in Figure~\ref{fig:SpmSnCurve}.
 Both $F_1$ and $F_2$ have genus one and so are elliptic curves.  
The elliptic curve corresponding to $F_1$ has rank zero and so therefore supports a finite number of rational points.
$F_2$ on the other hand has rank 1 and has an infinite number of rational solutions of which some may be integer. 
(Naively extrapolating the beta functions
 to the octonions ($\nu=8$) and the sedenions ($\nu=16$), we find further elliptic curves, of rank
one and zero respectively.)

 If a Gr\"obner basis is calculated for the polynomial ring given by the full set of one loop $\beta$-functions $\beta_{\hat \lambda}$, $\beta_g$, $\beta_h$ and 
also $8 || \lambda ||^2 - N$,
then for generic values of $n$ and $\mu$ the resulting extremality polynomial factorises, up to unimportant prefactors, as
\be
\label{GmnFormula}
G(m,n) = %m\, \nu^7 
\left( N - 4 \right)^3  \left( m \nu - 4 \right)^2 P_\nu(m,n) F_\nu(m,n) \ ,
\ee
with $P_\nu(m,n)$ given in equations \eqref{Rmn} and \eqref{bifundPnu}. The different factors are associated with the different fixed points discussed above.  
The factor $N-4$ arises because of the $O(N)$ symmetric fixed points when $g=h=0$.  The factor $m \nu - 4$
corresponds to the $M\!N$ theory arising for $g=0$.  The $ P_\nu(m,n) $ factor is associated with the bifundamental theory
which appears for $h=0$.

The results for the three new cases are discussed  further below.

\subsection{
\texorpdfstring{$Sp(m) \times {\mathcal S}_n$}{Sp(N) x Sn}}
\label{sec:SpNxSpv2}

New extremal solutions, if they exist, will correspond to integer points along the plane curve
\be
C = \{ (m,n ) : {\hat F}_4(m,n) = 0  \} \ .
\ee
Since ${\hat F}_4(m,n)$ is quadratic in $n$ it is possible to use the same Pell equation $x^2 - 6 y^2 = c^2$ 
that appeared before to compute  infinitely many rational solutions.
As we are interested in integer solutions, it is convenient to take a different approach.
The curve $C$ would ordinarily be genus one, but because of the presence of an ordinary double point at $(m,n) = (-1,2)$ it
is genus zero.
Note also that it has three lines along which it reaches infinity, corresponding to the three roots of
$m^3 - 10 m^2 n + m n^2$.  As a result, due to a theorem of Maillet \cite{Maillet1918, Maillet1919}, 
it will have only a finite
number of integer solutions.
  
 To find these integer solutions, we make use of a birational transformation that maps the complex plane
 into the curve $C$. (A general algorithm for solving such equations is presented in ref.\ \cite{POULAKIS2000573}, which we essentially follow here.)
 We use the following parametrisation:
 \be
m= \frac{13 + 14 S + S^2}{1 - 10 S + S^2}
 \ , \; \; \;
 n = \frac{2(85+9S-9S^2+S^3)}{(1+2S)(1-10S+S^2)} \ ,
\ee
where $S \in \mathbb{C}$.
Note that infinity is reached at three different finite values of $S$.  

To establish all integer solutions,
a necessary condition that the pair $(m,n)$ be integer is that $n$ alone be integer, and so we focus on
finding all rational values of $S$ such that $n$ is integer.
The fact that the denominator in the expression for $n$ is higher degree leads to a stronger necessary condition
than if we were to insist $m$ be integer.

Consider the homogenisation of the map for $n$,
\be
U(R,S) = 2(85 R^3+9S R^2-9S^2 R+S^3) \ , \; \; \;
W(R,S) = (R+2S)(R^2 - 10 S R + S^2) \ .
\ee
where now $n = U(1,S) / W(1,S)$.  
Note that
\begin{align}
37500 R^5 = {}& 
(222 R^2 - 35 R S + 2 S^2) U(R,S) + (-240 R^2 +34 R S - 2 S^2) W(R,S)  \ , \nn \\
37500 S^5 = {}&
(66714 S^2 + 26630 R S - 3371 R^2) U(R,S) \nn \\
\noalign{\vskip -3pt}
&{} + (-47964 S^2 + 118138 R S + 573070 R^2) W(R,S) \, .
\end{align}
If we find an integer solution for $n$, then $U(R,S)/W(R,S)$ must be integer for some coprime integers $R$ and $S$, which in turn implies
that $W(R,S)$ must divide $37500$.  The problem then reduces to finding all pairs $(R,S)$ such that $\mbox{gcd}(R,S) = 1$
and $W(R,S)$ is equal to a divisor $l$ of $37500=2^2\cdot 3 \cdot 5^5$.

Solving $W(R,S) = l$ there are pairs $(R,S)$ such that $(R+2S) = l_1$ and $R^2 - 10 S R + S^2=l_2$ where $l = l_1 l_2$ 
and $l_1$ and $l_2$ are integers.  The algorithm is implemented in Mathematica's Reduce function \cite{Mathematica}.
The 20 allowed $(R,S)$ pairs are 
\begin{align*}
&(1,0), (0,1), (1,-1),(10,1),(2,1),(1,2),(1,-2),(3,1),(8,1),(1,7), \\
&(1,12),(21,2),(1,-13),(23,1),(17,4),(7,9),(4,23),(61,7), (12,119), (208, 21) \ .
\end{align*}
Flipping the sign of both $R$ and $S$ is also a solution but for $-l$.  
These pairs do not all give integer solutions, but the ten that do give rise to the $(m,n)$ values
\begin{align}
(13, 170), (1,1),(0,-11),(-3,-2),&(-8,-47),(-63,-587),(13,2), \nn \\
(253,2542),(0,1),&(24,277) \ . 
\end{align}
The solutions with obvious physical interest are $(13,2)$, $(13,170)$, $(24, 277)$, and $(253, 2542)$.  
The solutions with negative values for $m$ do not seem to have corresponding fixed points for the 
$O(m) \times {\mathcal S}_n$
theories.  The solution $(1,1)$ is just the $O(4)$ model.

\subsection{
\texorpdfstring{$U(m) \times {\mathcal S}_n$}{U(m) x Sn}
}

The maximal factor $F_2(m,n)$
has nodes at $(m,n) = (-2,2)$ and also at infinity and so its genus is one instead of three.  The only other integer solutions
we were able to find were $(-5,-1)$ and $(1,5)$.  None of these produce saturating fixed point solutions to the beta functions.

We can repeat the analysis that was performed on the $Sp(N_1) \times U(N_2)$ example earlier.  There is a birational transformation
\begin{align}
n = \, & \frac{-9491323392 + 21570624 u - 7992 u^2 - 5 u^3 - 342144 v + 432 u v}{(936-u)(1368-u)^2} \ , \\
m = \, & \frac{504-u}{936-u} \ ,
\end{align}
 which puts the curve in Weierstrass form: 
\be
\label{weierstrass2}
v^2 = u^3 - 1881792 u + 981642240 \ .
\ee
Note this transformation pushes the point $(1,5)$ off to infinity and resolves the nodal points.  
Sage then tells us that this curve has rank 1 and so supports an infinite number of rational solutions.  
We repeat the same analysis that we used on the $Sp(N_1) \times U(N_2)$ described in section \ref{sec:StrTz}.
There is again a lattice of rational points $P = p P_1 + s T$ where $p \in {\mathbb Z}$ and $s \in \{0, 1\}$.
(We choose $u(T) = 720$ and $u(P_1) = 1368$.  The first few rational solutions are tabulated in appendix \ref{sec:rationalpoints}.)
We find that any integer solutions must have $|p|<9$.  Tabulating these points using Sage, indeed none of them
except for $(-5,-1)$ correspond to integer solutions in the original coordinate system.  Thus, we have found all possible integer solutions.

\subsection{
\texorpdfstring{$O(m) \times {\mathcal S}_n$}{O(m) x Sn}}

Now we look for integer solutions of the quartic polynomial $F_1(m,n) = 0$.
This quartic polynomial has a nodal point at $(m,n) = (-4,2)$ and another one at infinity and thus 
again its genus reduces to one.
 A birational transformation that maps the integer point $(-2,0)$ to infinity 
\begin{align}
u = {}& -\frac{384(2m-5)}{m+2} \ , \\
v = {}&\frac{6912(5 m^3 - m^2 n + 42 m^2 + 8 m n + 48 m - 16 n - 32)}{(m+4)(m+2)^2} \ ,
\end{align}
converts the elliptic curve to standard Weierstrass form
\be
v^2 = u^3 - 1437696 u + 495452160 = (u-960)(u-384)(u+1344) \ .
\ee
This curve has rank zero and thus a finite number of rational points. The torsion group is ${\mathbb Z}_2 \times {\mathbb Z}_4$, which means all the points can be enumerated in the form $P = s_1 T_1 + s_2 T_2$ where $s_1 \in \{0, 1 \}$ and $s_2  \in \{0, 1, 2, 3 \}$.
Taking $(u(T_1), v(T_1)) = (-1344,0)$ and $(u(T_2), v(T_2)) = (-192, -27648)$ there is the following table
\[
\begin{array}{|c|c|c|c|}
\hline
& (u,v) & (m,n) \\
\hline
T_1 & (-1344,0) & (-8,-2) \\
T_2 & (-192, -27648) & \infty \\
2T_2 & (960,0) & (0,-2) \\
3 T_2 & (-192, 27648) & (4,2) \\
T_1 + T_2 & (2112, 82944) & ( -\frac{4}{5},-\frac{22}{5}) \\
T_1 + 2 T_2 & (384,0) & (1,7) \\
T_1 + 3 T_2 & (2112, -82944) & (-\frac{4}{5},\frac{2}{5})
\\
\hline
\end{array}
\]
Note we have lost the nodal point $(-4,2)$ in the birational transformation and pushed the point $(-2,0)$ off to infinity.
In total, the six integer solutions of the original plane curve $F_1(m,n)=0$ are 
\be
(m,n) = (-8,-2), (-4,2),(-2,0),(0,-2),(1,7),(4,2) \ .
\ee
Of these integer solutions, only $(-2,0)$ and $(2,4)$ have fixed points which saturate the bound, and only (4,2) has an obvious interpretation as a field theory.  In fact, it is two copies of the O(4) model 

\section{Trifundamental Theories}
\label{sec:Trifund}
The discussion of bifundamental theories is naturally extended to fields with three indices.
The symmetry groups are then $G_1 \times G_2 \times G_3$.

\subsection{Trifundamental unitary theory: 
\texorpdfstring{$U(N_1)\times U(N_2)\times U(N_3)$}{U(N1) x U(N2) x U(N3)}
}

We now advance to theories of fields carrying three indices. The first family of such theories we will study is the family of quantum field theories of complex fields $\phi^{abc}$ carrying three indices that each transform under their own unitary group.\footnote{Recent literature on tri-unitary theories includes Refs.~\cite{benedetti2019phase,pascalie2019large,benedetti2020sextic,pascalie2021correlation}.} The theories contain three quartic operators, which are depicted in row four of Table~\ref{tab:operators}. Explicitly, the potential can be expressed as 
\begin{align}
8\,V(\phi) &= \lambda_0 \, \phi^{\ra1 \bb1 \gc1} \bar \phi_{\ra1 \bb1 \gc1} \phi^{\ra2 \bb2 \gc2} \bar \phi_{\ra2 \bb2 \gc2} 
+ \lambda_1 \, \phi^{\ra1 \bb1 \gc1} \bar \phi_{\ra1 \bb1 \gc2} \phi^{\ra2 \bb2 \gc2} \bar \phi_{\ra2 \bb2 \gc1} \\
&+\lambda_2 \, \phi^{\ra1 \bb1 \gc1} \bar \phi_{\ra1 \bb2 \gc1} \phi^{\ra2 \bb2 \gc2} \bar \phi_{\ra2 \bb1 \gc2} 
+\lambda_3\,  \phi^{\ra1 \bb1 \gc1} \bar \phi_{\ra2 \bb1 \gc1} \phi^{\ra2 \bb2 \gc2} \bar \phi_{\ra1 \bb2 \gc2} \ . \nonumber
\end{align}
The total number of real scalars is
\be
N=2N_1N_2N_3 \, .
\ee
The coupling $\lambda_0$ corresponds to the interaction with maximal $O(N)$ symmetry 
and $\lambda_1, \lambda_2, \lambda_3$ are related by permutation symmetry.
The resulting one-loop beta functions are given by 
\begin{align}
\beta_{0}
= {}&-\lambda_0
+\
\big(
(N+8) \lambda_0^2 +  
 6 \lambda_1^2  +  6 \lambda_2^2 + 6 \lambda_3^2 + 4 N_1 \lambda_2 \lambda_3
  + 4 N_2 \lambda_1 \lambda_3 + 4 N_3 \lambda_1 \lambda_2 
 \nn \\
 &\hskip 1.5cm{} + 4 \big ( (N_1 N_2+N_3)  \lambda_3 +  (N_1N_3+N_2)  \lambda_2  
  +  (N_2N_3+N_1) \lambda_1 \big ) \lambda_0  \big)\,, \nn 
\\
\beta_{1}
= {} &-\lambda_1
+ 2
\big(
6 \lambda_0 \lambda_1 + (N_2 N_3+N_1) \lambda_1^2  + 4 \lambda_2 \lambda_3+ 
2 N_2 \lambda_1 \lambda_2 +   2 N_3 \lambda_1 \lambda_3
\big)\,, \nn
\\
\beta_{2}
={}&-\lambda_2
+2
\big(6 \lambda_0 \lambda_2 +  (N_1 N_3+N_2) \lambda_2^2 + 4 \lambda_1 \lambda_3 + 2 N_1 \lambda_1 \lambda_2 +  2 N_3 \lambda_2 \lambda_3\big)\,,
\nn \\
\beta_{3}
= {} &-\lambda_3
+ 2
\big(6 \lambda_0 \lambda_3 +  ( N_1 N_2+N_3) \lambda_3^2
+ 4 \lambda_1 \lambda_2 + 
 2 N_1 \lambda_1 \lambda_3 + 2 N_2 \lambda_2 \lambda_3 
\big)\,.
\end{align}
The sum of squares of couplings evaluates as follows: 
\begin{align}
 || \lambda ||^2
= {}&
3\, N \Big(
(N+2)\big ( \lambda_0^2 + \lambda_1^2   + \lambda_2^2 +   \lambda_3^2 \big )\nn \\
  &{}+  4\big ( (N_1 N_2+N_3) \lambda_3 
      +   (N_1N_3+N_2 ) \lambda_2 
   + (N_2N_3+N_1) \lambda_1 \big ) \lambda_0  \nn \\
&{}   + 4\big (  (N_2N_3+N_1) \lambda_2\lambda_3 + (N_1N_3+N_2) \lambda_1 \lambda_3
  + ( N_1 N_2 + N_3) \lambda_1 \lambda_2\big )  \Big)\,.
\end{align}
By computing the Gr\"obner basis with respect to the beta functions and the polynomial $8||\lambda ||^2-N$, we obtain the extremality condition $0=G_{UUU}$, where\footnote{$G_{UUU}$ can in principle be calculated by determining the Gr\"obner basis as a function of variables $N_1$ and $\lambda_0$ to $\lambda_3$, leaving $N_2$ and $N_3$ as generic parameters, but computationally this procedure is quite demanding. A more practical way is to compute the Gr\"obner basis for many different fixed values of $N_2$ and $N_3$ and then use this data as input to fix all the coefficients in $G_{UUU}$.}
\begin{align}
\label{GUUU}
G_{UUU}={}&
P_U(N_1N_2N_3)^2
\,P_{UU}(N_1N_2,N_3)
\,P_{UU}(N_1N_3,N_2)
\,P_{UU}(N_2N_3,N_1)\nn \\
&{}\times P_{UUU}(N_1,N_2,N_3)\,.
\end{align}
The first factor is indicative of the fact that the tri-unitary model contains the $U(2)$, or $O(4)$, extremal fixed point. The second to fourth factors are given in terms of the polynomial $P_{UU}$ defined in equation \eqref{Pbifund} and represent the fact that the bi-unitary theories and their extremal fixed points are contained inside the tri-unitary model. Finally, there is the last factor, the maximal factor $P_{UUU}$. This polynomial is quite lengthy and bears witness to the explosion in complexity that transpires as we gradually consider more generic classes of theories. The polynomial is symmetric in its three arguments and is of order 18 in each argument and of order 36 in products of arguments. In total the polynomial has $802$ terms. The absolute values of the coefficients have a median value of about $4.9\cdot 10^{6}$ and a mean value of roughly $2.4\cdot 10^{11}$. We write down the exact form of the polynomial in Appendix~\ref{appendix}.

\subsection{Likely absence of new extremal fixed points}
\label{UUUabsence}

Abstractly considered, the maximal factor $P_{UUU}$ has infinitely many integer roots. This can be seen by setting one of its arguments to one, in which case the polynomial factorises as
\begin{align}
P_{UUU}(N_1,N_2,1)
=\,&(N_1N_2-5)^2\,P_{UU}(N_1,N_2)^2\,
\big((N_1-N_2)^2(N_1N_2+11)-256)\big)^4\,.
\end{align}
Since $P_{UU}(N_1,N_2)$ has infinitely many integer roots, it follows immediately that the same applies to $P_{UUU}(N_1,N_2,1)$. However, when any of the group sizes equals one, the three-index theory degenerates to a simpler theory, and the beta functions must not be treated as independent objects. Therefore, theories with $N_1$, $N_2$, or $N_3$ equal to one do not properly belong to the class of theories we are studying in this section.

Physically, the allowed values of $N_1$, $N_2$, and $N_3$ are the integers greater than one. By explicit computer checks we find that $P_{UUU}(N_1,N_2,N_3)$ has no roots for $2 \leq N_1,N_2\leq 10\,000$. In other words, if there is a theory in the tri-unitary family with an extremal fixed point, at least two of the group dimensions must be greater than $10^4$. But such an extremal fixed point is highly unlikely to exist, as we will now argue.

Since $P_{UUU}(X,Y,Z)$ is completely symmetric in $X$, $Y$, and $Z$, to argue that it has no integer roots, it suffices to argue that it has no integer roots with $X\leq Y \leq Z$. Our line of reasoning is similar to the arguments adduced in Subsection~\ref{UOabsence}, but slightly more involved since we are now dealing with a polynomial in three variables.

Whenever one argument is parametrically larger than the others, or if all three arguments are parametrically large in the same parameter, then $P_{UUU}(X,Y,Z)$ has no roots.  This can be seen by introducing some large number $M$ and expanding,
\begin{align}
P_{UUU}(X,Y,MZ) &= X^{6}Y^{6}Z^{18}M^{18}\Big(1+\mathcal{O}(1/M)\Big)\,,
\\
P_{UUU}(MX,MY,MZ) &= X^{12}Y^{12}Z^{12}M^{36}\Big(1+\mathcal{O}(1/M)\Big)\,,
\end{align}
and observing that in each case the leading term is strictly positive. The only domain with large arguments where roots are possible is the domain with two arguments comparable to each other but parametrically large compared to the third argument. Indeed expanding $P_{UUU}(X,MY,MZ)$ in large $M$, the leading part, which scales as $M^{24}$, contains a number of terms of either sign.

Solving the equation $0=P_{UUU}(X,MY,MZ)$ for $X$ to leading order in $M$ produces eight asymptotic solution surfaces for real-valued $X$. Of these eight surfaces, three can immediately be discarded because for these surfaces $X$ does not lie in the positive range $X>2$ that is physically pertinent. We can then proceed to consider each of the remaining five solution surfaces in turn and estimate the probability to find integer roots on the surface.

As an example, one of these asymptotic solution surfaces is given by $X=X_s(Y,Z)$ with
\begin{align}
X_s(Y,Z) = (5-2\sqrt{6})\frac{Z}{Y}\,.
\end{align}
For integer-valued $Y$ and $Z$, $X_s$ is not an integer. But $X_s$ is the asymptotic value determined from only the leading terms of $P_{UUU}(X,MY,MZ)$ in $M$, and it is possible that $X$ is integer-valued on the solution surface for the full polynomial. For fixed integer values of $Y$ and $Z$, let $X^{\text{(int)}}$ be the nearest integer to $X_s$. Then $P_{UUU}(X^{\text{(int)}},Y,Z)$ is also integer-valued, and, treated as a random variable in $Y$ and $Z$, the range of possible values it can range over is roughly given by
\begin{align}
\label{UUUrange}
&\text{Range}\Big[P_{UUU}(X^{\text{(int)}},Y,Z)\Big] \nn
\\[5pt] 
& {}= 
\underset{\delta \in(-1/2,1/2)}{\text{max}}P_{UUU}(X_s+\delta,Y,Z)
\,\,\,-\underset{\delta \in(-1/2,1/2)}{\text{min}}P_{UUU}(X_s+\delta,Y,Z)\,,
\end{align}
and the probability $\text{Pr}_s(Y,Z)$ to have an integer root at $Y$ and $Z$ is roughly one over this range.\footnote{This probability assignment fails if $P_{UUU}(X,Y,Z)$ factorises at special values of any one of its arguments, in which case zeros become more frequent. For fixed values of $Z$ greater than two and up to $10\,000$, we have explicitly checked that $P_{UUU}(X,Y,Z)$ does not factorise, except for the case $Z=5$, where it factorises into $(XY-1)$ times a large polynomial. This slight factorisation does not appreciably change the likelihood of finding integer roots.} Let us use the letter $r$ to denote the ratio of $Z$ to $Y$, i.e.\ $r=Z/Y$. In order to have $X_s > 1$, we must impose $r\geq r_0 \equiv 1/(5-2\sqrt{6})  \approx 9.9$. To estimate the expected number of integer roots on the solution sheet for $Y$ greater than the value $Y_0=10^4$, we sum over the probabilities for each allowed value of $Y$ and $Z$:\footnote{In principle, we get a slight overestimate for $\left<n\right>_s$ by letting the sum over $Z$ extend out to infinity, since in that case we enter the domain where one argument of $P_{UUU}$ dominates the two others, so that no roots are possible.}
\begin{align}
\left<n\right>_s\sim \sum_{Y>Y_0}\sum_{Z\geq r_0Y}\text{Pr}_s(Y,Z)\,.
\end{align}
At the cost of a small relative error, we can convert the sum over $Z$ into an integral,
\begin{align}
\label{nsFormula}
\left<n\right>_s\sim \sum_{Y>Y_0}Y\int_{r_0}^\infty dr\,\text{Pr}_s(Y,rY)\,.
\end{align}
By a crude numerical estimate of the above expression we arrive at an estimate $\left<n\right>_s\sim 1\cdot 10^{-106}$. 

A more careful analysis than the one sketched here, would likely change the estimated value by a couple of orders of magnitude, but the expected value would still be extremely close to zero. And performing similar estimates for the other solution surfaces too yields minuscule values.\footnote{We get rough estimates of $1\cdot 10^{-100}$, $6\cdot 10^{-101}$, $3\cdot 10^{-95}$, and $1\cdot 10^{-102}$ for the expected number of integer roots on the other surfaces.} The smallness of $\left<n\right>_s$ in each case ultimately hinges on the high degree of the polynomial $P_{UUU}$. The high degree leads to a wide range of values the polynomial can range over and thereby to a tiny probability to have a root. For the probability scales as 
$\text{Pr}_s(Y,rY) \sim Y^{-24}$, which means that in performing the sum over $Y$ in \eqref{nsFormula}, we pick up a factor of
\begin{align}
\sum_{Y>Y_0} \frac{1}{Y^{23}}
=\zeta(23)-\sum_{Y=1}^{Y_0} \frac{1}{Y^{23}}
\end{align}
For $Y_0=10^4$, this sum evaluates to about $5\cdot 10^{-90}$. The remaining orders of magnitude that separate this number from the even smaller estimate quoted above, mostly hail from the large coefficients of $P_{UUU}$.

In conclusion, the expected number of integer roots assumes such a tiny value that the $U(N_1)\times U(N_2) \times U(N_3)$ familiy of theories almost certainly contains no new extremal fixed points.

\subsection{Trifundamental orthogonal theory: 
\texorpdfstring{$O(N_1)\times O(N_2)\times O(N_3)$}{O(N1) x O(N2) x O(N3)}
}
\label{triOrthogonal}

We now turn to the last family of quantum field theories that we study in this note: the quartic theories of three-index fields $\phi^{abc}$ where each index transforms under its own orthogonal group.\footnote{For recent results on tri-orthogonal theories, see Refs.~\cite{Giombi,benedetti2019line,bonzom2019diagrammatics,avohou2019counting,benedetti2020hints,benedetti2020conformal,benedetti2020s,benedetti2021trifundamental,benedetti2022f,bednyakov2021six,Jepsen:2023pzm}.} 

A general potential can be written as a sum involving five couplings
\begin{align}
8\,V(\phi) = {}& \lambda_0\, \phi_{\ra1 \bb1 \gc1} \phi_{\ra1 \bb1 \gc1} \phi_{\ra2 \bb2 \gc2}  \phi_{\ra2 \bb2 \gc2} 
+ \lambda_1\, \phi_{\ra1 \bb1 \gc1}  \phi_{\ra1 \bb1 \gc2} \phi_{\ra2 \bb2 \gc2}  \phi_{\ra2 \bb2 \gc1} \nonumber \\
&+\lambda_2\, \phi_{\ra1 \bb1 \gc1} \phi_{\ra1 \bb2 \gc1} \phi_{\ra2 \bb2 \gc2}  \phi_{\ra2 \bb1 \gc2} 
+\lambda_3 \,\phi_{\ra1 \bb1 \gc1}  \phi_{\ra2 \bb1 \gc1} \phi_{\ra2 \bb2 \gc2}  \phi_{\ra1 \bb2 \gc2} \nonumber\\
 & + \lambda_4\,  \phi_{\ra1 \bb1 \gc1}  \phi_{\ra1 \bb2 \gc2} \phi_{\ra2 \bb1 \gc2}  \phi_{\ra2 \bb2 \gc1} \ .
\end{align}
While we have tried to make the couplings clearer by color coding the indices, the reader may still prefer the graph depiction 
in  row five of Table~\ref{tab:operators}. The total number of real scalar fields is 
\be
N= N_1 N_2 N_3 \, .
\ee
The coupling $\lambda_0$ corresponds to the $O(N)$ symmetric theory while $\lambda_1,\lambda_2,\lambda_3$
are associated with the same graph and are related by a permutation symmetry, and $\lambda_4$ corresponds
to the fields attached to the vertices of a tetrahedron. The five beta functions are given at one-loop level by
\begin{align}
\beta_{0}=-\lambda_0+ {} &
(8 + N) \lambda_0^2 + 3 \lambda_1^2 + 3 \lambda_2^2  +   3 \lambda_3^2  + 
 2 (1 + N_1+N_2 N_3) \lambda_0 \lambda_1
\nn  \\ \nonumber &+ 
 2 (1 + N_2 + N_1 N_3) \lambda_0 \lambda_2 + 2 (1 + N_3 + N_1 N_2) \lambda_0 \lambda_3 \nn \\
 &{} + 
 2 N_1 \lambda_2 \lambda_3  + 2 N_2 \lambda_1 \lambda_3 
+ 2 N_3 \lambda_1 \lambda_2 + 2 (N_1 + N_2 + N_3) \lambda_0 \lambda_4 \nn \\
&{} +  
 2( \lambda_1   + \lambda_2  +   \lambda_3) \lambda_4\,,  \nn
\\[5pt] \nonumber
\beta_{1}=-\lambda_1+ {} &
12 \lambda_0 \lambda_1 + (4 + N_1+ N_2 N_3) \lambda_1^2 + 
 2 (1 + N_2) \lambda_1 \lambda_2 + 
 2 (1 + N_3) \lambda_1 \lambda_3 
  \\  &+
  4 \lambda_2 \lambda_3 +  4 (\lambda_2  +    \lambda_3) \lambda_4
+ 2 (N_2 + N_3) \lambda_1 \lambda_4 
 + (2 + N_1) \lambda_4^2 \,,
 \nn \\[5pt]
\beta_{2}=-\lambda_2+ {} & 
12 \lambda_0 \lambda_2+  (4 + N_2 + N_1 N_3) \lambda_2^2  + 
 2 (1 + N_1) \lambda_1 \lambda_2 +
 2 (1 + N_3) \lambda_2 \lambda_3  \nn \\
 &{} + 4 \lambda_1 \lambda_3  + 4( \lambda_1 + 
 \lambda_3 )\lambda_4  +  2 (N_1 + N_3) \lambda_2 \lambda_4 + (2 + N_2) \lambda_4^2
\,,
\nn \\[5pt]  \nonumber
\beta_{3}=-\lambda_3+ {} & 
12 \lambda_0 \lambda_3 + 
  (4 + N_3 + N_1 N_2) \lambda_3^2 +  
 2 (1 + N_1) \lambda_1 \lambda_3 + 
 2 (1 + N_2) \lambda_2 \lambda_3 
 \nn   \\ 
 &{}+ 4 \lambda_1 \lambda_2 + 4 (\lambda_1  +  \lambda_2) \lambda_4 + 
 2 (N_1 + N_2) \lambda_3 \lambda_4 
 + (2 + N_3) \lambda_4^2
\,, \nn
\\[5pt] \nonumber
\beta_{4}=-\lambda_4+  {} &  12 \lambda_0 \lambda_4 
+ 4 \lambda_1 \lambda_2 + 4 \lambda_1 \lambda_3 +
 4 \lambda_2 \lambda_3  \nn \\
 &{} +2  \big ( (1 + N_1) \lambda_1 
 + (1 + N_2) \lambda_2    + (1 + N_3) \lambda_3 \big )  \lambda_4
\,.
\end{align}
The sum of squares of couplings is then
\begin{align}
||\lambda||^2={}& 3 N \Big(
 (2 + N) \lambda_0^2 
+   2 (1 + N_1 + N_2 N_3) \lambda_0 \lambda_1+   2 (1 + N_2 + N_1 N_3) \lambda_0 \lambda_2 
 \nn \\
 &
     +    2 (1 + N_3 + N_1 N_2) \lambda_0 \lambda_3
 +  2 (N_1 + N_2 + N_3) \lambda_0 \lambda_4 
\nn \\
&{}  
 + \tfrac12 (3 + N_1+ N_2 N_3 + N ) \lambda_1^2
+ \tfrac12(3 + N_2 + N_1 N_3 + N) \lambda_2^2 
\nn  \\ 
  & {}
  +   \tfrac12(3 + N_3 + N_1 N_2 + N) \lambda_3^2
+     (1 +  N_1 + N_2 + 2 N_3 + N_1 N_2) \lambda_1 \lambda_2  
\nn  \\
  & {}
   +    (1 + N_1 + 2 N_2 + N_3 + N_1 N_3) \lambda_1 \lambda_3 
 +  (1 + 2N_1 + N_2 +  N_3+ N_2 N_3 ) \lambda_2 \lambda_3 
 \nn  \\ 
  & {}
  +    (2 + N_2 + N_3 + N_1 N_2 + N_1 N_3) \lambda_1 \lambda_4 
\nn  \\  
  & {}
+  (2 + N_1 +N_3+ N_1 N_2  + N_2 N_3) \lambda_2 \lambda_4 
\nn  \\[-3pt]  
  & {}
+     (2 + N_1 + N_2+N_1 N_3  + N_2 N_3) \lambda_3 \lambda_4
 + \tfrac12  (2 + N_1 + N_2 + N_3 + N ) \lambda_4^2\Big)\,.
\end{align}
From the Gr\"obner basis associated to the beta functions and also the polynomial $8||\lambda ||^2-N$ we obtain the extremality condition $G_{OOO}=0$, with 
\begin{align}
\label{GOOO}
G_{OOO}={}& 
P_O(N_1N_2N_3)^3
\,P_{OO}(N_1N_2,N_3)
\,P_{OO}(N_1N_3,N_2)
\,P_{OO}(N_2N_3,N_1) \nn \\
&{}\times P_{OOO}(N_1,N_2,N_3)\,.
\end{align}
The maximal factor $P_{OOO}(N_1,N_2,N_3)$ is too long to write down, even in an appendix, but it is given explicitly in an ancillary Mathematica file. The polynomial is symmetric in its arguments and is of order 58 in each argument and of order 108 in products of the arguments. The polynomial has $85\,807$ terms, and the absolute values of the coefficients have a median value of approximately $9.3\cdot 10^{24}$ and a mean value of about $4.7\cdot 10^{45}$. 

\subsection{Likely absence of new extremal fixed points}

Arguments almost identical to those we adduced for the tri-unitary theories in Subsection~\ref{UUUabsence} can be made in this case in an attempt to conclude that the irreducible extremality polynomial for the tri-orthogonal theories also has no integer roots within the physically relevant range of arguments. But the kind of probabilistic arguments we presented assign equal probability for the polynomial to assume any value within varying bounds. A situation where this assumption certainly fails is when the polynomial factorises. In such circumstances composite values are more frequent than prime values, and zero values too become more frequent. And in fact there are special values where the irreducible extremality polynomial $P_{OOO}(N_1,N_2,N_3)$ does factorise, namely when one of the group sizes is two: 
\begin{align}
\label{POOO2}
&P_{OOO}(N_1,N_2,2)=
P_{UU}(N_1,N_2)\, P_{UO}(N_1,N_2) \, P_{UO}(N_2,N_1) 
\times \text{very large polynomial}\,,
\end{align}
where the very large polynomial contains 1596 terms, is of degree 62 in total and of degree 48 in each variable, with the absolute values of the coefficients having a mean, median, and maximum of about $1.2\cdot 10^{37}$, $2.7 \cdot 10^{25}$, and $1.3 \cdot 10^{39}$.
Since $P_{2}(N_1,N_2)$ has infinitely many integer roots, so too does $P_{OOO}(N_1,N_2,N_3)$. These roots do correspond to true, physical, extremal fixed points, but they do not furnish new examples. The extremal fixed points signified by the roots of $P_{OOO}(N_1,N_2,2)$ due to zeroes in the first three factors in the right-hand side of \eqref{POOO2} are identical to the respective extremal fixed points in the bi-unitary and mixed bifundamental theories, with the value $N_3=2$ for the $O(N_1)\times O(N_2)\times O(N_3)$ 
model providing the doubling of fields necessary to relate real fields to complex fields. 

So we have encountered infinitely many integer roots, but none of them are new. They all correspond to extremal fixed points present already in bifundamental theories. Does $P_{OOO}(N_1,N_2,N_3)$ contain any additional roots, indicative of novel extremal fixed points? It appears unlikely. Extremal fixed points peculiar to the tri-orthogonal theories would correspond to roots of the ``very large polynomial" in \eqref{POOO2} or to roots where all three group sizes are greater than two. By an explicit computer check, sweeping over values of $N_1$ and $N_2$ and solving in each case for all integer values for $N_3$, we find that there are no such roots with $N_1,N_2\leq 1000$, meaning that at least two group sizes must exceed one thousand to have an extremal fixed point.\footnote{%
 We have also checked for negative values of $N_i$ in the chance the vanishing may suggest symplectic examples that saturate the bound.  We found no such examples.
}
 And we have also explicitly checked for the first one thousand values for $N_3$ greater than two that $P_{OOO}(N_1,N_2,N_3)$ does not factorise,\footnote{%
To be more precise, $P_{OOO}(N_1,N_2,7)$ actually does factorise into $(XY-1)^3$ times a very large polynomial, but this slight factorisation doesn't appreciably impact estimates of likely integer roots for arguments greater than one. There are  factorisations for other linear constraints on the $N_i$.  For example setting two of the $N_i$ equal, a ${\mathbb Z}_2$ permutation symmetry  can be imposed that reduces the number of couplings to four.  Although our search was far from exhaustive, we did not find any interesting factorisations with this approach. 
} 
and there is no reason to believe the polynomial factorises at yet higher values. This means we can now carry out the same type of analysis as in Subsection~\ref{UUUabsence}.

\sloppy As for the tri-unitary model, it can be checked that the leading terms in  $P_{OOO}(N_1,N_2,N_3)$ are all positive when one argument is parametrically larger than the others, or when all three arguments are large with respect to the same parameter:
\begin{align}
P_{OOO}(X,Y,MZ) &= X^{14}Y^{14}Z^{58}M^{58}
\Big(1+{\mathcal O}(1/M)\Big)\,,
\\
P_{OOO}(MX,MY,MZ) &=
6561X^{36}Y^{36}Z^{36}M^{108}\Big(
1+{\mathcal O}(1/M)
\Big)\,.
\end{align}
The asymptotic regime where integer roots are possible is again the regime where two arguments are parametrically large. Expanding $P_{OOO}(X,MY,MZ)$ at large $M$ produces leading terms that are sign indefinite and scale as $M^{72}$. Solving the equation $0=P_{OOO}(X,MY,MZ)$ for $X$ at large $M$ yields 16 real-valued asymptotic solution curves, but imposing $X>1$ and restricting ourselves without loss of generality to searching for integer roots with $Z\geq Y$, we are left with nine asymptotic solution surfaces. For each of these solution surfaces $s$, we can estimate the expected number of integer roots $\left<n\right>_s$ using formulas \eqref{UUUrange} and 
\eqref{nsFormula}, with $P_{OOO}$ substituted for $P_{UUU}$. Our estimates for $\left<n\right>_s$ are all less than $10^{-220}$, so we deem new extremal fixed points within this family of theories to be unlikely.\footnote{Specifically, the estimates for $\left<n\right>_s$ we arrive at for the seven solution surfaces are given by $6\cdot 10^{-249}$, $9\cdot 10^{-266}$, $1\cdot 10^{-250}$, $2\cdot 10^{-223}$, $1\cdot 10^{-259}$, $2\cdot 10^{-223}$, $2\cdot 10^{-225}$, $5\cdot 10^{-263}$, and $2\cdot 10^{-267}$, but again these values can easily change by a couple of orders of magnitudes depending on the details of how the estimation is carried out.} The specific minuteness of the estimates in this case hails from the fact that we assess the probability for an integer root Pr$_s(Y,Z=rY)$ at large $Y$ to scale as Pr$_s(Y,rY)\sim Y^{-72}$, so that, performing the double sum over $Y$ and $Z$, each estimate of $\left<n\right>_s$ is accompanied by a factor of 
\begin{align}
\sum_{Y>Y_0} \frac{1}{Y^{71}}
=\zeta(71)-\sum_{Y=1}^{Y_0} \frac{1}{Y^{71}}\,,
\end{align}
which, for $Y_0=1000$, comes out to about $1\cdot 10^{-212}$.

\section{Discussion}
\label{sec:Discussion}
We have introduced the term extremal fixed points to denote such RG fixed points in the $4-\epsilon$ expansion that saturate the Rychkov-Stergiou bound and in consequence undergo a saddle-node bifurcation and possess a marginal quartic deformation. A necessary condition to have such a fixed point is that the group sizes of the theory furnish an integer root to what we have dubbed the extremality polynomial. There is an odd mix of symmetry and statistics at play that govern which models have bifurcation nodes.
As the number of couplings in our examples grows, the degree of the extremality polynomial grows as well, and statistics
makes the appearance of these saturating fixed points very unlikely.
At the same time, the extremality polynomial always factors, and the factors of lower degree in our examples always correspond to theories with fewer couplings and more symmetry.  Thus symmetry can enforce the appearance of these special solutions even if statistical reasoning suggests the opposite. For symmetry induces correlations in the distribution of the values of the extremality polynomial, but statistical arguments that treat the distribution as pseudo-random are valid only when such correlations are absent and for polynomials that do not factorise further.

Let us briefly review our bestiary of examples. 

\begin{itemize}

\item
{\bf One coupling:} These theories have $O(N)$ symmetry.  The extremality polynomial is linear, and the particular case $N=4$ supports a saturating fixed point.

\item
{\bf Two couplings:} Our three examples with two couplings were the $O(N_1) \times O(N_2)$, $U(N_1) \times U(N_2)$, and
$Sp(N_1) \times Sp(N_2)$ bifundamental theories.  The extremality polynomial factored into a linear piece corresponding
to the $O(N)$ theories and a maximal factor of degree two in $N_1$ and $N_2$.  This maximal factor in each
case supported an infinite number of saturating solutions, in correspondence with integer solutions of the Pell equation.

\item
{\bf Three couplings:}  We looked at five examples that contain three couplings, $Sp(N_1) \times O(N_2)$, $U(N_1) \times O(N_2)$, and a generalisation of the $M\!N$-theories with symmetry group $Sp(m) \times {\mathcal S}_n$, $U(m) \times {\mathcal S}_n$, and $O(m) \times {\mathcal S}_n$.  
In these borderline cases, 
the extremality polynomial factored into lower degree pieces corresponding to the $O(N)$ and bifundamental
models just described and also a maximal factor of degree three or four.  In fact the maximal factor could be identified
as a restriction of the full model to a two-coupling system.
This maximal component did not allow for an infinite number of saturating solutions but sometimes afforded a few sporadic
examples.  The $Sp(m) \times {\mathcal S}_n$ could be analyzed using a birational transformation, where we found just four
saturating solutions.  The maximal factor for $U(m) \times {\mathcal S}_n$ and $Sp(N_1) \times O(N_2)$ were rank one elliptic curves.
They support an infinite number of rational solutions but only a finite number of integer solutions by Siegel's Theorem. 
In fact, none of these integer solutions were interesting for us, corresponding to degenerate small 
rank limits of the models.
The maximal factor of $O(m) \times {\mathcal S}_n$  was a rank zero elliptic curve and thus supported only a finite number
of rational solutions.  None were interesting physically however.
(Most elliptic curves have rank zero or one.)
Finally, the maximal factor of $U(N_1) \times O(N_2)$ was a genus two curve which remarkably supported a single new saturating solution, $U(2) \times O(34)$.

\item
{\bf Four and more couplings:}
We considered trifundamental unitary and orthogonal  theories with four and five couplings respectively. The extremality polynomial factored to give linear and quadratic pieces corresponding the $O(N)$ and bifundamental theories above. Additionally, the extremality polynomial for the tri-orthogonal $O(N_1)\times O(N_2)\times O(N_3)$ theories contained, when one of the group factors was $O(2)$, the extremal fixed points of the bi-unitary and mixed unitary-orthogonal bifundamental theories. But for neither trifundamental theory did the maximal factor support any interesting new saturating solutions. 

\end{itemize}

There are many possible quiver-type generalisations of the models considered here such as mixed tri-fundamental theories or theories with several bifundamental operators.  These theories are all likely to involve more couplings.  For instance,
as indicated in Figure~\ref{tab:operators}, mixed trifundamental  theories contain six or nine quartic singlets. 
As we see from the pattern above, the extremality polynomial will explode in complexity as the number of couplings grow.  
Indeed, our current computational techniques are unlikely to succeed with these examples.  It will simply take too long to compute the Gr\"obner basis.  
But in regards to these extremal fixed points, our prediction is that the pattern above will repeat.  The extremality polynomial will factor into a number of smaller degree pieces corresponding to previously studied cases.  The maximal factor will then be of such high degree, it is unlikely to have any integer solutions.  

That said, there is an intriguing possibility here that some of these more complicated models may have extremality polynomials with simple factors we have not come across before.   There might be cases where factorisation leads us to a new simpler theory, perhaps even with an infinite number of fixed points that saturate the bound.
A suggestive accident in our example set also encourages a more thorough exploration of these quiver type theories.
In all our examples where the maximal factor of the extremality polynomial was a plane curve of genus zero, integer solutions
were furnished by a variant of the Pell equation, $x^2 - 6 y^2 = c^2$ where $c$ depended on the details of the theory but the six was universal.  Are there models leading to a version of the Pell equation with some other number that is not six appears?  Is there reason why six appeared in our examples?  We do not know the answer to these questions. Moreover, the fact that we studied theories whose symmetry groups were given by direct products of smaller groups entailed that $N$ was a composite number. It may be interesting to investigate if there are extremal fixed points where $N$ is a prime number greater than five.

\begin{figure}
\begin{center}
\includegraphics[scale=0.65]{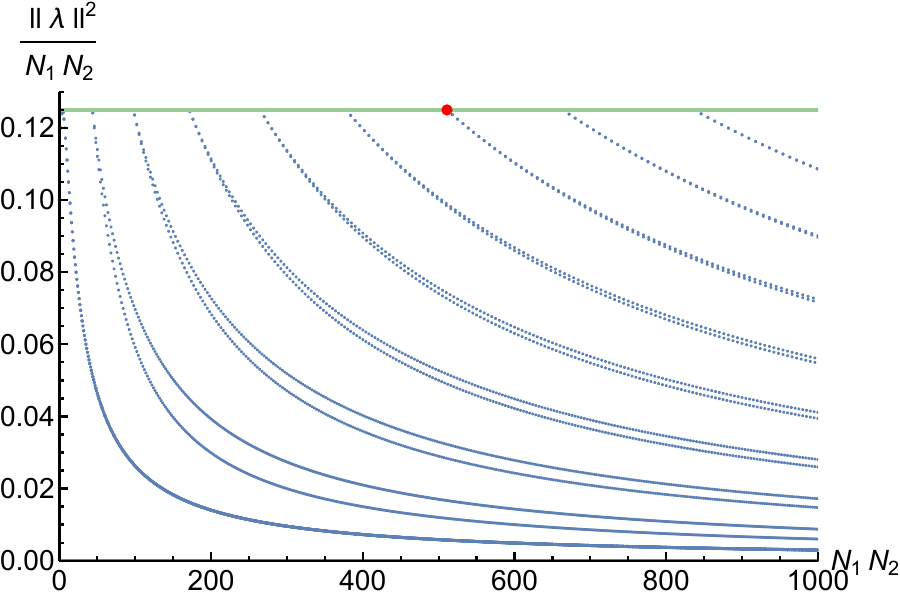}
\end{center}
\caption{
 The value of $\frac{|| \lambda ||^2}{N}$ for the $O(N_1) \times O(N_2)$ theories.
 The red dot is the (7,73) theory which sits on the bifurcation node.  Note however the existence of a number
 of other fixed points 
 which sit very close to bifurcation nodes.  For example there are a pair of fixed points with $N_1 =8$ and $N_2 = 83$ which
 have $\frac{|| \lambda ||^2}{N} = 0.124985908$ and $0.124985913$. 
\label{fig:bifurcation}
}
\end{figure}

We would be remiss without saying a few words about the physical significance or possible lack thereof of these extremal fixed points.
This extremality calculation is valid at leading order in the epsilon expansion.  Plausibly higher order corrections will make this
marginal operator slightly relevant or slightly irrelevant and move the theory away from the bifurcation point.\footnote{See ref.~\cite{Reehorst:2024vyq} for recent non-perturbative results on bifurcations in the $O(2)\times O(N)$ model using the bootstrap.}  The bifurcation point
will still formally exist, but for a critical value of $N_c$ that is shifted away from an integer by an epsilon 
dependent quantity.\footnote{%
In the case of the $O(N)$ model, the shift in the critical value at $N=4$ due to higher loop corrections can be computed to fifth order in $\epsilon$ through the use of the beta functions of ref.~\cite{Bednyakov:2021ojn}. We find that 
\begin{align*}
N_c =\, & 4-2\epsilon+5\frac{6\zeta(3)-1}{12}\epsilon^2
  - 
 5.87431 \epsilon^3 + 16.827 \epsilon^4 - 
 56.6219 \epsilon^5 + O(\epsilon^6)\,.
\end{align*}
}
The topology of the solutions is expected to remain however.  
One way to visualise the more general setting is to plot the value of 
$|| \lambda ||^2$ for all fixed points and all possible values of the integer parameters $N_i$.  For instance, figure \ref{fig:bifurcation} shows the result for the $O(N_1) \times O(N_2)$ theories where the $x$-axis is the product $N_1 N_2$.  
Each blue dot is the value of $||\lambda ||^2$ at a fixed point.  The plot reveals  a line of bifurcation nodes that happens to pass through integer points every once in a while.  Because the fixed points that are close to saturation happen for $N_1 \ll N_2$,  the plot separates into a number of
upside down v-shaped patterns, each pattern associated with a small integer value of $N_1$.  Despite the fact that we work to leading order in epsilon, the topological pattern made by the points in figure \ref{fig:bifurcation} is likely to persist for larger values of $\epsilon$.  

We hope that some of the number theory results and techniques that we used in this work may be useful more generally in looking at
renormalisation group flows in the epsilon expansion. 
The Pell equation and Siegel's Theorem \cite{Siegel} are not among the standard mathematical tools familiar to most physicists, and the methods of refs. \cite{stroeker1994solving,stroeker2003computing} have not been used in the physics literature before. 

Unlike, for example, the analysis in \cite{Witten:2007kt} where marginal deformations of the 3$d$ gravitational action are shown to be absent because the cosmological constant times Newton constant squared must take on discrete values, in our case there is no deep reason for the absence of extremality in theories with more than three operators. The presence of extremal fixed points is an occurrence by all means possible, but with the odds overwhelming stacked against by statistical properties of integer distributions.
The primary focus of this paper has been perturbation theory in the $\epsilon$ expansion, but we hope it may serve to illustrate a wider lesson. There are facts in physics which come about, not because underlying principles force them to be so, but merely by the preponderant weight of circumstance.

\section*{Acknowledgments}
We would like to thank Carl Frederik Nyberg Brodda, John Cremona, Netan Dogra, Peter Jossen, and James Newton for discussions and correspondence about integer solutions to Diophantine equations.
We thank Andy Stergiou for discussions during the early stages of this project and for confirming many of our results.
We also would like to thank Slava Rychkov for comments on the manuscript.
C.~H.\ was supported by a Wolfson Fellowship from the Royal Society and by the
U.K. Science and Technology Facilities Council Grant ST/P000258/1.
C.~B.~J. \ is supported by Korea Institute for Advanced Study (KIAS) Grant PG095901.
Y.~O. \ is supported in part by the Israel Science Foundation Excellence Center, the US-Israel Binational Science Foundation, and the Israel Ministry of Science.

\appendix

\section{Breaking apparent bounds}
\label{App:BoundBreaking}

We here accumulate a number of results pertaining to the eigenvalue equations for quadratic and quartic operators at fixed points in the $\epsilon$ expansion.

\subsection{Quadratic operators}

Consider a solution $v_{ij}$ to the eigenvalue equation for $\phi^2$:
\begin{align}
\mu\,v_{ij} = \lambda_{ijkl}\,v_{kl}\,,
\label{eq:muv}
\end{align}
where $v_{ij}$ is a symmetric matrix, which we take to have unit normalisation,
\begin{align}
\label{eq:vnorm}
v_{ij}\,v_{ij}=1\,.
\end{align}
If we contact \eqref{eq:muv} with $v_{ij}$, we get
\begin{align}
\label{eq:muApp}
\mu  = v_{ij}\,\lambda_{ijkl}\,v_{kl}\,.
\end{align}
Applying $\beta_{ijkl}=0$ (or equivalently $3\, \lambda \vee \lambda = \lambda$) to the right-hand side results in
\begin{align}
\mu = v_{ij}\,\lambda_{ijmn}\,\lambda_{klmn}\,v_{kl}
+v_{ij}\,\lambda_{ikmn}\,\lambda_{jlmn}\,v_{kl}
+v_{ij}\,\lambda_{ilmn}\,\lambda_{jkmn}\,v_{kl}\,.
\end{align}
Using \eqref{eq:muv} and \eqref{eq:vnorm}, the first term on the right-hand side is seen to equal $\mu^2$. The second and third terms are identical, and if we define
\begin{align}
\label{eq:kappaDef}
\kappa_{j\,lmn} = v_{ij}\,\lambda_{ilmn}\,,
\end{align}
then we arrive at the equation
\begin{align}
\label{eq:finalmu}
\mu = \mu^2+2\, \kappa_{j\,kmn}\,\kappa_{k\,jmn}\,.
\end{align}
From the definition \eqref{eq:kappaDef} we see that $\kappa$ is fully symmetric in its last three indices while the first index is distinct. For this reason, the second term on the right-hand side of \eqref{eq:finalmu} is not an absolute square and so it can in some cases be negative, entailing that $\mu$ can in fact be negative.

An instance where the apparent bound $\mu>0$ is violated occurs in the mixed bifundamental $U(N_1)\times O(N_2)$ theory for $N_1=3$ and $N_2=2$. For these values of $N_1$ and $N_2$, the beta functions given in equations \eqref{mixedBetas} with $\nu=2$ have a fixed point at
\begin{align}
\lambda= \frac{1}{44}\,,
\hspace{20mm}
g= \frac{1}{22}\,,
\hspace{20mm}
h= -\frac{1}{22}\,.
\end{align}
Equivalently, the fixed point is realised in the $O(N_1)\times O(N_2) \times O(N_3)$ model, whose beta functions are listed in Section~\ref{triOrthogonal}, for $N_1=3$ and $N_2=N_3=2$. In this model the fixed point is situated at
\begin{align}
\lambda_0=\frac{1}{44}\,,
\hspace{15mm}
\lambda_1=\frac{1}{11}\,,
\hspace{15mm}
\lambda_2=\lambda_3=0\,,
\hspace{15mm}
\lambda_4=-\frac{1}{11}\,.
\end{align}
The fixed point has a degenerate eigenvalue $\mu=-\frac{1}{22}$, and it can be explicitly verified that \eqref{eq:finalmu} is satisfied with $\kappa_{j\,kmn}\,\kappa_{k\,jmn}=-\frac{23}{968}$.

For another example, let us consider the extremal $U(2) \times O(34)$ theory. For this theory $N=136$ giving 9316 symmetric quadratic scalars.
The anomalous  dimensions and corresponding degeneracies under $O(34)$ and $U(2)$
are then 
\begin{align}
& \tfrac12 (1,1)\, , \ \  \tfrac16 (1,3)\, , \ \   \tfrac{20}{21} (1,6)\, , \ \
\tfrac{2}{21}(594,1)\, , \ \  \tfrac{2}{63}(594,3)\, , \ \ \tfrac{1}{126}(594,6)\, , \nn \\
& -\tfrac{1}{14}(561,1)\, , \ \  0(561,2)\, , \ \ -\tfrac{1}{42}(561,3)\, , 
\end{align}
where the symmetric representations of $O(34)$ and $U(2)$ have dimensions 1,594 and 1,3,6 
respectively and the antisymmetric representations have corresponding dimensions 561 and 1,2,3.
The total dimension is
\begin{align}
(594+1)(1+3+6) + 561(1+2+3) = 9316\,.
\end{align}
The dimensions for the eigenvalues thus all fit into representations of the symmetry group and there is just one singlet $\phi^2$, which is why we are confident this fixed point does not factorise. 
Like the previous example, this one too has negative eigenvalues.

\subsection{Quartic operators}

The anomalous dimensions of the quartic operators are the 
stability matrix eigenvalues $\{\kappa\}$, obtained at lowest order by solving \eqref{keig}.
We normalise $v_{ijkl}$ such that
\begin{align}
||v ||^2 = 1 \ .
% v_{ijkl}\,v_{ijkl}=1\,.
\end{align}
By contracting \eqref{keig} with $v_{ijkl}$, we find that
\begin{align}
\kappa+1=6\,v_{ijxy}\,\lambda_{xykl}\,v_{klij}\,.
\end{align}
Applying $\beta_{ijkl} = 0$ to this equation, we find that
\begin{align}
\kappa+1=
6\big(
v_{ijxy}\,\lambda_{xyzw}\,\lambda_{zwkl}\,v_{klij}
+2\,v_{ijxy}\,\lambda_{xkzw}\,\lambda_{zwyl}\,v_{klij}
\big)\,.
\end{align}
In diagrammatic form, clarifying the index contractions,
\begin{align}
\label{kappa1eq}
\kappa+1=
6
\begin{matrix}
\\[-12pt]
\text{
\scalebox{0.9}{
\begin{tikzpicture}
\draw [very thick] plot [smooth, tension=1] coordinates { (0,1) (0.5,1.1) (1,1)};
\draw [very thick] plot [smooth, tension=1] coordinates { (0,1) (0.5,0.9) (1,1)};
\draw [very thick] plot [smooth, tension=1] coordinates { (0,0) (0.5,0.1) (1,0)};
\draw [very thick] plot [smooth, tension=1] coordinates { (0,0) (0.5,-0.1) (1,0)};
\draw [very thick] plot [smooth, tension=1] coordinates { (0,0) (0.1,0.5) (0,1)};
\draw [very thick] plot [smooth, tension=1] coordinates { (0,0) (-0.1,0.5) (0,1)};
\draw [very thick] plot [smooth, tension=1] coordinates { (1,0) (1.1,0.5) (1,1)};
\draw [very thick] plot [smooth, tension=1] coordinates { (1,0) (0.9,0.5) (1,1)};
\filldraw[white] (0,0) circle (2.8pt);
\filldraw[white] (0,1) circle (2.8pt);
\filldraw[white] (1,0) circle (2.8pt);
\filldraw[white] (1,1) circle (2.8pt);
\filldraw[black] (0,0) circle (2pt);
\filldraw[gray] (0,1) circle (2pt);
\filldraw[black] (1,0) circle (2pt);
\filldraw[gray] (1,1) circle (2pt);
\node at (-0.2,-0.2)  {$\lambda $};
\node at (1.2,-0.2)  {$\lambda $};
\node at (-0.2,1.2)  {$v $};
\node at (1.2,1.2)  {$v $};
\end{tikzpicture}
}}
\end{matrix}
+
12
\begin{matrix}
\\[-12pt]
\text{
\scalebox{0.9}{
\begin{tikzpicture}
\draw [very thick] plot [smooth, tension=1] coordinates { (0,1) (0.5,1.1) (1,1)};
\draw [very thick] plot [smooth, tension=1] coordinates { (0,1) (0.5,0.9) (1,1)};
\draw [very thick] plot [smooth, tension=1] coordinates { (0,0) (0.5,0.1) (1,0)};
\draw [very thick] plot [smooth, tension=1] coordinates { (0,0) (0.5,-0.1) (1,0)};
\draw [very thick] (0,1)--(1,0);
\draw [very thick] (0,0)--(0,1);
\draw [very thick] (1,0)--(1,1);
\draw [very thick] (1,1)--(0,0);
\filldraw[white] (0,0) circle (2.8pt);
\filldraw[white] (0,1) circle (2.8pt);
\filldraw[white] (1,0) circle (2.8pt);
\filldraw[white] (1,1) circle (2.8pt);
\filldraw[black] (0,0) circle (2pt);
\filldraw[gray] (0,1) circle (2pt);
\filldraw[black] (1,0) circle (2pt);
\filldraw[gray] (1,1) circle (2pt);
\node at (-0.2,-0.2)  {$\lambda $};
\node at (1.2,-0.2)  {$\lambda $};
\node at (-0.2,1.2)  {$v $};
\node at (1.2,1.2)  {$v $};
\end{tikzpicture}
}}
\end{matrix}
\end{align}
The first diagram on the right-hand side is an absolute square and so it must be positive. But it is possible for the second term to be negative such that the total sum is negative and $\kappa$ is less than one.

Let us check lack of positivity in an example. Take the $U(N_1)\times O(N_2)$ model with $N_1=5$ and $N_2=2$. In this case the beta functions in \eqref{mixedBetas} with $\nu=2$ have a fixed point at
\begin{align}
\lambda= \frac{1}{108}\,,
\hspace{20mm}
g= \frac{1}{27}\,,
\hspace{20mm}
h= -\frac{1}{27}\,.
\end{align}
Equivalently, we find the same fixed point in the $O(N_1)\times O(N_2) \times O(N_3)$ model with $N_1=5$ and $N_2=N_3=2$. In this case, for the beta functions given in Section~\ref{triOrthogonal}, the fixed point is located at
\begin{align}
\lambda_0=\frac{1}{108}\,,
    \hspace{16mm}
\lambda_1=\frac{2}{27}\,,
\hspace{16mm}
\lambda_2=\lambda_3=0\,,
\hspace{16mm}
\lambda_4=-\frac{2}{27}\,.
\end{align}
At the fixed point there is a degenerate eigenvalue with $\kappa = -\frac{28}{27}$, as we show in the following subsection. Picking one of the eigenvectors with this eigenvalue, it can be seen that \eqref{kappa1eq} is satisfied as
\begin{align}
-\frac{28}{27}+1 = 6\cdot \frac{43}{8748}+12\cdot \left(-\frac{97}{17\,496}\right)\,.
\end{align}

\subsection{Stability matrix eigenvalues in bifundamental theories with
\texorpdfstring{$N_2=2$}{N2 equal 2}
and
\texorpdfstring{$N_1=n$}{N1 equal n}
}

Here we describe some results for the bifundamental theories discussed in Section~\ref{sec:UO}
for $N_2=2$
when the fixed points have an enhanced symmetry. 

Starting from the $U(n)\times O(2)$ case 
for  low $n$ there are four fixed points. Two are the trivial fixed point with all
couplings zero and  the $O(4n)$ fixed point when $h=g=0$, $\lambda_{H*} = 1/(4(n+2))$. Two more
arise for $h=-g$  when the potential takes the form in \eqref{Vbi}. The fixed points are given by
\be
(\lambda_{D*},g_{D*}) = \tfrac{1}{4(n+4)}(1,1)\, , \qquad (\lambda_{B*},g_{B*}) = \tfrac{1}{4(n^2+2)}(1,n-1)\, .
\ee
The first is a decoupled theory while the second is the nontrivial biconical fixed point obtained from \eqref{Vbi};
they coincide when $n=2$.
For $n> 2 ( 3 + \sqrt{6})\approx 10.9$ there is a fixed point with $h=g$ which corresponds to the 
$O(2n) \times O(2)$
theory with $\nu=1, \ N_2=2$ and $N_1 \to 2n$. When $n>2(5+3\sqrt{2})$ there are
fixed points with $h=0$ and $U(n)\times U(2)/U(1)$ symmetry  with $\nu=2, \ N_2=2$. 

The eigenvalues of the stability matrix at the fixed points are readily obtained. Focusing on fixed points with $\nu=2$, $N_2=2$, and $g = -h \neq 0$, the nontrivial eigenvectors have $\delta h = - \delta g$ or $(n+1)\delta h = (n-1) \delta g$. The eigenvalues obtained for these eigenvectors with $h=-g$ are
\begin{align}
 \kappa_D=  {}&   \tfrac{n-2}{n+4} \, , \hskip 1.8cm {\tilde \kappa}_D=    - \tfrac{n+2}{n+4} \, ,
\nn \\
\kappa_B =  {}& - \tfrac{(n-1)(n-2)}{n^2+2}  \, , \ \ {\tilde  \kappa}_B =   - \tfrac{(n-1)(n+2)}{n^2+2}   \, .
\end{align}
The first corresponds to two  decoupled theories  with $O(2n)$ symmetry with the potential at the
fixed point as in \eqref{Vbi} with $g=\lambda$.
The eigenvalue $\tilde \kappa_B$ is ${<-1}$ if $n>4$. For $n=5$,  ${\tilde  \kappa}_B = - \frac{28}{27}$.

For each of the additional fixed points which emerge for $n>2(5+3\sqrt{2})\approx 18.5 $ with $U(2)\times U(n)$ symmetry and $h=0$, there are two stability matrix eigenvalues corresponding to eigenvectors with $ \delta h = 0$ and another eigenvalue corresponding to an eigenvector with $(\delta \lambda, \delta g, \delta h) \sim (1, 1, -3)$.
The eigenvalues corresponding to the eigenvector $(1,1,-3)$ are given by $\partial \beta_h/\partial h$ at the two fixed points:
 \be
{\tilde  \kappa}_\pm=\tfrac{2(n-1)(n+2)}{(n+2)(2n-5) \,\pm \, 3 \sqrt{n^2-20n + 28}}\, .
 \ee
 For $n=19$, ${\tilde \kappa}_\pm$  take the rational values $\frac{14}{13}, \ \frac{21}{19}$ and are in general $>1$.
 For large $n$, ${\tilde \kappa}_+ \sim 1 + \frac{18}{n^2}, {\tilde \kappa}_- \sim 1 + \frac{3}{n}$.

For the $Sp(n) \times O(2)$ theory then
with the constraints $h=0$ or $h=\pm g$ there are just two couplings and the associated stability matrix
 has two eigenvalues. One is always $1$ and the other for each fixed point is
 \begin{align}
 h = 0 \; : \; \; \;
 \kappa_{1,\pm} = {}& \pm \tfrac{2(2n-1)\sqrt{n^2 -18\,n +21}}{(n+1)(4n-9) \,\pm \,3 \sqrt{n^2 -18\,n +21}}\, , \nn \\
  h = g \; : \; \; \;
 \kappa_{2,\pm} = {}& \pm \tfrac{2(2n-1)\sqrt{n^2 -6\,n + 3}}{(n+1)(4n-3) \,\pm \,3 \sqrt{n^2 -6\,n + 3}}\, ,\nn \\
   h = -g \; : \; \; \;
 \kappa_{3,\pm} = {}& \pm \tfrac{2(2n-1)\sqrt{n^2 -10\,n + 7}}{(n+1)(4n-5) \,\pm \,3 \sqrt{n^2 -10\,n + 7}} \, .
 \end{align}
However, in computing eigenvalues of the stability matrix for the full three coupling theory there is an additional eigenvector, which corresponds to breaking of the extended symmetry, and which for the three pairs of fixed points is oriented respectively along the directions in the space of couplings given by $(-3,1,5)$, $(3,-2n-1,2n-2)$, and $(-n,n,n-1)$.
 The associated eigenvalues with fixed point couplings and large $n$ behaviour are
 \begin{align}
 \tkappa_{1,\pm} = {}&  \tfrac{2(n+1)(2n-1)}{(n+1)(4n-9)\,\pm \,3 \sqrt{n^2 -18\,n +21}}\, , \qquad
 \tkappa_{1+} \sim 1 + \tfrac{1}{n}\, , \quad \tkappa_{1-} \sim 1 + \tfrac{5}{2n}  \, ,   \nn \\
 \tkappa_{2,\pm} = {}& - \tfrac{2(n+1)(2n-1)}{(n+1)(4n-3) \,\pm \,3 \sqrt{n^2 -6\,n + 3}}\, ,\quad \
 \tkappa_{2+} \sim -1 + \tfrac{1}{2n}\, , \ \ \tkappa_{2-} \sim - 1 - \tfrac{1}{n}  \, ,  \nn \\
 \tkappa_{3,\pm} = {}& - \tfrac{2(n+1)(2n-1)}{(n+1)(4n-5) \,\pm \,3 \sqrt{n^2 -10\,n + 7}}\, , 
 \quad 
 \tkappa_{3+} \sim -1 - \tfrac{9}{2n^2}\, , \ \ \tkappa_{3-} \sim - 1 - \tfrac{3}{2n}  \, .
 \end{align} 
 Except in one case the eigenvalues are outside the usual range $[-1,1]$.

\section{The case of 
\texorpdfstring{$Sp(m) \times {\mathbb Z}_2$}{Sp(m) x Z2}
symmetry}
\label{sec:closerlook}

We take a closer look at a theory with two fields, $A^a$ and $B^b$, 
which form vectors over the quaternions so each  transforms in a fundamental representation of $Sp(m)$.  
There are several quartic monomials consistent with the $Sp(m)$ symmetry which add together to give the potential
\begin{align}
8\, V (A,B) = {}& \lambda_1 (A^a \overline A_a)  (A^b \overline A_b)  
+ \lambda_2 (B^a \overline B_a)  (B^b \overline B_b)
+2 \lambda_3  (A^a \overline A_a)  (B^b \overline B_b) \nn \\
& {} +  \lambda_4 \big [ (A^a \overline B_a) (B^b \overline A_b) + (B^a \overline A_a) (A^b \overline B_b)\big ]
\nn \\[-9pt]
\\[-9pt] \nn
&{} + \lambda_5 \left[  (A^a \overline B_a) (A^b \overline B_b)  +  (B^a \overline A_a) (B^b \overline A_b)  \right] \nonumber \\
& {}  + \lambda_6  (A^a \overline A_a) \big ( (A^b \overline B_b)+  (B^b \overline A_b) \big ) 
+ \lambda_7 (B^a \overline B_a) \big ( (B^b \overline A_b)+ (A^b \overline B_b) \big ) \ .  \nn
\end{align}
For each coupling the field contributions are real. If $\lambda_1=\lambda_2$ and
$\lambda_5=\lambda_6 = \lambda_7 =0$ then this corresponds to \eqref{Vthree} with $n=2, \, \nu=4$
where $\lambda_3= \lambda, \ \lambda_4=g, \ \lambda_1 = \lambda+ g +h$.
If $\lambda_5=\lambda_6=\lambda_7=0$ the symmetry is enhanced to
$Sp(m)\times Sp(1)$.

The one-loop $\beta$-functions that arise from this potential are 
\begin{align}
    \beta_1 ={} & -\lambda _1 +4(m+2) \lambda _1{\!}^2 + 4m \lambda _3{\!}^2 + 
    4 \,\lambda _3 (2 \lambda _4-\lambda _5)
   +4\, \lambda _4^2+4\, \lambda _5^2-4 \,\lambda _4 \lambda _5 \nn \\
   & + 2(m+2)\lambda _6{\!}^2\ ,\nn  \\
   \beta_ 2 ={} & -\lambda _2+ 4(m+2) \lambda _2{\!}^2 + 4m \lambda _3{\!}^2 + 
    4 \,\lambda _3 (2 \lambda _4-\lambda _5)
   +4\, \lambda _4{\!}^2+4\, \lambda _5{\!}^2-4 \,\lambda _4 \lambda _5 \nn \\
   &{} + 2(m+2)\lambda _7^2\ , \nn \\
\beta_3 = {} & -\lambda _3+ 4 \lambda _3{\!}^2  + 2 (2m+1)(\lambda_1+\lambda _2) \lambda_3
+ 2(\lambda_1+\lambda _2) (2 \lambda _4- \lambda _5) \nn \\
&{} +4\,  \lambda _4{\!}^2+4 \, \lambda _5{\!}^2  - 4\, \lambda_4 \lambda_5 
+ \lambda_6{\!}^2+ \lambda _7{\!}^2   +2(m+1)  \lambda _6 \lambda _7  \ , \nn \\
  \beta_4 = {}& - \lambda _4+
  2 (\lambda _1+\lambda_2) \lambda _4+8 \lambda _3 \lambda _4+4(m-1) \lambda _4{\!}^2  
    + 4(m+1) \lambda _5{\!}^2 +4\, \lambda _4 \lambda _5 \nn \\
    &{} + \tfrac12(2m+3)(\lambda _6{\!}^2 +\lambda _7{\!}^2 ) + \lambda _6 \lambda_7\ , \nn \\
      \beta_5 = {}& - \lambda _5+
  2 (\lambda _1+\lambda_2) \lambda_{5}+8 \, \lambda _3 \lambda _5    
  + 4\, \lambda _5{\!}^2 +8m\, \lambda _4 \lambda _5 
    + \tfrac12(2m+3)(\lambda _6{\!}^2 +\lambda _7{\!}^2 ) + \lambda _6 \lambda_7\ ,\nn  \\
      \beta_6 ={}& -\lambda _6+4(m+2)( \lambda _1+\lambda_5) \lambda _6 
   +6( \lambda _3 \lambda_6 +  \lambda _4 \lambda _7)
   + 2(2m+1)( \lambda_3 \lambda _7  +\lambda _4 \lambda_6 ) \ , \nn \\
   \beta_7 ={}& -\lambda _7+4(m+2)( \lambda _2+\lambda_5) \lambda _7 
   +6( \lambda _3 \lambda_7 +  \lambda _4 \lambda _6)
   + 2(2m+1)( \lambda_3 \lambda _6  +\lambda _4 \lambda_7 ) \ . 
  \end{align}
The couplings are redundant since the field redefinition $A \to A \cos \theta  + B \sin \theta $ and 
$B \to B \cos \theta - A \sin \theta$
induces the transformation
\begin{align}
 &{\lambda} \to 
 { R_\theta}\,  {\lambda} \, , \quad
 \beta (\lambda )= {R_\theta}\,  {\beta}( R_\theta{\!}^{-1} { \lambda} )\, ,
 \\[5pt]
 &{\lambda} = 
\left ( \begin{smallmatrix} \lambda_1 \\  \lambda_2 \\  \lambda_3 \\ \lambda_4 \\  \lambda_5 \\ \lambda_6 \\ 
 \lambda_7 \end{smallmatrix} \right )   \, , \quad 
   {\beta} (\lambda) =
\left ( \begin{smallmatrix} \beta_1 \\  \beta_2 \\  \beta_3 \\ \beta_4 \\  \beta_5 \\ \beta_6 \\ 
 \beta_7 \end{smallmatrix} \right )   \, ,
 \end{align}
with $R_\theta$ defining a representation of $SO(2)$ given explicitly by 
 \be 
 R_\theta = \left(
 \begin{smallmatrix}
 c^4 &~~ s^4 & 2 c^2 s^2 & 2 c^2 s^2 & 2 c^2 s^2 & -2 c^3 s & -2 c s^3 \\
 s^4 &~~ c^4 & 2 c^2 s^2 & 2 c^2 s^2 & 2 c^2 s^2 & 2 c s^3 & 2 c^3 s \\
 c^2 s^2 & ~~ c^2 s^2 & c^4+s^4 & -2 c^2 s^2 & -2 c^2 s^2 & c s \left(c^2-s^2\right) & -c s
   \left(c^2-s^2\right) \\
 c^2 s^2 & ~~ c^2 s^2 & -2 c^2 s^2 & c^4+s^4 & -2 c^2 s^2 & c s \left(c^2-s^2\right) & -c s
   \left(c^2-s^2\right) \\
 c^2 s^2 & ~~ c^2 s^2 & -2 c^2 s^2 & -2 c^2 s^2 & c^4+s^4 & c s \left(c^2-s^2\right) & -c s
   \left(c^2-s^2\right) \\
 2 c^3 s & ~a -2 c s^3 & -2 c s \left(c^2-s^2\right) & -2 c s \left(c^2-s^2\right) & -2 c s
   \left(c^2-s^2\right) & c^2 \left(c^2-3 s^2\right) & s^2 \left(3 c^2-s^2\right) \\
 2 c s^3 & ~ -2 c^3 s & 2 c s \left(c^2-s^2\right) & 2 c s \left(c^2-s^2\right) & 2 c s
   \left(c^2-s^2\right) & s^2 \left(3 c^2-s^2\right) & c^2 \left(c^2-3 s^2\right) \\
\end{smallmatrix}
\right)
\ee
for $c = \cos \theta$ and $s = \sin \theta$.

In general fixed points lie on one-dimensional circles under the action of $R_\theta$. To obtain
unique solutions it is necessary to restrict the couplings. It is possible to set 
$R_\theta \lambda_{ 1} =R_\theta \lambda_{ 2}$ by taking $\tan2\theta = \frac{\lambda_1-\lambda_2}
{\lambda_6+\lambda_7}$. A consistent truncation is to set $\lambda_6=\lambda_7=0$. If $\lambda_1=\lambda_2$ then this 
is preserved by $R_\frac{\pi}{4}$ and so there can be equivalent
fixed points.

Fixing $\lambda_6=\lambda_7=0$, 
when $m=13$, there are then 14  fixed points. Two have just $\lambda_1$ or
$\lambda_2$  non zero and involve a decoupled free theory
while  12 have $\lambda_1=\lambda_2$. 
Of the 12 two are just the trivial Gaussian fixed point and the one corresponding to maximal $O(104)$
symmetry.  Of the rest there are 3  pairs of fixed points corresponding to $\lambda_*$ and 
$R_{\frac\pi 4}\lambda_*$ where one of each pair has $\lambda_5=0$ and
so these fixed points are just those considered in section \ref{sec:MultiConical} for $\nu =4$.
For the two physically equivalent fixed points corresponding to the bifurcation when $m=13$
\begin{align}
\lambda_{*1} =\frac{1}{252} \left ( \begin{smallmatrix} 1 \\  1 \\  1 \\ 5 \\  0 \\ 0\\ 0 \end{smallmatrix} \right ) \, ,
\qquad
 \lambda_{*2} = R_{\frac\pi 4} \lambda_{1*}=
 \frac{1}{504}\left ( \begin{smallmatrix} 7 \\  7 \\  -3 \\ 5 \\  -5 \\ 0\\ 0 \end{smallmatrix} \right ) \, , 
\end{align}
the Rychkov-Stergiou bound is saturated in each case.
The nonzero eigenvalues of the stability matrix and associated eigenvectors for the first fixed point are
\begin{align}
1, \left ( \begin{smallmatrix} 1 \\  1 \\  1 \\ 5 \\  0 \\ 0\\ 0 \end{smallmatrix} \right ) \, , \quad
\frac23,  \left ( \begin{smallmatrix} 0 \\  0 \\  0 \\ 0 \\  0 \\ 1\\ 1 \end{smallmatrix} \right ) \, , \quad
-\frac{11}{21},  \left ( \begin{smallmatrix} 1 \\  -1 \\  0 \\ 0 \\  0 \\ 0\\ 0 \end{smallmatrix} \right ) \, , \quad
\frac{10}{9},  \left ( \begin{smallmatrix} 0 \\  0 \\  0 \\ 1 \\  2 \\ 0\\ 0 \end{smallmatrix} \right ) \, , \quad
-\frac{10}{9},  \left ( \begin{smallmatrix} 24 \\  24 \\ - 25 \\ 1 \\  0 \\ 0\\ 0 \end{smallmatrix} \right ) \, ,
\label{eig1}
\end{align}
with in addition two zero modes
\be
0, \left ( \begin{smallmatrix} 0 \\  0 \\  0 \\ 0 \\  0 \\ 1\\ -1\end{smallmatrix} \right ) \, , \qquad
0, \left ( \begin{smallmatrix} 4 \\  4 \\  4 \\ -1 \\  0 \\ 0\\ 0 \end{smallmatrix} \right ) \, .
\label{eig2}
\ee
In \eqref{eig1} the eigenvalue 1 has an eigenvector proportional to the fixed point couplings, as expected.  
In \eqref{eig2} the first zero mode corresponds to an eigenvector $L \, \lambda_{*1}$ 
with 
\be 
L= \frac {\rm d} {{\rm d} \theta}R_\theta \Big |_{\theta =0} =
\left ( \begin{smallmatrix} 0 & ~ 0& ~ 0 &  ~ 0&~ 0 & -2& ~ 0 \\  0 & ~ 0 & ~ 0& ~0 & ~0 & ~ 0 & ~2 \\
 0& ~ 0&~ 0& ~0 & ~ 0& ~ 1& -1 \\ 0& ~ 0& ~0& ~0& ~0& ~ 1& -1 \\ 0&~  0& ~0&~ 0 & ~0& ~ 1& -1  \\
  2 & ~ 0 & -2& -2& -2& ~ 0 & ~ 0 \\  0& -2 & ~2&~ 2&~  2& ~ 0&~ 0  
  \end {smallmatrix} \right )\,,
\ee
corresponding to the generator for an infinitesimal $SO(2)$ rotation. 
The associated zero mode operator is the divergence of the vector current operator 
for the $O(2)$ symmetry which is broken at this fixed point. This is expected to pick up a positive anomalous dimension at order $\epsilon^2$ \cite{Rychkov:2018vya}.  

In \eqref{eig1} there are both eigenvalues greater than one and less than minus one.  The eigenvalue $\frac{10}{9} $ is associated with a  perturbation of the couplings
breaking the additional symmetry present when 
$\lambda_5=\lambda_6  = \lambda_7=0$, while  $-\frac{10}{9}$ is not. 

\clearpage

\section{Extremality polynomial for the
\texorpdfstring{$U(N_1)\times U(N_2)\times U(N_3)$}{U(N1) x U(N2) x U(N3)}
model}
\label{appendix}
In this appendix we list the largest factor $P_{UUU}(X,Y,Z)$ of the extremality polynomial $G_{UUU}(X,Y,Z)$, given in equation \eqref{GUUU}, for the tri-unitary $U(N_1)\times U(N_2)\times U(N_3)$ model, where we now denote the arguments as $X$, $Y$, and $Z$ rather than $N_1$, $N_2$, and $N_3$. 

Since the polynomial is rather lengthy but completely symmetric in its three arguments, we will reduce the amount of equation space by breaking $P_{UUU}(X,Y,Z)$ into pieces related by permutations of arguments:
\begin{align}
&\hspace{50mm}P_{UUU}(X,Y,Z)
=
\\[3pt]
&P_{UUU}^{(1)}(X,Y,Z)
+\Big(P_{UUU}^{(2)}(X,Y|Z)
+P_{UUU}^{(2)}(X,Z|Y)
+P_{UUU}^{(2)}(Y,Z|X)\Big)
\nonumber\\&\hspace{25mm}
+\Big(P_{UUU}^{(3)}(X|Y|Z)
+P_{UUU}^{(3)}(X|Z|Y)
+P_{UUU}^{(3)}(Y|X|Z)
\nonumber\\[3pt]
&\hspace{28mm}
+P_{UUU}^{(3)}(Y|Z|X)
+P_{UUU}^{(3)}(Z|X|Y)
+P_{UUU}^{(3)}(Z|Y|X)\Big)\,,
\nonumber
\end{align}
where the three functions appearing on the right-hand side are displayed below. 

The first piece, which is completely symmetric in its three arguments, is given by
\begin{align}
&\hspace{40mm}P_{UUU}^{(1)}(X,Y,Z)=
\\[5pt]
&\hspace{5mm}
62882616180736 - 59215250980864 X Y Z + 16690949652480 X^2 Y^2 Z^2 
\nonumber\\[5pt]&
- 
 2103340105728 X^3 Y^3 Z^3 + 189739388416 X^4 Y^4 Z^4 + 
 54572904128 X^5 Y^5 Z^5 
 \nonumber\\[5pt]&
 + 18464644993 X^6 Y^6 Z^6 + 
 2269544648 X^7 Y^7 Z^7 - 34291889 X^8 Y^8 Z^8 
\nonumber \\[5pt]&
 - 
 12324576 X^9 Y^9 Z^9 - 749169 X^{10} Y^{10} Z^{10} - 13720 X^{11} Y^{11} Z^{11} 
 +
  X^{12} Y^{12} Z^{12}\,.
\nonumber
\end{align}

The second type of term, which is unchanged under interchange of its first two arguments, is given on the following page.

\begin{align}
&\hspace{47mm}P_{UUU}^{(2)}(X,Y|Z)=
\\&
-3543666786304 X^2 Y^2 + 279916249088 X^4 Y^4 + 5407925760 X^6 Y^6 
 \nonumber\\&
+ 
 523346112 X^8 Y^8 
 + 875208 X^{10} Y^{10} + 225 X^{12} Y^{12} + 
 2386478235648 X^3 Y^3 Z 
  \nonumber\\&
 - 237898461184 X^5 Y^5 Z 
 - 1888280448 X^7 Y^7 Z + 87053856 X^9 Y^9 Z
  \nonumber\\&
 + 306402 X^{11} Y^{11} Z - 
 1299227607040 Z^2 
 - 180876017664 X^4 Y^4 Z^2 
  \nonumber\\&
 + 5840348864 X^6 Y^6 Z^2 - 1697080464 X^8 Y^8 Z^2 + 
 6668847 X^{10} Y^{10} Z^2 
  \nonumber\\&
 + 25572 X^{12} Y^{12} Z^2 + 
 4369206476800 X Y Z^3 + 136694684672 X^5 Y^5 Z^3
  \nonumber\\&
 + 10971858400 X^7 Y^7 Z^3 
 - 286399524 X^9 Y^9 Z^3 - 
 447192 X^{11} Y^{11} Z^3 
  \nonumber\\&
 + 266556407808 Z^4 - 
 1121030832128 X^2 Y^2 Z^4 
 - 47052522680 X^6 Y^6 Z^4 
  \nonumber\\&
 + 
 3208394447 X^8 Y^8 Z^4 - 26791324 X^{10} Y^{10} Z^4 - 
 59871 X^{12} Y^{12} Z^4
  \nonumber\\&
 - 166467731456 X Y Z^5 - 
 492850434048 X^3 Y^3 Z^5 - 13620305182 X^7 Y^7 Z^5 
  \nonumber\\&
 + 
 411404272 X^9 Y^9 Z^5 
 - 1002590 X^{11} Y^{11} Z^5 - 2684354560 Z^6 + 
 64354328576 X^2 Y^2 Z^6 
  \nonumber\\&
 + 17931770816 X^4 Y^4 Z^6 
 - 
 2014175732 X^8 Y^8 Z^6 + 31047311 X^{10} Y^{10} Z^6 
  \nonumber\\&
 + 9872 X^{12} Y^{12} Z^6 - 5880414208 X Y Z^7 
 - 15609079808 X^3 Y^3 Z^7 
  \nonumber\\&
 + 9784731520 X^5 Y^5 Z^7 - 170997924 X^9 Y^9 Z^7 + 
 1598304 X^{11} Y^{11} Z^7 
 + 268435456 Z^8 
   \nonumber\\&
 + 2413494272 X^2 Y^2 Z^8 
 + 
 1492153216 X^4 Y^4 Z^8 + 6316736876 X^6 Y^6 Z^8 
  \nonumber\\&
 - 6757576 X^{10} Y^{10} Z^8 + 32897 X^{12} Y^{12} Z^8 - 138412032 X Y Z^9 + 
 55175168 X^3 Y^3 Z^9 
   \nonumber\\&
 - 1162615456 X^5 Y^5 Z^9 + 
 925299234 X^7 Y^7 Z^9 - 159326 X^{11} Y^{11} Z^9 
   \nonumber\\&
 + 
 13803520 X^2 Y^2 Z^{10} 
 + 285972224 X^4 Y^4 Z^{10} - 
 168105119 X^6 Y^6 Z^{10}
   \nonumber\\&
 + 82075104 X^8 Y^8 Z^{10} + 
 740 X^{12} Y^{12} Z^{10} 
 - 18335744 X^3 Y^3 Z^{11} - 22344960 X^5 Y^5 Z^{11} 
   \nonumber\\&
 - 37787712 X^7 Y^7 Z^{11} + 3859036 X^9 Y^9 Z^{11} 
 + 409600 X^2 Y^2 Z^{12} + 
 4534400 X^4 Y^4 Z^{12} 
   \nonumber\\&
 - 4018216 X^6 Y^6 Z^{12} - 2404689 X^8 Y^8 Z^{12} 
 + 
 71780 X^{10} Y^{10} Z^{12} - 174080 X^3 Y^3 Z^{13} 
   \nonumber\\&
 + 358528 X^5 Y^5 Z^{13} - 
 1124958 X^7 Y^7 Z^{13} 
 - 78288 X^9 Y^9 Z^{13} - 30 X^{11} Y^{11} Z^{13} 
   \nonumber\\&
 + 
 960 X^4 Y^4 Z^{14} + 200481 X^6 Y^6 Z^{14} - 57332 X^8 Y^8 Z^{14} 
 + 
 303 X^{10} Y^{10} Z^{14}
   \nonumber\\&
 - 10880 X^5 Y^5 Z^{15} + 14984 X^7 Y^7 Z^{15} - 
 1060 X^9 Y^9 Z^{15} + 256 X^4 Y^4 Z^{16} 
  \nonumber\\&
 - 1172 X^6 Y^6 Z^{16} + 
 303 X^8 Y^8 Z^{16} + 32 X^5 Y^5 Z^{17} - 30 X^7 Y^7 Z^{17} + X^6 Y^6 Z^{18}\,.
 \nonumber
\end{align}

Finally the third type of term, without any symmetry, is given on the following page.

\begin{align}
&\hspace{47mm}P_{UUU}^{(3)}(X|Y|Z)=
\\&
-644760993792 Y^2 Z^4 + 210288017408 X Y^3 Z^5 + 
 11708399616 Y^2 Z^6 
  \nonumber\\&
  + 5206380544 Y^4 Z^6 
 + 40747871744 X^2 Y^4 Z^6 - 
 289259520 X Y^3 Z^7 
  \nonumber\\&
 - 279541760 X Y^5 Z^7 + 
 13805273216 X^3 Y^5 Z^7 
 - 125829120 Y^2 Z^8 
  \nonumber\\&
 - 241377280 Y^4 Z^8 - 
 5178665472 X^2 Y^4 Z^8 - 638974464 Y^6 Z^8 
 \nonumber\\&
 + 733591936 X^2 Y^6 Z^8 - 
 3143287208 X^4 Y^6 Z^8 + 379453440 X Y^3 Z^9 
  \nonumber\\&
 + 174346240 X Y^5 Z^9 
 + 
 359937152 X^3 Y^5 Z^9 + 13668480 X Y^7 Z^9
  \nonumber\\&
 - 339111936 X^3 Y^7 Z^9 - 
 451653090 X^5 Y^7 Z^9 
 - 10485760 Y^2 Z^{10} 
  \nonumber\\&
 - 192512 Y^4 Z^{10} - 
 102693376 X^2 Y^4 Z^{10} + 14474752 Y^6 Z^{10} 
 \nonumber\\&
 + 70111872 X^2 Y^6 Z^{10} + 
 113175056 X^4 Y^6 Z^{10} - 21162048 Y^8 Z^{10} 
  \nonumber\\&
 + 59467448 X^2 Y^8 Z^{10} 
 - 
 59284576 X^4 Y^8 Z^{10} - 78848136 X^6 Y^8 Z^{10} 
  \nonumber\\&
 + 4931584 X Y^3 Z^{11} - 
 9576448 X Y^5 Z^{11} 
 - 6370688 X^3 Y^5 Z^{11} 
  \nonumber\\&
 + 3290752 X Y^7 Z^{11} - 
 19976448 X^3 Y^7 Z^{11} 
 + 69221540 X^5 Y^7 Z^{11} 
   \nonumber\\&
 - 5500800 X Y^9 Z^{11} + 
 16163646 X^3 Y^9 Z^{11} - 12116798 X^5 Y^9 Z^{11} 
   \nonumber\\&
 - 
 5719780 X^7 Y^9 Z^{11} 
 + 262144 Y^4 Z^{12} - 188928 X^2 Y^4 Z^{12} - 
 219648 Y^6 Z^{12} 
   \nonumber\\&
 + 1509248 X^2 Y^6 Z^{12} - 6569200 X^4 Y^6 Z^{12} 
 + 
 337344 Y^8 Z^{12} - 448296 X^2 Y^8 Z^{12} 
  \nonumber\\&
 - 3605375 X^4 Y^8 Z^{12} + 
 8145584 X^6 Y^8 Z^{12} - 14040 Y^{10} Z^{12} 
 - 406692 X^2 Y^{10} Z^{12}
  \nonumber\\&
 + 
 1404972 X^4 Y^{10} Z^{12} - 968680 X^6 Y^{10} Z^{12} 
 - 
 319520 X^8 Y^{10} Z^{12} 
 - 51200 X Y^5 Z^{13} 
   \nonumber\\&
 + 409728 X^3 Y^5 Z^{13} + 
 11904 X Y^7 Z^{13} 
 + 238720 X^3 Y^7 Z^{13} - 238108 X^5 Y^7 Z^{13} 
 \nonumber\\&
 + 
 28416 X Y^9 Z^{13} + 13342 X^3 Y^9 Z^{13} - 467184 X^5 Y^9 Z^{13} + 
 636098 X^7 Y^9 Z^{13} 
  \nonumber\\&
 - 2430 X Y^{11} Z^{13} 
 - 14658 X^3 Y^{11} Z^{13} + 
 59876 X^5 Y^{11} Z^{13} - 44636 X^7 Y^{11} Z^{13} 
  \nonumber\\&
 - 5858 X^9 Y^{11} Z^{13} - 
 10240 X^2 Y^4 Z^{14} 
 - 4416 X^2 Y^6 Z^{14} - 38600 X^4 Y^6 Z^{14} 
  \nonumber\\&
 - 
 7824 X^2 Y^8 Z^{14} + 66512 X^4 Y^8 Z^{14} - 60332 X^6 Y^8 Z^{14} 
 + 
 2031 X^2 Y^{10} Z^{14} 
  \nonumber\\&
 - 3848 X^4 Y^{10} Z^{14} - 9119 X^6 Y^{10} Z^{14} + 
 14840 X^8 Y^{10} Z^{14} + 1536 X^3 Y^5 Z^{15} 
 \nonumber\\&
 - 2144 X^3 Y^7 Z^{15} - 
 1218 X^5 Y^7 Z^{15} - 228 X^3 Y^9 Z^{15} + 2462 X^5 Y^9 Z^{15} 
  \nonumber\\&
 - 
 2626 X^7 Y^9 Z^{15} 
 + 168 X^4 Y^6 Z^{16} - 113 X^4 Y^8 Z^{16} + 
 100 X^6 Y^8 Z^{16} 
 + 2 X^5 Y^7 Z^{17}\,.
 \nonumber
\end{align}

\noindent
\section{First few rational points of 
\texorpdfstring{$P_{SpO}(N_1, N_2)$}{PSpO(N1,N2)}
and
\texorpdfstring{$F_2(m,n)$}{F2(m,n)}
}
\label{sec:rationalpoints}

\noindent
Rational solutions of the elliptic curve (\ref{Weierstrassform}) with $n \leq 3$.  
 Only the torsion point $T$ is integral in the original $(N_1, N_2)$ coordinate system. 
\begin{figure}[H]
\[
\begin{array}{|c|c|c|c|}
\hline P = n P_1 + s T & (x,y) & (N_2, N_1)_+ & (N_2, N_1)_- \\
\hline
T  & (-34992,0) & (4,1) & - \\
\pm P_1 & (-11664, \pm 7558272) & ( \frac{8}{5}, \frac{7}{10} ) & ( 7,\frac{1}{4} ) \\
\pm P_1 + T & (151632 , \mp 60466176) & (-\frac{7}{5}, -\frac{4}{5} ) & (-\frac{1}{2}, -\frac{7}{2} )\\
\pm 2 P_1  & (28512, \mp 10450944) & ( -\frac{7274}{1153}, -\frac{758}{1153}) & ( \frac{1898}{1535}, \frac{3446}{1535} ) \\
\pm 2 P_1 +T & (\frac{1644624}{49}, \pm \frac{3869835264}{343}) & ( \frac{1516}{1153} , \frac{3637}{1153} ) & (-\frac{6892}{1535} , -\frac{949}{1535} )\\
\pm 3 P_1  & (\frac{176907888}{961}, \pm \frac{2388905239680}{29791}) & ( -\frac{7397383}{11167103} , -\frac{147645313}{44668412}) & (-\frac{73582856}{51728251} , -\frac{88857223}{103456502} ) \\
\pm 3 P_1  + T& (-\frac{136457136}{9025} ,\mp \frac{6236299994112}{857375} ) & (\frac{147645313}{22334206} , \frac{7397383}{22334206}) &
( \frac{88857223}{51728251} , \frac{36791428}{51728251}) \\
\hline
\end{array}
\]
\end{figure}

 \noindent
Rational solutions of the elliptic curve (\ref{weierstrass2}) with $q \leq 3$.  
 Only the point $-2P_1$ is integral in the original $(m,n)$ coordinate system.
 \begin{figure}[H]
\[
\begin{array}{|c|c|c|c|}
\hline P = q P_1 + s T & (u,v) & (n,m)_+ & (n,m)_- \\
\hline
T  & (720,0) & (\frac{1}{3}, -1) & - \\
\pm P_1 & (1368, \pm 31104) & \infty & \infty \\
\pm P_1 + T & (216 , \pm 24192) & (-\frac{59}{5}, \frac{2}{5} ) & ( \frac{4}{5}, \frac{2}{5} )\\
\pm 2 P_1  & (864, \mp 864) & (-\frac{193}{49}, -5 ) & (-1,-5)\\
\pm 2 P_1 +T & (-1548, \mp 13608) & ( -\frac{7867}{5589}, \frac{19}{23} ) & (-\frac{187}{69}, \frac{19}{23})\\
\pm 3 P_1  & (\frac{87768}{49}, \mp \frac{19875456}{343}) & (\frac{381853}{1164},\frac{146}{97}) & (\frac{412}{291},\frac{146}{97}) \\
\pm 3 P_1  + T& (\frac{3736}{9}, \mp \frac{445312}{27}) & (\frac{1149628}{1315277},\frac{50}{293}) & (-\frac{3071}{293},\frac{50}{293}) \\
\hline
\end{array}
\]
\end{figure}

\bibliographystyle{JHEP}
\bibliography{literature}

\end{document}